\documentclass[5p,final,authoryear,times]{elsarticle}
\usepackage[titletoc,toc,title]{appendix}
\usepackage{graphicx}
\usepackage{verbatim,times,rotating,float,setspace,natbib,amsmath}

\DeclareRobustCommand{\rchi}{{\mathpalette\irchi\relax}}
\newcommand{\irchi}[2]{\raisebox{\depth}{$#1\chi$}} 

\journal{NeuroImage}

\begin{document}

\begin{frontmatter}

\title{Promise and pitfalls of g-ratio estimation with MRI}

\author[1,2]{Jennifer S. W. Campbell\corref{*}}
\author[1]{Ilana R. Leppert}
\author[1]{Sridar Narayanan}
\author[1]{Mathieu Boudreau}
\author[2]{Tanguy Duval}
\author[2,5]{Julien Cohen-Adad}
\author[3]{G.~Bruce Pike}
\author[2,4]{Nikola~Stikov}

\address[1]{Montreal Neurological Institute, McGill University, Montreal, QC, Canada}
\address[2]{NeuroPoly Lab, Institute of Biomedical Engineering, Polytechnique Montr\'{e}al, Montr\'{e}al, QC, Canada}
\address[3]{University of Calgary, Calgary, AB, Canada}
\address[4]{Montreal Heart Institute, Universit\'{e} de Montr\'{e}al, Montr\'{e}al, QC, Canada}
\address[5]{Functional Neuroimaging Unit, CRIUGM, Universit\'{e} de Montr\'{e}al, Montr\'{e}al, QC, Canada}

\cortext[*]{jennifer.campbell@mcgill.ca}

\begin{abstract}

The fiber g-ratio is the ratio of the inner to the outer diameter of
the myelin sheath of a myelinated axon.  It has a limited dynamic
range in healthy white matter, as it is optimized for speed of signal
conduction, cellular energetics, and spatial constraints.  \emph{In
  vivo} imaging of the g-ratio in health and disease would greatly
increase our knowledge of the nervous system and our ability to
diagnose, monitor, and treat disease. MRI based g-ratio imaging was
first conceived in 2011, and expanded to be feasible in full brain
white matter with preliminary results in 2013.  This manuscript
reviews the growing g-ratio imaging literature and speculates on
future applications. It details the methodology for imaging the
g-ratio with MRI, and describes the known pitfalls and challenges in
doing so.

\end{abstract}

\end{frontmatter}

\section{Introduction}

The g-ratio is an explicit quantitative measure of the relative myelin
thickness of a myelinated axon, given by the ratio of the inner to the
outer diameter of the myelin sheath.  Both axon diameter and myelin
thickness contribute to neuronal conduction velocity, and given the
spatial constraints of the nervous system and cellular energetics, an
optimal g-ratio of roughly 0.6-0.8 arises
\citep{Rushton51,Waxman75,Chomiak2009a}.  Spatial constraints are more
stringent in the central nervous system (CNS), leading to higher
g-ratios than in peripheral nerve \citep{Chomiak2009a}.  Study of the
g-ratio \emph{in vivo} is interesting in the context of healthy
development, aging, learning, and disease progression and treatment. In
demyelinating diseases such as multiple sclerosis (MS), g-ratio
changes and axon loss occur, and the g-ratio changes can then
partially recover during the remyelination phase \citep{Albert07}.
The possibility that the g-ratio is dependent on gender during
development, driven by testosterone differences, has recently been
proposed \citep{Paus2009a} and investigated
\citep{Pesaresi2015a,Perrin09}.  Possible clinical ramifications of a
non-optimal g-ratio include ``disconnection'' syndromes such as
schizophrenia \citep{Paus2009a}, in which g-ratio differences have
been reported \citep{Uranova2001a,Du2014a}.

The g-ratio is expected to vary slightly in healthy neuronal
tissue. The relationship between axon size and myelin sheath thickness
is close to, but not exactly, linear \citep{Berthold83}, with the
nonlinearity more pronounced for larger axon size
\citep{Hildebrand78}, where the g-ratio is higher \citep{Graf1984}.  During
development, axon growth outpaces myelination, resulting in a
decreasing g-ratio as myelination catches up \citep{Schroder88}.
There is relatively little literature on the spatial variation of the
g-ratio in healthy tissue.  Values in the range 0.72-0.81 have been
reported in the CNS of small animals (mouse, rat, guinea pig, rabbit)
\citep{Benninger2006a,Duval2016c}.  Other primary pathology and
disorders may lead to an abnormal g-ratio.  These include
leukodystrophies and axonal changes, such as axonal swelling in
ischemia.

There are many outstanding questions in demyelinating disease that
could be best answered by imaging the g-ratio \emph{in vivo}.  For
example, in MS, disease progression is still the topic of active
research. Most histopathological data are from patients at the latest
stages of the disease.  Therapies designed to promote remyelination
are being developed to augment immunomodulatory and immunosuppressive
treatments.  Since remyelination of chronically demyelinated axons
would be neuroprotective, this may help slow progression in MS.
Detailed longitudinal study of the extent of remyelination can
therefore aid in choosing avenues for therapy.  While techniques exist
for measurement of the g-ratio \emph{ex vivo} \citep{Graf1984}, measurement of the
g-ratio \emph{in vivo} is an area of active research.

Many MR imaging contrasts are sensitive to changes in the
  g-ratio, in that they are sensitive to changes in, e.g., the total
  myelin content or total fiber content.  The purpose of the g-ratio
  imaging formulation is to decouple the fiber density from the
  g-ratio, such that a more complete and specific picture of the
  microstructural detail can be achieved.  In recent work
  \citep{Stikov2011,Stikov2015a}, it has been shown that the
  combination of an MRI marker that is sensitive to the myelin volume
  fraction (MVF) and an MRI marker that is sensitive to the
  intra-axonal volume fraction or axon volume fraction (AVF) is
  sufficient to compute a g-ratio for each voxel, i.e., an
  \emph{aggregate} g-ratio, without explicit estimation of axon
  diameter and myelin sheath thickness.  The aggregate g-ratio is a
  function of the ratio of the MVF to the AVF.  g-Ratio imaging does
  not involve the acquisition of a novel contrast, but is a specific
  computation using the parameters (MVF and AVF) extracted from
  existing contrasts.  The challenge then becomes how to estimate the
MVF and the AVF precisely and accurately with MRI.  The
g-ratio imaging framework, coupled with independent
microstructural measures such as axon diameter
\citep{Assaf2008a,Zhang2011a}, comprises the field of \emph{in vivo
  histology} of white matter.  The ultimate goal is to describe
microstructure in detail on a scale much finer than an imaging voxel,
as a distribution or single aggregate value for each the voxel.

\section{Methodology}

\subsection {\label{g-ratio} The g-ratio formulation}
\label{calculation}

\begin{figure}
 \begin{center}
   \includegraphics[width=0.5\textwidth]{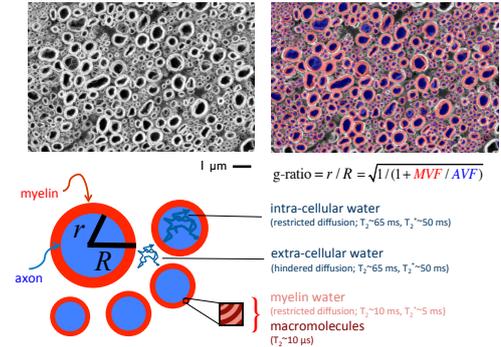}
   \caption{\label{gcartoon}Original (top left) and segmented (top
     right) scanning electron micrograph showing axons of white matter, the
     intra-axonal space (blue), and the myelin (red).  The segmentation
     was performed with AxonSeg \citep{Zaimi2016a}. In the SEM
     image, the myelin appears white because of osmium
     preparation. The fiber g-ratio is the ratio of the inner to the
     outer radius of the myelin sheath surrounding an axon.  The
     aggregate g-ratio can be expressed as a function of the myelin
     volume fraction (MVF) and the axon volume fraction (AVF).  The
     myelin macromolecules, myelin water, and intra- and extra-axonal
     water compartments all have distinct properties, which can be
     exploited to generate MRI images from which the respective
     compartment volume fractions can be estimated.}
\end{center}
\end{figure}

As previously defined, the g-ratio is the ratio of the inner to the
outer diameter of the myelin sheath of a myelinated axon (see
Fig. \ref{gcartoon}).  It has been shown in recent work
\citep{Stikov2011,Stikov2015a} that the aggregate g-ratio can be
expressed as a function of the myelin volume fraction and the axon
volume fraction, and hence can be estimated without explicit
measurement of these diameters:

\begin{equation}\label{geq}
g=\sqrt{\frac{1}{1+MVF/AVF}}.
\end{equation}

This formulation applies to any imaging modality (e.g., electron
microscopy (EM) and scanning electron microscopy (SEM), where the MVF
and AVF can be measured after segmentation of the image - see
Fig.\ref{gcartoon}), but it is of particular interest to be able to
estimate the g-ratio \emph{in vivo}.  MRI provides us with several
different contrast mechanisms for estimation of these volume
fractions, and given MVF$_{MRI}$ and AVF$_{MRI}$, we can estimate
g$_{MRI}$.  Hereon, we sometimes refer to g$_{MRI}$ as ``the
  g-ratio'', but note that it is derived from MRI images with certain
  contrasts sensitive but not equivalent to the MVF and AVF. We forego
  the subscript MRI on the MVF and AVF acronyms, but it should be
  clear from context when these are MR estimates. Estimation of these
quantities is discussed in the next sections.

\subsection {\label{AVF} Axon volume fraction}

Diffusion MRI is particularly well suited to aid in the estimation of
the axon volume fraction.  As we will see later, it cannot estimate the
absolute AVF, because there is very little signal from the myelin in a
typical diffusion MRI experiment.  However, the relative AVF can be
estimated, because diffusion MRI is sensitive to the displacement
distribution of water molecules moving randomly with thermal energy,
and this displacement distribution is affected by the cellular
structure present in the tissue.  As the molecules impinge on the
cellular membranes, organelles, and cytoskeleton, the displacement
distribution takes on a unique shape depending on the environment.
Intra-axonal diffusion is said to be \emph{restricted}, resembling
free Gaussian diffusion at short diffusion times, but departing
markedly from Gaussianity at longer times, where the displacement
distribution is limited by the pore shape. There will be a sharp drop
in the probability of displacement beyond the cell radius.
Extra-axonal diffusion is said to be \emph{hindered}, resembling free
Gaussian diffusion, but with a smaller variance due to impingement of
motion.

Many diffusion models exist for explicit estimation of the relative
cellular compartment sizes.  These include neurite orientation density
and dispersion imaging (NODDI) \citep{Zhang2012a}, the composite
hindered and restricted model of diffusion (CHARMED) \citep{Assaf05a},
diffusion basis spectrum imaging (DBSI) \citep{Wang2011a}, restriction
spectrum imaging (RSI) \citep{White2013a}, white matter tract integrity
(WMTI) from diffusion kurtosis imaging (DKI) \citep{Fieremans2011a},
temporal diffusion spectroscopy \citep{Xu2014a},  microscopic
anisotropy obtainable from multiple pulsed field gradient MRI
\citep{Shemesh2012a,Zhou13a,avram2013a}, the spherical mean technique
\citep{Kaden2016a}, the distribution of anisotropic microstructural
environments in diffusion-compartment imaging (DIAMOND)
\citep{Scherrer2016a}, and many others
\citep{Rokem2015a,Stanisz1997a}.  It is also possible to perform NODDI
with relaxed constraints (NODDIDA) \citep{Jelescu2016a}, and to do
this calculation analytically (LEMONADE) \citep{Novikov2016a}.

Another approach, termed the apparent fiber density (AFD), uses high
diffusion weighting to virtually eliminate the hindered diffusion
signal, leaving only intra-axonal water \citep{Raffelt2012a}. It has
been used to estimate the relative axon volume fraction of
different fiber populations in a voxel.  A modification, termed the
tensor fiber density (TFD), can be performed with lower diffusion
weighting \citep{Reisert2013a}.

The simplest diffusion MRI models do not differentiate between the
tissue compartments.  For instance, the diffusion tensor
\citep{Basser1994a} models the entire displacement distribution as an
anisotropic Gaussian function. The parameters defining this function
will change if the intra-axonal volume fraction changes, but to what
extent is it practical to extract meaningful quantitative compartment
volume fractions from the tensor?  Recently, a framework called
NODDI-DTI has been developed, in which the proximity of DTI-based
parameters to the computed NODDI parameters is assessed
\citep{Edwards2016a}, given certain assumptions.  The FA and the mean
diffusivity (MD) are highly correlated in straight, parallel fiber
bundles, and will change with changing AVF, leading to estimates of
the relative intra-axonal volume.  However, this formulation is
probably an oversimplification of the microstructural situation, and
more detailed modeling is a better choice to ensure specificity to
white matter fibers.

All of the diffusion models above are potential candidates for use in
g-ratio imaging.  They have strengths and weaknessess that will not be
detailed here. The original full-brain g$_{MRI}$ demonstration
\citep{Stikov2015a} employed the NODDI model of diffusion.  It was
chosen because of its suitability in the presence of complex subvoxel
fiber geometry, including fiber divergence, which may occur to a
significant scale in almost all imaging voxels \citep{Ghosh2016a}, and
its suitability on clinical scanners with relatively low gradient
strength.  Having a fast implementation of the model fitting with
numerical stability is important for large studies, hence, the
convex-optimized AMICO implementation is beneficial
\citep{Daducci2015a}.

While diffusion MRI is a modality of choice for imaging
microstructure, it can only measure the displacement distribution of
water molecules that are visible in a diffusion MRI experiment.  This
limits us to water that is visible at an echo time (TE) on the order
of 50-100 ms, and therefore excludes water that is trapped between the
myelin bilayers, which has a T$_2$ on the order of 10 ms.  Hence,
the estimates provided by these models are of the intra-axonal volume
fraction \emph{of the diffusion visible volume}.  Myelin does not
figure in the models.  Given, e.g., the NODDI model outputs, a
complementary myelin imaging technique must be used to estimate the
absolute axon volume fraction. The AVF is given by

\begin{equation}
 AVF=(1-MVF)(1-v_{iso})v_{ic}\,,                  
\label{AVFfrNODDI}
\end{equation}

\noindent where $v_{iso}$ and $v_{ic}$ are the isotropic and restricted
volume fractions from the NODDI model, and the MVF is obtained from one of many possible
myelin mapping techniques, examples of which are discussed below.

Diffusion contrast may not be our only window onto the axon volume
fraction.  Recent work has shown that it is possible to disambiguate
the myelin, intra-axonal, and extra-axonal water compartments using
complex gradient echo (GRE) images \citep{Sati2013a,Wu2016a}.  The myelin
water is separable from the combined intra- and extra-axonal water
using multicomponent T$_2^*$ reconstruction, providing a myelin marker
(see below).  However, incorporation of the phase of the GRE images
potentially allows us to separate all three compartments based on frequency
shifts.  Challenges include the fact that the frequency shift is
dependent on the orientation of the axon to the main magnetic field
B$_0$.  When the axon is oriented perpendicular to B$_0$,
the myelin water will experience a positive frequency shift, the
intra-axonal water a negative frequency shift, and the extra-axonal
water will not experience a frequency shift.

Note that the AVF as defined by these diffusion MRI models is specific
to white matter.  While it makes sense to define an axon volume
fraction in grey matter, the models in general cannot distinguish
between axons and dendrites.  The NODDI model's $v_{ic}$ parameter,
for example, is ``neurite density'', i.e., the density of all cellular processes that
can be assumed to have infinitely restricted diffusion in their
transverse plane.  Hence, g$_{MRI}$ from such MRI data is undefined in grey matter.

\subsection {\label{FVF} Fiber volume fraction}

The fiber volume fraction (FVF), or fiber density, is the sum of the
AVF and the MVF.  Can diffusion MRI, or any other MRI contrast
mechanism, measure the total fiber volume fraction itself?  Clearly,
GRE images have potential, as discussed above.  Is diffusion imaging
sensitive to the FVF, as opposed to the AVF?  While myelin water is
virtually invisible in diffusion MRI, diffusion MRI is not insensitive
to myelin.  First, the ratio of the intra- to extra-axonal diffusion
MRI visible water in a voxel will change as the myelin volume fraction
in that voxel changes.  In early work on the fiber g-ratio, it
  was shown that assuming a simple white matter model of straight,
  parallel cylinders, the fractional anisotropy (FA) of the diffusion
  tensor, which weighs both intra-axonal and extra-axonal anisotropy,
  is proportional to the total fiber volume fraction, with a quadratic
  relationship \citep{Stikov2011}. The NODDI parameter $v_{ic}$ also
  changes with demyelination, even if all axons remain intact. 
Second, diffusion acquisitions are heavily T$_2$ weighted, and T$_2$
is myelin-sensitive.  The total diffusion weighted signal thus
decreases as myelin content increases.  However, to robustly quantify
myelin volume fraction, it is necessary to add a second contrast
mechanism, even if it is additional T$_2$ weighted images, to the
scanning protocol.  This is discussed below.

Despite the nomenclature, as noted above, even the Apparent Fiber
Density and Tensor Fiber Density are in fact relative axon densities.
They would provide a relative FVF only if the g-ratio is constant.  In
a recent study of the g-ratio \citep{Mohammadi2015a}, the TFD was
equated with the FVF, not the AVF, for input to the g-ratio formula.
The g-ratio is a function of the ratio of the MVF to the AVF
(Eq. \ref{geq}), or alternately, the ratio of the MVF to the FVF:

  \begin{equation}\label{geqFVF}
g=\sqrt{1-MVF/FVF}.
\end{equation}

\noindent This means that conclusions reached about the variation of
the g-ratio found by equating the TFD with the FVF will be robust in
the above case.  Absolute g-ratios in the above case were calibrated to have a
mean of 0.7 in healthy white matter.

We note that if diffusion MRI were capable of estimating the absolute
FVF (or AVF) as well as the ratio of the intra-axonal to extra-fiber water, the
g-ratio could immediately be estimated from these two quantities,
without further myelin imaging.  This has yet to be done robustly, and
it is therefore preferable to use a more robust independent myelin
marker.

\subsection {\label{MVF} Myelin volume fraction}

There are many different contrasts and computed parameters that are
sensitive to myelin \citep{Laule2007review}.  The possible sources of
signal from the myelin compartment are the ultra-short T$_2$ protons
in the macromolecules of the myelin sheath itself (T$_2\sim10 \mu s$)
and the short T$_2$ water protons present between the phospholipid
bilayers (T$_2\sim10 ms$, see Fig. \ref{gcartoon}).  Most MRI contrast
mechanisms are sensitive to myelin content, but few are specific, for
reasons that are detailed in section \ref{cal}.  The myelin
phospholipid bilayers create local Larmor frequency variations for
water protons in their vicinity due to diamagnetic susceptibility
effects.  This results in myelin content modulated transverse
relaxation times T$_2^*$ \citep{Hwang2010a} and T$_2$, and
longitudinal relaxation time T$_1$.  It has been shown that
  macromolecular content is the dominant source of variance in T$_1$ in the brain
  \citep{rooney2007a}.  The local Larmor field shift (fL) and the
susceptibility itself ($\rchi$) can be computed as well
\citep{Liu2015a}. Ultra-short TE (UTE) imaging can be used to
image the protons tightly bound to macromolecules 
  \citep{Wilhelm2012a,Du2014a}.  An alternate approach to isolating
the myelin compartment is magnetization transfer (MT) imaging, where
the ultra-short T$_2$ macromolecular proton pool size can be estimated
by transfer of magnetization to the observable water pool.

MT based parameters sensitive to macromolecular protons include the
magnetization transfer ratio (MTR) \citep{Wolff1991a}, the MT
saturation index (MT$_{sat}$) \citep{Helms2008,Helms2010erratum}, the
macromolecular pool size (F) from quantitative magnetization transfer
\citep{Sled2001a,Yarnykh2002a,Ramani2002a}, single-point two-pool
modeling \citep{Yarnykh2012a}, and inhomogeneous MT
\citep{Varma2015a}.

Alternately, the myelin water can be imaged with quantitative
multicomponent T$_2$ \citep{MacKay1994a} or T$_2^*$
\citep{Du2007a,Alonso2016a} relaxation, which yields the myelin water
fraction (MWF) surrogate for myelin density.  Variants include
gradient and spin echo (GRASE) MWF imaging
\citep{Does2000a,Prasloski2012a}, linear combination myelin imaging
       \citep{JonesCK2004a,Vidarsson2005a}, T$_2$ prepared MWF
      imaging \citep{Oh2006a}, multi-component driven equilibrium
      single point estimation of T$_2$ (mcDESPOT) \citep{Deoni2008a},
      direct visualization of the short transverse relaxation time
      component via an inversion recovery preparation to reduce long
      T$_1$ signal (ViSTa) \citep{Oh2015a},  and a fast adiabatic
        T$_2$-prep and spiral readout approach (FAST-T2)
        \citep{Nguyen2016a}.  Other alternate approaches exploiting
      myelin-modulated relaxation times include combined contrast
      imaging (T$_{1W}$/T$_{2W}$) \citep{Glasser2011a} or independent
      component analysis \citep{Mangeat2015a}.  Proton density is also
      sensitive to macromolecular content, and the proton-density
      based macromolecular tissue volume (MTV) \citep{Mezer2013a} has
      been used as a quantitative myelin marker.

While these MRI measures have been shown to correlate highly with
myelin content
\citep{Thiessen2013a,Schmierer2007a,Gareau2000a,Laule2006a,Mottershead2003a},
they have not been incorporated in a specific tissue model in a manner
similar to the diffusion signal, and hence some calibration is
needed. This is still a topic of research.  Caveats of improper
calibration of the MVF are discussed in section \ref{cal}.

\vspace{1 cm}

 Above, we have discussed imaging techniques for both
 diffusion-visible microstructure and myelin.  Any multi-modal modal
 imaging protocol with contrasts such as these, sensitive to the axon
 and myelin volume fractions, is sensitive to the g-ratio (e.g.,
 \citep{Molina-Romero2016a,Nossin-Manor2015a,DeSantis16a,Bells2011a}).
       The purpose of the explicit g-ratio formulation is to
         create a measure that is \emph{specific} to the g-ratio.  It
         provides us with a metric that is independent of the fiber
         density, which none of the MRI contrasts sensitive to the MVF
         and AVF separately accomplish. It is interesting to ask
       whether we could use a technique such as deep learning
         \citep{Bengioreview2009} to estimate the g-ratio, skipping
       explicit modeling completely.

{
 \subsection {\label{acq} MRI acquisition protocol: an example}

Given the large number of MRI contrasts that could potentially be used
in the g-ratio imaging framework, we cannot prescribe an exact
protocol for g-ratio imaging here.  However, in the following
sections, we illustrate several important points about g-ratio imaging
using experimental data acquired at our site.  The following describes
the acquisition protocol we have used.

We acquired data from healthy volunteers and from multiple sclerosis
patients.  These data were acquired on a Siemens 3T Trio MRI scanner
with a 32 channel head coil.  A T$_{1W}$ structural MPRAGE volume with
1 mm isotropic voxel size was acquired for all subjects.  For
diffusion imaging, the voxel size was 2 mm isotropic.
For most experiments, the NODDI diffusion protocol consisted of 7 b=0
s/mm$^2$, 30 b=700 s/mm$^2$, and 64 b=2000 s/mm$^2$ images, 3x slice
acceleration, 2x GRAPPA acceleration, all acquired twice with AP-PA
phase encode reversal.  For the other experiments, as detailed below
when they are introduced, the slice acceleration and phase encode
reversal were not employed.  For a dataset optimized for diffusion
tensor reconstruction, a dataset with 99 diffusion encoding directions
at b=1000 s/mm$^2$  and 9 b=0 s/mm$^2$ images was acquired.

For magnetization transfer images, we also used 2 mm iso-tropic voxels
to match the diffusion imaging voxel size. For MTR, one 3D
non-selective PD-weighted RF-spoiled gradient echo (SPGR) scan was
acquired with TR=30 ms and excitation flip angle $\alpha=5^{\circ}$,
and one MT-weighted scan was acquired with the same parameters and an
MT pulse with 2.2 kHz frequency offset and 540$^{\circ}$ MT pulse flip
angle.  For MT$_{sat}$ computation, these same MT-on and MT-off scans
were used, with one additional T$_1$-weighted scan with TR=11 ms and
excitation flip angle $\alpha=15^{\circ}$.  For qMT computation,
10-point logarithmic sampling of the z-spectrum from 0.433-17.235 kHz
frequency offset was acquired, with two MT pulse flip angles for each
point, 426$^{\circ}$ and 142$^{\circ}$, and excitation flip angle
$\alpha=4.5^{\circ}$.  The qMT acquisition was accelerated with 2x
GRAPPA acceleration.  Additional scans for correction of the maps
included B$_1$ field mapping using the double angle technique
 \citep{Boudreau2017a,Stollberger1996a}, with 60$^{\circ}$ and
120$^{\circ}$ flip angles, B$_0$ field mapping using the two-point
phase difference technique, with TE$_1$/TE$_2$ = 4.0/8.48 ms, and
T$_1$ mapping using the variable flip angle technique
\citep{Fram1987a}, with flip angles 3$^{\circ}$ and 20$^{\circ}$.
Additional T$_2$-FLAIR and PD$_W$ images were acquired for
the MS subjects to aid in lesion segmentation.

\section{The pitfalls: outstanding technical challenges}

In this section, we discuss pitfalls and outstanding issues in g-ratio
imaging.  These include the challenge of combining multiple different
contrast mechanisms, limitations in diffusion modeling, specificity
and calibration of myelin markers, and the limitation of estimating a
single g-ratio metric in the presence of subvoxel heterogeneity.
Experimental results are included in these sections to illustrate
these problems.  In the experiments, we focus on the NODDI and DTI
models and MT-based myelin imaging because they were used in the early
g-ratio imaging publications \citep{Stikov2011,Stikov2015a}, but many
of these concerns apply to any chosen AVF and MVF markers.

\subsection{Confounds in multi-modal image acquisition}

The computation of the g-ratio metric includes several pre-processing steps,
including distortion and field inhomogeneity correction, that deserve
further discussion.  The MT-based contrasts are acquired with
spin-warp acquisition trains, and the diffusion-based contrasts are
acquired with single-shot EPI. When any acquisition details are
changed, the distortions in the images change, and co-registration of
voxels for voxelwise quantitative computations becomes more difficult.

The blip-up blip-down phase encode strategy (section \ref{acq}) allows
for precise correction of susceptibility-induced distortion in the
diffusion images \citep{Andersson2003a}.  Lack of correction for this
distortion leads to visible bands of artifactually high g$_{MRI}$ near
tissue-CSF interfaces (see, e.g, \citep{Campbell14a,Cercignani2016b}).
This was illustrated by Mohammadi \emph{et al.} \citep{Mohammadi2015a} (see
Fig. \ref{Siawoosh_misreg}). Uncorrected diffusion MRI data leads to
g-ratios in the vicinity of unity at the edge of the genu of the
corpus callosum, caused by voxels where the AVF is artifactually high
(containing little or no CSF), and the MVF low (because the correctly
localized voxels actually contain CSF).  The white matter - CSF
boundary is a region of obvious misregistration, but much of the
frontal lobe suffers from susceptibility induced distortion, and would
therefore have incorrect g-ratios.

\begin{figure}
  \begin{center}
   \includegraphics[width=0.5\textwidth]{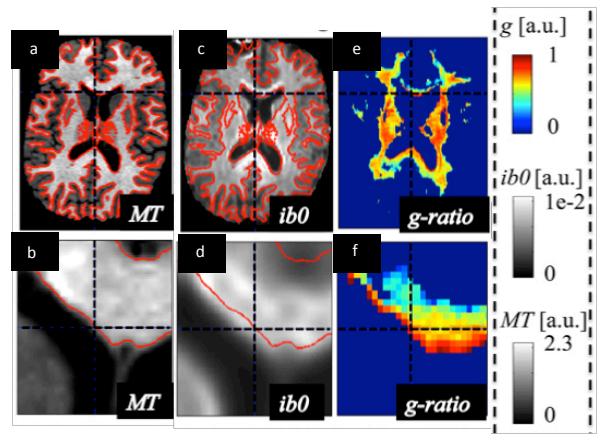}
   \caption{\label{Siawoosh_misreg} Misregistration artifact due to
     susceptibility-induced distortion in diffusion weighted images.
     At left is an MT$_{sat}$ image with white matter outlined in red,
     for one slice (a) and a cropped region at the genu (b).  In the
     center (c,d) is an original EPI diffusion scan with no diffusion
     weighting and contrast inverted (ib0).  The misregistration with
     the MT$_{sat}$-defined white matter boundary is marked.  At right
     (e,f) is the g-ratio computed with these contrasts. Uncorrected
     diffusion MRI data leads to g-ratios in the vicinity of unity at
     the edge of the genu, caused by voxels where the AVF is
     artifactually high (containing little or no CSF), and the MVF low
     (because the correctly localized voxels actually contain CSF).
     Reproduced from \citep{Mohammadi2015a}. }
  \end{center}
\end{figure}

Multi-modal imaging protocols are a powerful tool for investigation of
microstructure.  We have thus far discussed combining multiple images
with partially orthogonal contrasts in order to estimate the g-ratio.
However, problems such as the above misregistration issue arise.  Can
a single acquisition train provide multiple contrasts?  One such
approach was described recently for simultaneous mapping of myelin
content and diffusion parameters \citep{DeSantis16a}.  It consists of
an inversion-recovery preparation before a diffusion weighted
sequence, allowing for fitting of a model that incorporates both T$_1$
(a myelin marker \citep{Stuber2014a}) and axonal attributes.  This
approach is conceptually extensible to other myelin-sensitive
preparations or modifications of a diffusion weighted sequence, such
as quantitative T$_2$ estimation \citep{Kim2016a} or MT preparation
\citep{Gupta2003a}.

 Can the g-ratio be estimated using a single contrast mechanism?  This
 could also offer inherent co-registration, as well as possibly
 increasing the acquisition speed.  As discussed in section \ref{AVF},
 analysis of complex GRE images may lead us to a technique for
 estimating the MVF and AVF volume fractions from one set of images.

 Making a multi-modal imaging protocol short enough for the study of
 patient populations and use in the clinic is a considerable
 challenge.  In section \ref{cal}, we investigate the MT saturation
 \citep{Helms2008} as a more time efficient replacement for qMT.  There
 exist other short MT-based approaches, such as single-point two-pool
 modeling \citep{Yarnykh2012a} and inhomogeneous MT
 \citep{Varma2015a}.  Another approach could be to use compressed
 sensing \citep{Lustig2007a} for MT-based acquisitions
   \citep{McLean2017a}.  Recent advances in myelin water fraction
   imaging \citep{Nguyen2016a} may make MWF estimation more
   efficient, and GRE-based myelin water fraction approaches
 \citep{Alonso2016a} may also offer a faster approach for estimating
 the MWF, with the possibility, as mentioned above, to eliminate the
 diffusion imaging part of the protocol.

  Diffusion imaging has benefited from many acceleration approaches in
  recent years, including parallel imaging, which can also be used in
  the myelin mapping protocols, slice multiplexing
  \citep{setsompop12a}, and hardware advances such as the CONNECTOM
  gradient system.

\subsection{Diffusion modeling}

\subsubsection{Model parameters}
\label{params}

The diffusion MRI post-processing techniques described in section
\ref{AVF} give a range of outputs.  Some are physical quantities (such
as the diffusion displacement distribution; kurtosis), while some are
parameters of detailed biological models (such as the intra-axonal
volume fraction). Models are valuable, but the user has to be aware of
the assumptions made.

The parameter space in existing models ranges from three free
parameters in NODDI to six \citep{Assaf05a}, twenty three
\citep{Jespersen2010a}, and thirty one \citep{Novikov2016a,Jelescu2015a} in
other models,  with the maximum dependent on acquisition details.
Recent analysis hypothesizes that the lower number of free parameters
in, e.g., NODDI and CHARMED, may be matched to the level of complexity
possible on current clinical systems \citep{Ferizi2015a}, while high
gradient strength, high b-values, and more b-shells may be necessary
for more complex models \citep{Jelescu2015a}, and would make them more
optimal.  This is a general problem with multi-exponential models when
diffusion weighting is weak \citep{Kiselev2007a}. On standard MR
systems, relaxing the constraints on fixed parameters has been shown
to lead to degeneracy of solutions \citep{Jelescu2016a}.
Regularization approaches such as the spherical mean technique (SMT)
\citep{Kaden2016a} have been employed in an attempt to make the
problem less ill-posed.

One of the fixed parameters in the NODDI model is the parallel
diffusivity in the intra- and extra-axonal space, both set to the same
fixed value.  Other models explicitly model these as unequal; for
instance, WMTI assumes that the intra-axonal diffusivity is less than
or equal to the extra-axonal diffusivity.  The actual values are
unknown, however simulations have shown that the assumption of equal
parallel diffusivities leads to a 34-53\% overestimation of the
intra-axonal compartment size if the diffusivities are in fact unequal
\citep{Jelescu2015a}, with the intra-axonal diffusivity either greater
than or less than the extra-axonal diffusivity.  If the fixed
diffusivities are incorrect, the intra-axonal compartment size
estimated by the NODDI model will be non-zero even if there is no
anisotropy \citep{Lampinen2017a}.

Independent of whether the intra- and extra-axonal parallel
diffusivities are equal, another source of this bias is the tortuosity
model \citep{Szafer1995a} employed by many models, including NODDI,
DIAMOND, and the SMT.  This model computes the perpendicular
extra-axonal diffusivity as a function of the diffusion-visible
intra-axonal volume fraction of the non-CSF tissue (v$_{ic}$ in the
NODDI model). This tortuosity estimate is bound to be inaccurate
because the tortuosity is expected to vary as the absolute fiber
volume fraction of the non-CSF tissue, not the diffusion-visible fiber
volume fraction \citep{Jelescu2015a}.  These two quantities are
very different, as the myelin and axon volume fractions are almost
equal in healthy tissue \citep{Stikov2015a}.  The MVF could be
explicitly included in the equation, and would be expected to result
in a myelin volume dependent reduction in $v_{ic}$.  However, in
healthy tissue, where the FVF should scale roughly as the v$_{ic}$
parameter, the tortuosity model of Szafer {\emph et al.} does not
appear to hold when applied to experimental data with independent
estimates of the parallel and perpendicular extra-axonal diffusivities
\citep{Novikov2016a}.  The model is probably not correct for varying
packing density on the sub-voxel scale \citep{Novikov2011a}; it has
been shown to depend on the packing arrangement and breaks down for
tight axon packing \citep{Fieremans2008a,Novikov2012a}.  This might
explain the discrepancy between model and experiment mentioned above,
because the geometry of axonal packing can vary considerably for a
given average volume fraction.

Another fixed parameter in the NODDI model is the T$_2$ relaxation time
of all tissue, assumed to be the same, even in CSF. This leads to an
overestimation of v$_{iso}$, which can be corrected
\citep{Bouyagoub2016a} given T$_2$ estimates from, e.g., a T$_2$ mapping
technique such as mcDESPOT \citep{Deoni2008a}.

\subsubsection{Geometry in diffusion imaging}

Diffusion MRI is exquisitely sensitive to fiber geometry.  The
fractional anisotropy may be more sensitive to geometry than to any
microstructural feature \citep{Hutchinson2016a}.  Hence,
microstructural models must be careful to take geometry (crossing,
splaying, curving, microscopic packing configuration) into account.  A
typical diffusion imaging voxel is roughly 8 mm$^3$, while the axons
probed by microstructural models are on the order of one micron.

Early work on the fiber g-ratio investigated the human corpus
callosum, where it was assumed that the white matter fibers were
effectively straight and parallel \citep{Stikov2011}.  The use of this
model that assumes straight, parallel fibers suffers from several
problems.  First, the regions of the brain where this model can be expected to hold at
all are very limited, as there are crossing or splaying fibers in up
to 95\% of diffusion MRI voxels in parenchyma
\citep{Jeurissen2010a,Behrens07a,Ghosh2016a}, and curvature is almost
ubiquitous at standard imaging resolution.  Even the axons of callosal
fibers are not straight and parallel, with splay up to 18$^{\circ}$
\citep{Ronen2014a,Mollink2016a}.  Second, the model assumes a
relatively uniform, if random, packing of axons on the scale of the
MRI voxel.  Due to the nonlinear nature of the FA, it will depend
strongly on the packing geometry.  If two voxels, one with densely
packed axons and one with sparsely packed axons, are combined into
one, the FA for that voxel will be less than the average of the two
original voxels, whereas the fiber density will be the average of the
fiber densities.  Third, FA is in practice acquisition and b-value
dependent.

\noindent {\bf\emph{3.2.2.1 Experiments: AVF from NODDI in the presence of crossing fibers}}

The NODDI model that has been used in several g-ratio imaging studies
to date assumes there is a single fiber population with potential
splay or curvature, but does not explicitly model crossings.  To what
extent does the fiber dispersion model of NODDI handle crossing fiber
bundles?  We have employed the diffusion MRI simulator dSim
\citep{Sveinsson2011a} to investigate this question
\citep{Campbell14a}.  We simulated realistic axonal packing
\citep{Aboitiz1992a} in voxels with straight, parallel fibers and with
two equal size bundles of straight fibers crossing at 90$^{\circ}$
(See Fig.\ref{sim}).  Fiber volume fractions were set equal for both
configurations and were varied from 0.3 to 0.7.  g-Ratios were varied
from 0.7 to 0.9.  The diffusion weighted signal was generated, and the
NODDI model parameters computed using the NODDI Matlab toolbox
\citep{NODDIurl}.  The FVF was computed from the NODDI parameters
using the known MVF.  The computed FVF was $3.8\pm0.3\%$ lower in the
crossing fiber case for the NODDI-based FVF.  This demonstrates that
the NODDI model, while not explicitly designed for crossing fibers,
gives acceptable results in this case, and can be used for full-brain
g-ratio estimation at standard voxel size with significant subvoxel
fiber crossing, with only a small decrease in the estimated FVF due to
partial volume averaging of fiber orientations.

\begin{figure}
 \begin{center}
 \includegraphics[width=0.5\textwidth]{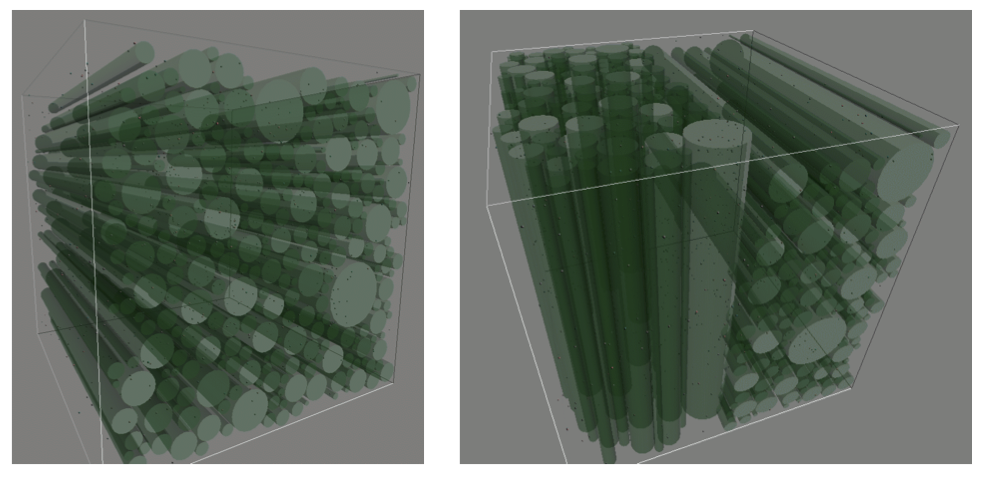}
 
 \caption{\label{sim}Simulated fibers in straight, parallel
   configuration (left) vs. crossing (right), with equal fiber volume
   fraction and similar distributions of axon diameter and position.
   The NODDI model underestimates the FVF by $3.8\pm0.3\%$ in the
   crossing fiber case, whereas the DTI model underestimates the FVF
   by $58.7\pm3.2\%$.}
 \end{center}
 \end{figure}

\noindent {\bf\emph{3.2.2.2 Experiments: Comparison of DTI and NODDI for FVF estimation}}

\label{DTIvsNODDI}

NODDI works optimally with diffusion MRI measurements made on at least
two shells in q-space, i.e., two different nonzero b-values, although
recent work has proposed solutions for single shell data, at least
where certain assumptions can be made about the tissue, or where high
b-values are used \citep{GrussuISMRM14,MagnollayISMRM14}.  In
contrast, the diffusion tensor can be robustly fitted and the fiber
volume fraction inferred \citep{Stikov2011} (see section \ref{FVF})
using a much more sparsely sampled, single shell dataset. Many
research programs have large databases of single-shell diffusion data,
often with limited angular sampling of q-space as well. It is
therefore of interest to explore to what extent such data, using the
diffusion tensor model, can be used in investigation of the g-ratio.

In the simulations described above, we also computed the diffusion
tensor using in-house software \citep{mincdiffusionurl}. The FVF
(FVF$_{DTI}$) was computed from the FA using the quadratic
relationship determined from previous simulations \citep{Stikov2011}.
As expected, the FA is not a predictor of FVF in the presence of
crossing fibers: the computed DTI-based FVF was $58.7\pm3.2\%$ lower
in the crossing fiber case compared to the parallel fiber case.

To compare NODDI and DTI \emph{in vivo}, diffusion and qMT data were
acquired as described in section \ref{acq} for one healthy volunteer,
without slice acceleration or phase encode reversal. The qMT data were
processed with in-house software and the NODDI parameters as described
above.  Additionally, the diffusion tensor was calculated using the
b=1000 s/mm$^2$ diffusion shell.  The AVF, MVF, and g$_{MRI}$ were
computed voxelwise from the diffusion and qMT data as described in
section \ref{calculation}: the NODDI-based FVF (FVF$_{NODDI}$) is the
sum of the MVF and the AVF computed from Eq. \ref{AVFfrNODDI}, and
FVF$_{DTI}$ was calculated from the fractional anisotropy of the
diffusion tensor using the quadratic relationship \citep{Stikov2011}.
The corpus callosum was skeletonized on the FA image and a voxel-wise
correlation between the FVF computed from DTI and from NODDI and qMT
was performed for these voxels.  The coefficient of proportionality
between F and MVF was determined from previous EM histological
analysis \citep{Stikov2015a}.

Fig. \ref{correlationscatter} shows the FVF computed using both NODDI
and the FA from DTI in the skeleton of the healthy human corpus callosum.  The
Pearson correlation coefficient between the FVF measured using the two
techniques was r=0.79, with
$\mbox{FVF}_{\mbox{NODDI}}=1.15*\mbox{FVF}_{\mbox{DTI}}+0.00$.  This
indicates a slight discrepancy between the FVF using NODDI compared to
DTI, and a reasonably high correlation between techniques on the
skeleton.

\begin{figure}
\begin{center}
  \includegraphics[width=0.45\textwidth]{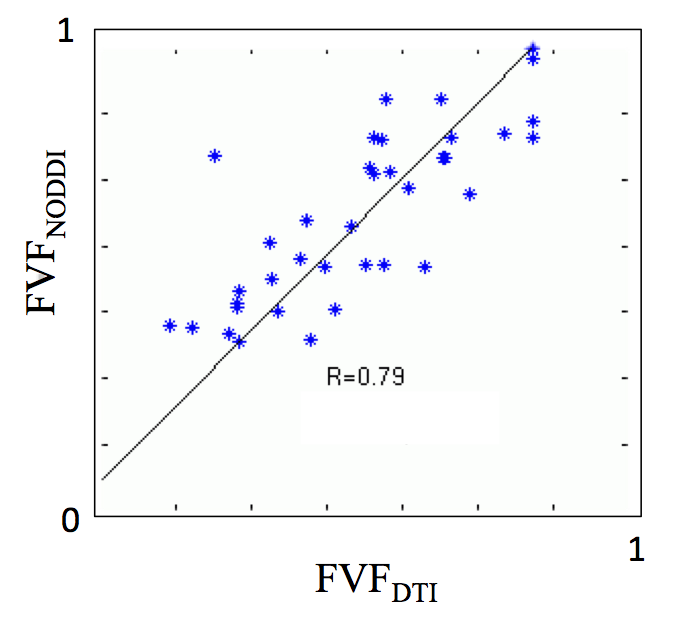}
  \caption{\label{correlationscatter} Correlation between DTI- and
    NODDI-derived fiber volume fraction on the skeleton of the corpus
    callosum.  FVF$_{DTI}$ was calculated from the fractional
    anisotropy of the diffusion tensor using a quadratic relationship
    \citep{Stikov2011}, and FVF$_{NODDI}$ is the sum of the MVF from qMT
    and the AVF computed from Eq. \ref{AVFfrNODDI}.}
\end{center}
\end{figure}

Possible explanations for the higher estimates using NODDI appear in
section \ref{params}, although without ground truth is is difficult to
say which approach is more accurate.  Additionally, because the FA
does not explicitly model compartments, it is subject to partial
volume effects.  While partial volume averaging with CSF will decrease
the FA, the FA-based quadratic FVF model appears to break down in this
case.  This effect could possibly be reduced by applying the free
water elimination technique \citep{Pasternak2009a} to obtain the correct FVF
for the non-CSF compartment and then scaling to reflect the partial
volume averaging with CSF afterward.  To conclude, while the FA is
generally a poor indicator of FVF, it may be a reasonable surrogate in
certain special cases when data are limited.

It is interesting to consider how useful imaging a cross-section of a
white matter fascicle may be, regardless of the model used.  If the
g-ratio can be assumed to be constant along an axon, measurement of a
cross-section is useful.  However, in many pathological situations,
such as Wallerian degeneration, it is of interest to study the entire
length of the axon.

\subsubsection{Restricted extracellular diffusion}

Most existing diffusion models assume that extra-axonal diffusion is
Gaussian, hindered by the structures present, but not restricted.
However, observation of tightly packed axons in microscopy (e.g.,
Fig.\ref{gcartoon}) indicates that the intra- and extra-axonal spaces
may not be as distinguishable as the models assume.  It is unclear to
what extent the extra-axonal diffusion is non-Gaussian.  If axons are
packed tightly together, is extra-axonal diffusion non-Gaussian?  It
is not clear whether the water mobility through the tight passageways
between fibers is distinguishable from the restricted diffusion within
spaces surrounded by contiguous myelin.  If signal from the
extra-axonal space is erroneously attributed to the intra-axonal
space, the model output will be incorrect.

Some models make no attempt to distinguish between intra- and
extra-cellular restricted diffusion, meaning the pore size estimates
may reflect a mixture of the two \citep{Ning2016a}.  Time-dependent
(i.e., non-Gaussian) diffusion has recently been observed in the
extra-axonal space using long \citep{Fieremans2016a,desantis2016b} and
short \citep{Xu2014a} diffusion times. This may be due to axon
varicosity, axonal beading, or variation in axonal packing
\citep{Fieremans2016a}.  In order for the intra- and extra-axonal
compartment fractions to be estimated, the diffusion MRI \emph{signal}
from these compartments must be distinguishable with the experimental
paradigm and modeling used.

\vspace{1 cm}

Diffusion modeling is an active field, and advances in the near future
will hopefully improve precision and accuracy of AVF estimates using
diffusion MRI.  Histological validation may aid in understanding the
strengths and limitations of these estimates.  At present, the
limitations of these models propagate to the g-ratio, as do the
limitations of MVF estimates, which are discussed below.

\subsection{MVF calibration}
\label{cal}

In this section, we investigate the effect of imperfect
  calibration of MRI markers of the MVF.  As an example, we present
  detailed experimental analysis of the magnetization transfer ratio
  (MTR), which is a commonly used myelin marker that is implemented on
  clinical scanners and would be a logical first choice for g-ratio
  imaging in the clinic.  Computing a precise and accurate MVF from
  the MTR is, however, challenging, as is detailed below.  We also
  investigate the performance of MT$_{sat}$ \citep{Helms2008} as a
  potential time-efficient replacement for qMT in the estimation of
  the MVF.  Care must be taken in the use of any myelin marker in
  quantitative g-ratio computation.

How do we make a quantitative estimate of the MVF from myelin
sensitive MRI markers?  Linear correlations have been shown between
the individual myelin sensitive metrics (such as F
\citep{Thiessen2013a,Schmierer2007a}, MTR
\citep{Gareau2000a,Schmierer2004a}, R$_1$ \citep{Mottershead2003a},
MWF \citep{Laule2006a}, and MTV \citep{Duval2016a}) with the MVF from
histology.  Given the linear correlations that have been established,
a logical first approximation is to assume a linear relationship
between the chosen myelin-sensitive metric and the MVF.  Then, using
the macromolecular pool size F as an example, the relationship is

\begin{equation}\label{MVFlin}
  MVF=cF+b,
\end{equation}

\noindent with $c$ and $b$ constants.  While a non-zero value for $b$
has been indicated by some studies \citep{Thiessen2013a,West2017d}, this may be an artifact due to the
inherent bias in linear regression.  The assumption of a linear
relationship hinges on the assumption that non-myelin macromolecular
content scales linearly with myelin content, but this relationship can
break down in disease, or even in healthy tissue.  In disease,
astrocyte scarring, glial cell processes, and inflammatory cell
swelling could all modulate the relationship between a marker of
macromolecular content and the MVF.  If the myelin and non-myelin
macromolecular content do scale linearly, as is assumed here, a
theoretical prior that $b=0$ is reasonable.  However, $b$ could
  reflect a fixed population of macromolecules that are uncorrelated
  with myelin content.  It would most likely be negative, i.e., F
  could be positive with MVF=0.

There is evidence that even if a simple scaling relationship exists
between F and MVF, it is dependent on acquisition and post-processing
details.  For instance, a recent study calibrated F at two different
sites, and found a different scaling factor for each
\citep{Cercignani2016b}.  These scaling factors in turn differ from
those obtained from other investigations
\citep{Stikov2015a,Dula10,Thiessen2013a}.  Hence, careful calibration for each
study must be performed.  Several studies have calibrated scaling
factors based on a given expected g-ratio in healthy white matter
\citep{Cercignani2016b,Mohammadi2015a}.  However, the g-ratio in healthy
white matter is not precisely known.

None of the myelin-sensitive MRI markers is 100\% specific to myelin,
and most are sensitive to myelin in a slightly different way.
Magnetization transfer contrast is specific to macromolecules, and
more specific to lipids than to proteins \citep{Kucharczyk1994}.
Macromolecules in the axon membrane itself, in neurofilaments within
the axons, and in glial cell bodies, will contribute to the MT signal,
with myelin constituting only 50\% of the macromolecular content in
healthy white matter \citep{Bjarnason2005a}.  Additionally, MT-based metrics
such as the magnetization transfer ratio will have residual contrast
from other mechanisms.  We expect the MTR contrast to vary linearly
with macromolecular content, but also with T$_1$
\citep{Vavasour2011a}.  T$_1$ has the opposite sensitivity to myelin
than does the MT effect \citep{Mottershead2003a}, meaning that these
effects work against each other, reducing the dynamic range and power
of MTR as a marker of myelin. Furthermore, T$_1$ is sensitive to iron
and calcium content, intercompartmental exchange, and diffusion, and
hence sensitive to axon size \citep{Harkins2016a} and axon count
\citep{Schmierer2008a}.  This means the relationship between MTR and
MVF may not be monotonic, and is certainly nonlinear.  This
nonlinearity is evident in published plots of MTR vs. F, e.g., that
shown by Levesque et al. \citep{Levesque2005}, and the lack of dynamic
range of MTR is also evident \citep{Levesque2005,Garcia2012}.  The
MT$_{sat}$ technique aims to remove the T$_1$ dependence in MTR.
Both MTR and MT$_{sat}$ depend on the offset frequency used in the
acquisition. ihMT shows promise as a more myelin-specific MT marker
due to its sensitivity to specific molecules in myelin that broaden
the z-spectrum asymmetrically, although it has recently been shown
that asymmetric broadening is not essential to generate a non-zero
ihMT signal \citep{Manning2016b}, and the technique suffers from low
signal.  qMT is the most comprehensive of the MT-based myelin markers,
although its use is impeded by long acquisition times, and its
parameters appear to be sensitive to the specific model and fitting
algorithm.  

Proton-density based techniques \citep{Mezer2013a} will, like MT, be
sensitive to all macromolecules, with a different weighting on these
macromolecules compared to the lipid dominated MT signal.
Relaxation-based myelin markers are also not 100\% specific to myelin.
The confounds with using T$_1$ directly were mentioned above, and the
dependence on iron and calcium concentration, intercompartmental
exchange and diffusion will also affect T$_2$.  T$_2^*$ is also
sensitive to iron concentration, as well as fiber orientation
\citep{Cohen-Adad2014a}.  Isolating the short T$_2$ or short T$_2^*$
compartment enhances specificity to myelin, but MWF estimates vary
nonlinearly with myelin content as the sheath thins and exchange and
diffusion properties are modulated
\citep{Levesque2009a,West2014a,Harkins2012a}.  Variants may
  suffer from reduced accuracy or precision, for example, the mcDESPOT
  technique has been shown to overestimate the MWF
  \citep{Bouhrara2016a} and lack precision \citep{Lankford2013a}.
  Combining T$_1$ and T$_2$ in various ways
  \citep{Glasser2011a,Mangeat2015a} may increase specificity, although
  this approach relies on myelin being the dominant source of contrast.  In the
  UTE technique, it is as yet unclear how to map the signal directly
  to myelin content.  In addition to these confounds, most of these
  myelin markers have recently been shown to have orientation
  dependence.  These include T$_2^*$, $\rchi$
  \citep{Rudko2014a,Liu2015a}, and T$_2$ of the macromolecular pool
  \citep{Pampel2015a}.

While these myelin imaging techniques are certainly powerful tools in
the study of healthy and diseased brain, can they be used reliably in
the g-ratio imaging framework?  As an illustration of the effects of
miscalibration of myelin markers, consider the following scenario
\citep{Campbell2016b}.  We investigate three MT-based myelin markers:
MTR, MT$_{sat}$, and macromolecular pool size F.  We assume a simple
linear scaling between our MRI marker and the MVF.  As described
  in previous work \citep{Stikov2015a}, we calibrate F using combined
  {\emph in vivo} MRI acquisition \emph{ex vivo} electron microscopy
  in the macaque.  We then calibrate MTR and MT$_{sat}$ to match the
mean F-based MVF in white matter.  We subsequently compute g$_{MRI}$,
using the NODDI model of diffusion and the MVF derived from the myelin
markers (Eq.s \ref{AVFfrNODDI},\ref{geq}).

\subsubsection{MVF calibration: Experimental Methods}

 MTR, MT$_{sat}$, qMT, and NODDI data were acquired for five healthy
 volunteers and one MS patient, as described in section \ref{acq}.
 Additionally, for two of the healthy subjects, we acquired one
   MT-on image with an offset frequency of 1.2 kHz, which is the
   standard MTR sequence used at our site. For the MS patient, the
 MT$_{sat}$ images were computed from the qMT MT-off and MT-on (MT
 pulse offset 2.732 kHz, flip angle 142$^{\circ}$) images and one
 additional T$_{1W}$ image with TE=3.3 ms, TR=15 ms, and excitation
 flip angle $\alpha=20^{\circ}$.  The diffusion images were
 preprocessed using FSL \citep{Smith2012a}, and the NODDI parameters
 were computed using the NODDI matlab toolbox \citep{NODDIurl}.  The
 qMT computation of the macromolecular pool size F was performed using
 in-house software \citep{Sled2001a,Cabana2015a}, including B$_0$ and
 B$_1$ correction.  MT$_{sat}$ was computed according to Helms
 \emph{et al.}  \citep{Helms2008,Helms2010erratum}, using the 2.2
   kHz offset frequency data, and also the 1.2 kHz offset frequency
   data where available. MTR was computed for both offset
 frequencies where available.  A semi-empirical B$_1$ correction was
 made \citep{Weiskopf13a} to correct for higher order B$_1$ effects.
 Binary segmentation of white and grey matter was performed using an
 in-house pipeline, using the MPRAGE image only. Lesion segmentation
 for the MS subject was performed with in-house software.

The combined MRI/histology dataset \citep{Stikov2015a} was used to scale
each myelin marker (MTR, MT$_{sat}$, and F) to give the MVF, with the
assumption of a linear relationship (Eq. \ref{MVFlin}) with intercept
b=0.  Correlations between the three myelin markers were computed in
brain parenchyma.  Percent differences were computed between healthy
white matter and healthy grey matter for each of the three myelin
markers.  The AVF was computed using Eq. \ref{AVFfrNODDI}, and
g-ratios were computed in the MS and healthy brains using
Eq. \ref{geq}.  Average g-ratios were computed in healthy white
matter, normal appearing white matter (NAWM), and MS lesions.

A theoretical computation was also performed, varying the mapping of
an arbitrary myelin metric to MVF using Eq. \ref{MVFlin}.  We separately varied
  the slope (c) and the intercept (b) for a range of fiber volume
  fraction values and mapped the computed g-ratio as a function of
  FVF.  When varying the slope, the intercept was fixed at the origin.

\subsubsection{MVF calibration: Results}

\begin{figure}
 \begin{center}
   \includegraphics[width=0.5\textwidth]{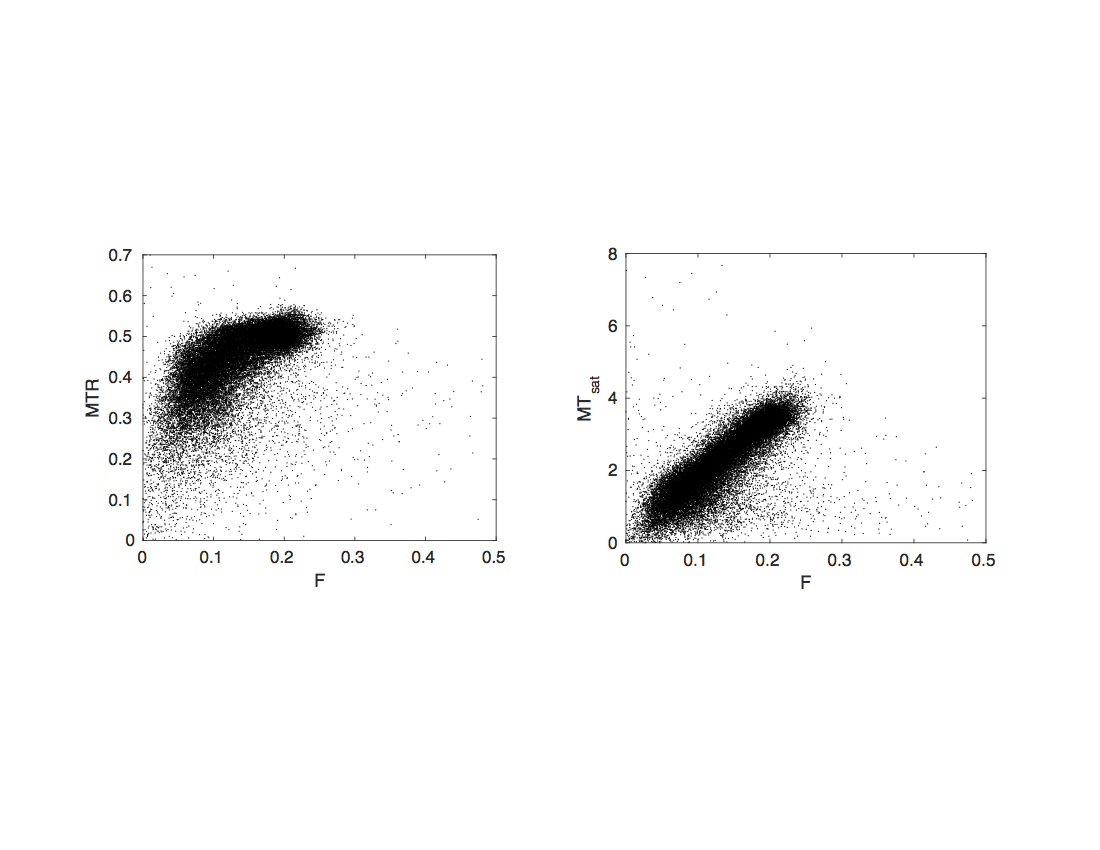}
   \caption{\label{MTR_MTsat_corrwF}MTR plotted versus F (left) and
     MT$_{sat}$ plotted versus F (right) in parenchyma for one
     subject at 2.2 kHz offset frequency.  The MTR vs. F plot shows a marked nonlinearity 
       (r=0.62 over five subjects; r=0.57 for the subject shown in
       this plot), as is expected. MT$_{sat}$ increases the
     linearity of the relationship (r=0.80 over five subjects;
       r=0.78 for the subject shown in this plot) and the dynamic
     range.}
\end{center}
\end{figure}

For the 2.2 kHz offset frequency \citep{Helms2008}, the average
correlation of MTR with F was r=0.62 (p$<$0.001), and of MT$_{sat}$
with F was r=0.80 (p$<$0.001) in parenchyma.  The relationship
  between MT$_{sat}$ and F is more linear than the relationship
  between MTR and F (p$<$0.0001). Fig. \ref{MTR_MTsat_corrwF} shows
plots of MTR versus F (left) and MT$_{sat}$ versus F (right) in
parenchyma for one subject at 2.2 kHz offset frequency.  Of note, the
plot of MTR versus F has a distinctive nonlinear shape, similar to
that seen in the literature \citep{Levesque2005}.  When T$_1$ effects
are reduced using MT$_{sat}$, the linearity and dynamic range
increase.  For the two subjects in which the lower, 1.2 kHz
  offset frequency was also used to compute both MTR and MT$_{sat}$,
  the average correlation of MTR with F was 0.55 and of MT$_{sat}$
  with F was 0.73.

In healthy brain, the percent difference between white and grey matter
was 15.02\% for MTR, 40.08\% for MT$_{sat}$, and 45.86\% for F. The
narrower dynamic range of the MVF derived from MTR can also be seen in
Fig. \ref{MVFs}, where grey matter has markedly higher values.  If
this simple scaling to obtain the MVF is used in the g-ratio formula,
the g-ratio in healthy white matter is relatively constant. However,
when lesions exist, the contrast using the different MVF markers is
very different.  In the MS patient, the mean g-ratio in normal
appearing white matter (NAWM) was 0.76 for all three MVF markers.  In
MS lesions, the mean g-ratio was 0.65, 0.80, and 0.80, for MTR,
MT$_{sat}$, and F, respectively.  Fig. \ref{gratios} shows the spatial
distribution of g-ratios in the MS patient for the three MVF markers.

\begin{figure}
 \begin{center}
   \includegraphics[width=0.5\textwidth]{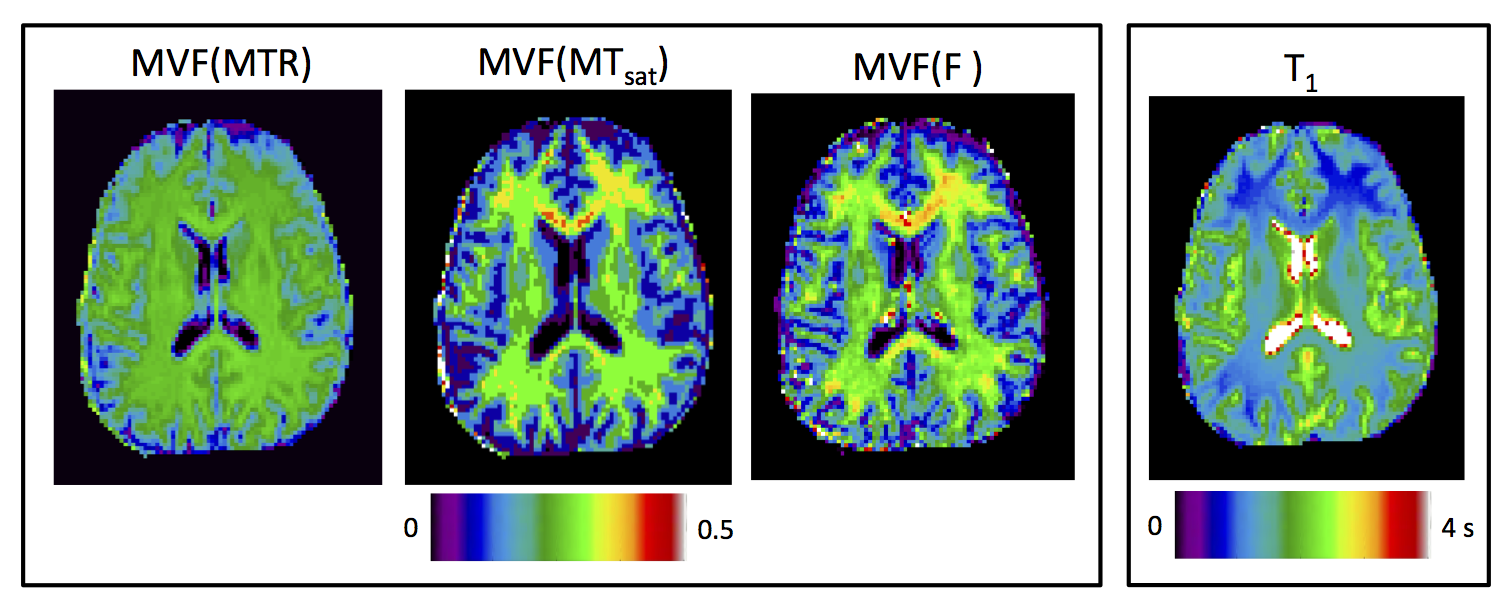}
   \caption{\label{MVFs}\emph{Left:} Plots of the MVF derived from (from left to right) MTR, MT$_{sat}$, and F, in healthy brain.  \emph{Right:} The MT$_{sat}$ calculation estimates a map of the apparent T$_1$ and removes the T$_1$ effects from the MTR map.  T$_1$ has the opposite contrast than does magnetization transfer, so these two effects work against each other in the MTR map, and the MT$_{sat}$ map therefore has more contrast and a greater dynamic range.  The inhomogeniety visible in the T$_1$ map is due to B$_1$ inhomogeniety, which is largely cancelled out in the MT$_{sat}$ computation \citep{Helms2008}, although further correction for residual B$_1$ effects is often performed \citep{Weiskopf13a}.}
\end{center}
\end{figure}

\begin{figure}
 \begin{center}
   \includegraphics[width=0.5\textwidth]{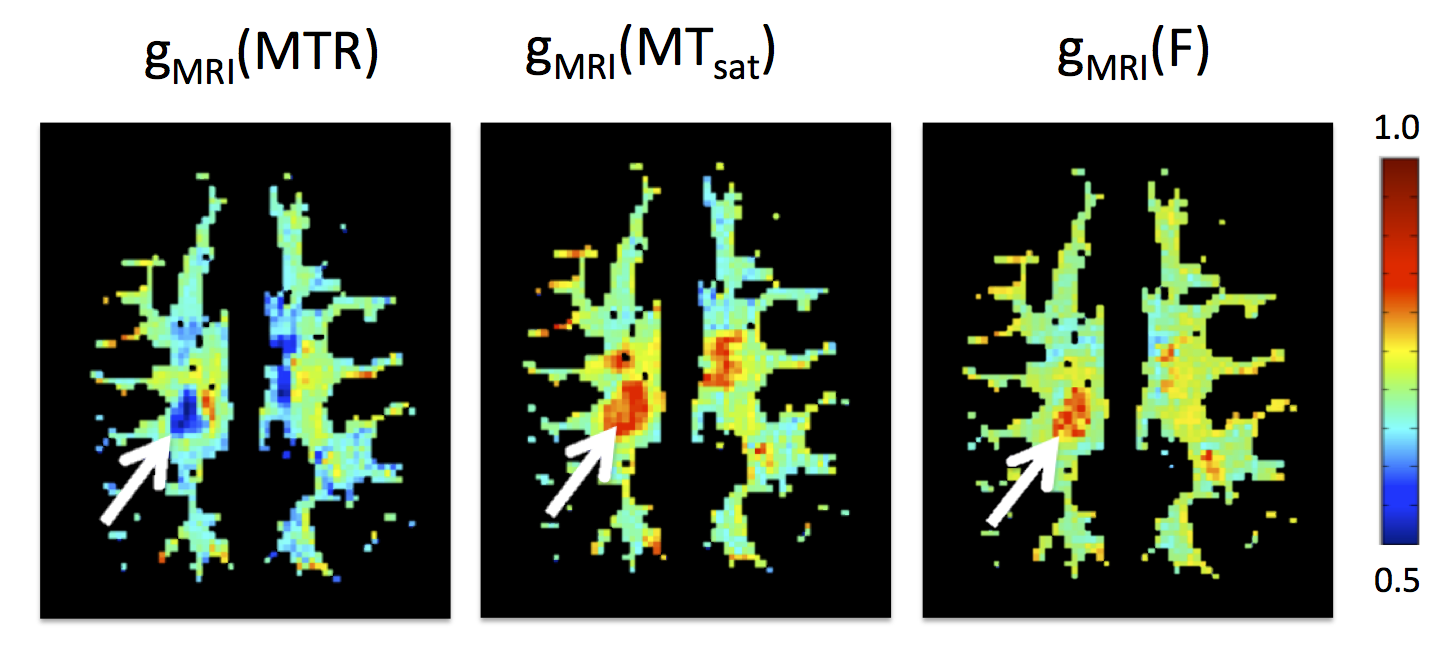}
   \caption{\label{gratios}Plots of g$_{MRI}$ computed using (from left to right) MTR, MT$_{sat}$, and F, in the MS patient.  The arrow indicates a lesion in which the apparent g-ratio is lower than in NAWM when using MTR, but higher than in NAWM when using MT$_{sat}$ and F.}
\end{center}
\end{figure}

\begin{figure}
 \begin{center}
   \includegraphics[width=0.5\textwidth]{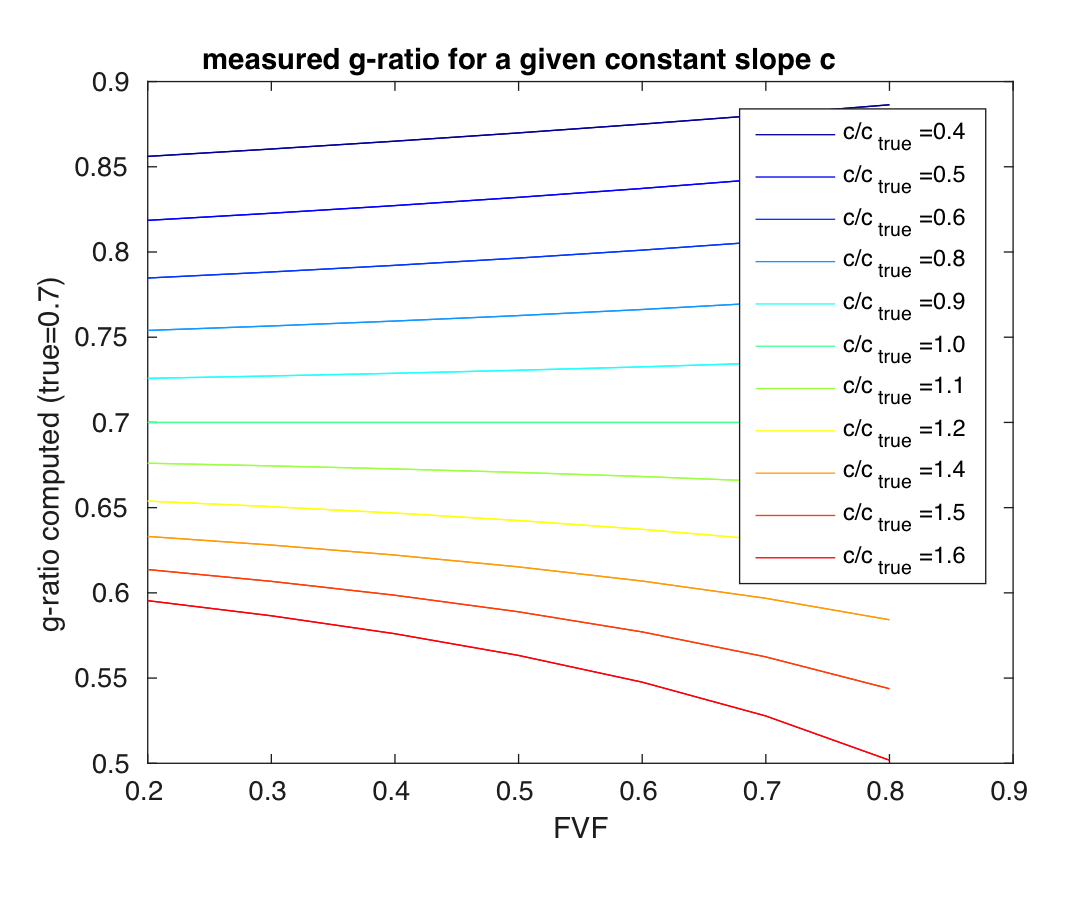}
    \includegraphics[width=0.5\textwidth]{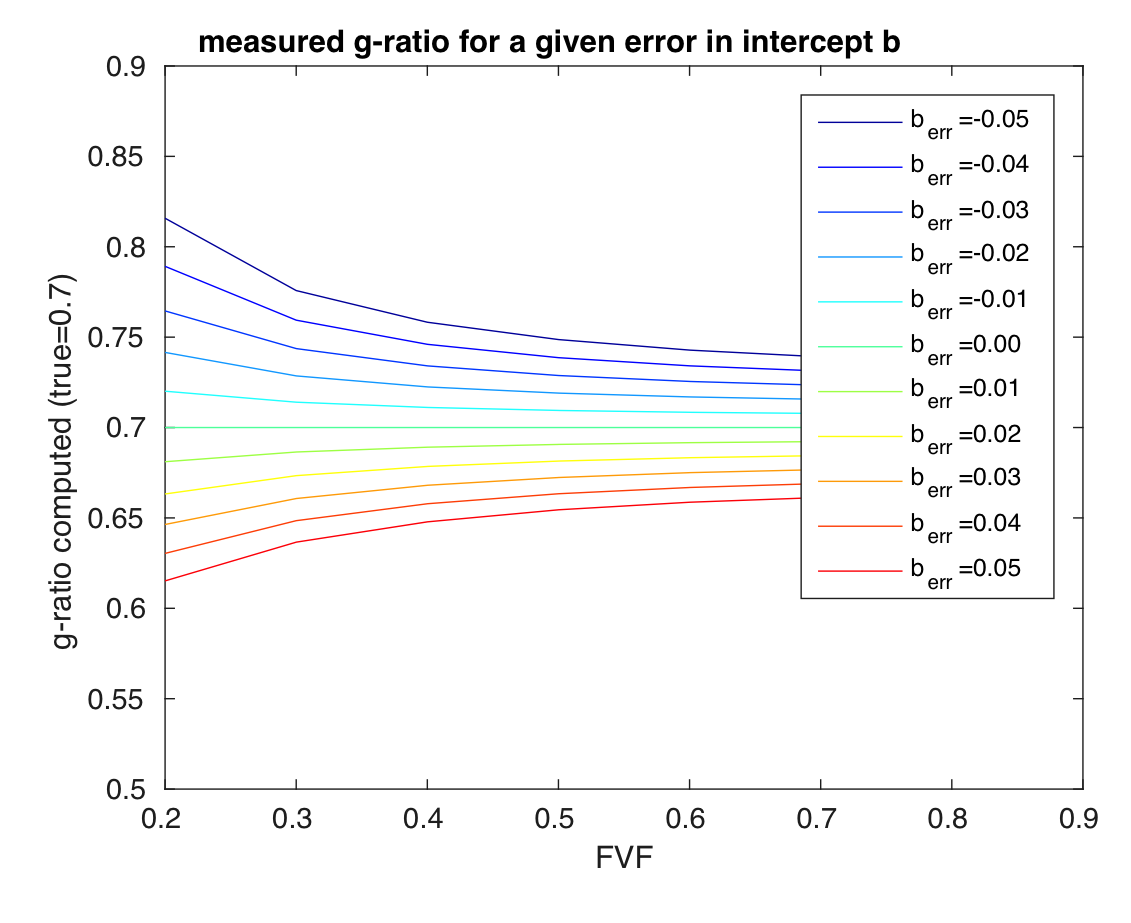}
   \caption{\label{MVFsim}Effect of having an improper slope (top) or intercept (bottom) in the relationship between an arbitrary myelin marker and the MVF, in the case where the (theoretical) relationship is in fact linear.  The plots show that the computed g-ratio becomes fiber density dependent, in addition to being incorrect.  }
\end{center}
\end{figure}

Fig. \ref{MVFsim} shows the theoretical effect of having an improper
slope (top) or intercept (bottom) in the relationship between an
arbitrary myelin marker and the MVF, in the case where the
(theoretical) relationship is in fact linear.  The plots show that the
computed g-ratio becomes fiber density dependent, in addition to being
incorrect.

\subsubsection{MVF calibration: Discussion}

The MTR is a commonly used myelin marker, however, due to T$_1$
sensitivity, it lacks dynamic range.  This results in unrealistic
g-ratios in MS lesions that are lower than in NAWM.  T$_1$ is one of
the possible MR-based myelin markers, but in the context of the MTR
experiment, it confounds the contrast, because the MT effect dominates
but is diminished by the T$_1$ contrast, which works against it.
MT$_{sat}$ correlates more highly with F, which is obtained from an
explicit qMT model designed to isolate the macromolecular tissue
content.  It is important to note, however, that this correlation may
be driven to some extent by the different B$_1$ sensitivities of the
techniques as the MTR was not corrected for B$_1$ induced variability
\citep{Volz2010a,Yarnykh2015a}.  Independent of this demonstration of
the potential of MT$_{sat}$ for myelin mapping, researchers have found
that MT$_{sat}$ may be more sensitive to tissue damage than MTR in
multiple sclerosis, with higher correlation with disability metrics
\citep{Lema2017a}.  MT$_{sat}$ has recently been used by other groups
in g-ratio imaging of healthy adults \citep{Mohammadi2015a}.

If the MVF is miscalibrated in this g-ratio imaging formulation, there
will be a residual dependence on fiber volume fraction in our
formulation.  This reduces the power of the g-ratio metric, which
ideally is completely decoupled from the fiber density.  Independent
of specificity of the myelin marker, if the myelin calibration is
inaccurate, this residual dependence on fiber volume fraction occurs.
It is clear that the g-ratio metric we will compute is g-ratio
\emph{weighted}, and the better the calibration, the more weighted to
the g-ratio it will be.  Until quantitative myelin mapping is
\emph{accurate}, the g-ratio metric will not be specific to the
g-ratio.

 There is evidence that fiber density drops precipitously in some MS
 lesions \citep{Stikov2015a}.  In Fig. \ref{gratios}, this was the
 case, and we can see that MTR does not drop enough, making MS lesions
 appear to have a lowered g-ratio instead of a higher g-ratio as
 expected.  Inspecting the bottom (red) curve in Fig. \ref{MVFsim}, we
 see that even if there is a linear relationship between the myelin
 marker of choice and the MVF, miscalibration leads to an apparent
 g-ratio metric that is elevated in regions of lower fiber density,
 and significantly lower in regions of healthy fiber density.  This
 occurs when in fact all of the fibers have the same g-ratio, and
 could easily be interpreted as hypomyelination in an MS subject or
 population.

As noted above, the problem of miscalibration of the myelin marker
exists independent of the specificity of the myelin marker.  However,
specificity is itself a major confound, as previously detailed.  In the
case of demyelinating disease, one must also consider that, with many
myelin markers, all myelin will affect the MR signal, even if it is
not part of an intact fiber. Research indicates that in MS, there is
acute demyelination followed by a period of clearance of myelin
debris, followed by effective remyelination.  During clearance,
remyelination can occur, but this myelin is of poor quality
\citep{Lampron2015a}. On the scale of an MRI voxel, there can be
myelin debris, poor remyelination, and higher quality remyelination.
The extent to which myelin debris affects the myelin volume estimates
may depend on the myelin mapping technique chosen.  It is also not
clear how well the estimates of MVF, and also AVF, behave at very low
fiber density.

One possible solution for MVF calibration is to calibrate the g-ratio
to a known value in certain regions of interest
\citep{Mohammadi2015a,Cercignani2016b}, as mentioned above. However,
care must be taken that this step is not adjusting for differences in
the diffusion part of the pipeline (e.g., 
different implementations of the diffusion model
\citep{Cercignani2016b}), and therefore still leaving a fiber density
dependence.  Additionally, the correct value in these regions of
interest must be known.
Calibration based on expected MVF would remove this sensitivity, but
is subject to error due to partial volume averaging of white matter
with other tissue.
If the relationship between the myelin-sensitive metric and the MVF is
not a simple scaling, such calibration will fail.  Particular care
needs to be taken when studying disease.

If the assumed relationship between the myelin marker and the MVF is
incorrect, the computed g-ratio will be incorrect.  Is it possible to
compute a g-ratio that is correct to within a scaling factor, and not
sensitive to the fiber density?  This would require that the AVF or
FVF be estimated independent of the MVF.  Simple models such as the
diffusion tensor, apparent fiber density \citep{Raffelt2012a}, and
tensor fiber density \citep{Reisert2013a}, are indicators of fiber or
axon density, but detailed modeling is most likely superior.
Consideration of contrasts other than diffusion MRI, such as
gradient-echo based approaches \citep{Sati2013a}, might also help with
this problem.  The g-ratio is a function of the ratio of the MVF to
the AVF, and a technique that measures this ratio directly would be
optimal.  However, due to the extremely short T$_2^*$ of myelin, GRE
based estimates would be of the myelin and axon \emph{water} fraction,
and hence would still need to be calibrated using the volumetric
occupancy of water in these tissues.

In summary, both specificity and accuracy are important for both AVF
and MVF estimation.  For both AVF and MVF estimation, more
sophisticated models may be required.  For example, we have thus far
ignored cell membranes.  The axon membrane should technically be
included in the AVF, and its volume is up to 4\% of the AVF
\citep{Sepehrband2015a}, but it would most likely be included in the
MVF using MT-based MVF estimation.  It is also important to keep
  in mind the different sensitivities of different MRI markers (e.g.,
  MWF from relaxometry vs. MT-based parameters), because these will
  not be expected to give identical g$_{MRI}$ metrics. The
  determination of which MRI markers are most optimal for g-ratio
  imaging is still the topic of active research
  \citep{Ellerbrock2017a}.

\subsection{g-Ratio distribution}

The g-ratio imaging paradigm extracts a single g-ratio metric per
voxel.  At typical imaging resolution feasible for the constituent MR
images, a voxel contains hundreds of thousands of axons.  As with
  the axon diameter, the g-ratio really applies to an individual axon,
  and takes on a broad distribution of values in tissue
  \citep{Graf1984} (see Fig. \ref{macaque_g_dist}, which shows the
g-ratio distribution in the macaque corpus callosum, measured using
electron microscopy).  The range of myelination includes some
unmyelinated axons within healthy white matter.  The g-ratio
distribution may broaden and become bi-modal in disease.  Even within
a single axon with intact myelin, the g-ratio may vary due to
organelle swelling.  Fiber bundles that cross within one voxel may
have different g-ratio distributions.  In development, some fibers
within one fiber bundle will fully develop, while others will be
pruned, resulting in an interim bimodal g-ratio distribution within
the fascicle.  The current MRI-based g-ratio framework will not be
able to distinguish these cases, as it reports only an intermediate
g-ratio value.  It is robust to crossing fibers, in that it will
report the same intermediate g-ratio value whether the separate
bundles cross or lie parallel to each other.  The broad g-ratio
distribution is in part a resolution problem, but the g-ratio is
expected to be heterogeneous on a scale smaller than we can hope to
resolve with MRI.

\begin{figure}
  \begin{center}
   \includegraphics[width=0.5\textwidth]{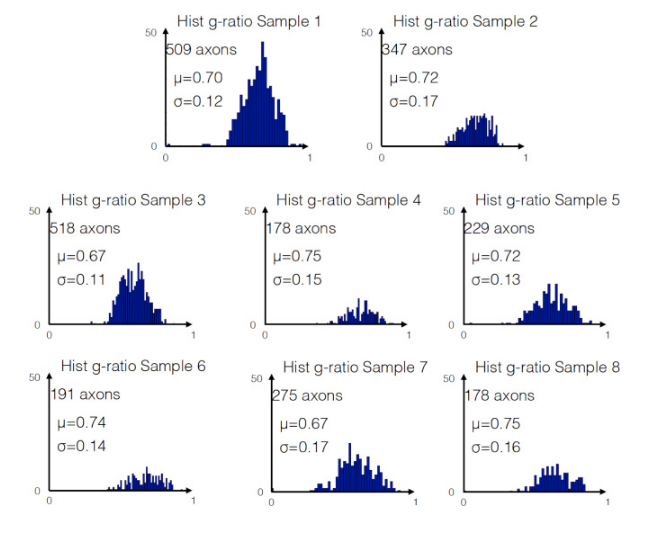}
   \caption{\label{macaque_g_dist} g-Ratio distributions from electron microscopy of the cynamolgus macaque corpus callosum, samples 1-8 from genu to splenium.  Reproduced from \citep{Stikov2015c}. }
  \end{center}
\end{figure}

The aggregate g-ratio we compute in the case of a distribution of
values is not precisely fiber- or axon-area weighted, but is close to
axon area weighted within a reasonable range of values
\citep{West15a}.  Larger axons will have a greater weight in the
aggregate g-ratio metric we measure.  Simply put, the aggregate
g-ratio is the g-ratio one would measure if
all axons had the same g-ratio.

In the case of an ambiguous g-ratio distribution, what techniques can
we use to infer what situation is occurring?  In multiple sclerosis,
for example, two possible scenarios probably occur frequently.  One is
patchy demyelination, on a scale much smaller than a voxel and smaller
than the diffusion distance, and the other is more extensively and
uniformly distributed thin myelin.  These two scenarios could give
rise to equal AVF, MVF, and aggregate g-ratio measurements.  One
possible way to differentiate these cases could be to look more
closely at parameters available to us from diffusion models.  It has
been shown that the extra-axonal perpendicular diffusivity is
relatively unchanged by patchy demyelination in a demyelinating mouse
model \citep{Jelescu2016b}, because diffusing molecules encounter
normal hindrance to motion on most of their trajectory, whereas the
axon water fraction is sensitive to this patchy demyelination. Hence,
the discrepancy between these two measures can be taken as a measure
of patchy demyelination.  Alternatively, one can scan subjects
longitudinally and infer disease progression.  From the ambiguous
timepoint described above, the axons in the patches that are
demyelinated may die, leaving a decreased AVF and MVF, and a return to
a near-healthy g-ratio.  In the case of uniformly thin myelin, the
remyelination may continue, leaving a near-healthy AVF, MVF, and
g-ratio.  Note that the g-ratio metric still does not distinguish
these pathologically distinct cases.  There are two unknowns - the
fiber density and the g-ratio (or, alternately, the MVF and the AVF),
and one must consider both to have a full picture of the tissue.
Looking at the time courses, one can hypothesize what the g-ratio
distribution was at the first timepoint.

It would be technically challenging to measure the g-ratio
distribution \emph{in vivo}.  Even with an estimate of a distribution
of diffusion properties, and an estimate of the distribution of a
myelin-sensitive metric, the g-ratio distribution is ill-defined.
However, several recent acquisition strategies may help us get closer
to this aim.  One approach is to take advantage of the distinguishable
diffusion signal between different fiber orientations. In the
IR-prepared diffusion acquisition described above \citep{DeSantis16a}, the model specifies
multiple fiber populations with distinct orientations, each with its
own T$_1$ value.  This means the diffusion properties, including the
restricted pool fraction (a marker of intra-axonal signal from the
CHARMED model), are paired with a corresponding T$_1$ for each fiber
orientation. Hence, a g-ratio metric could be computed for each fiber
orientation.  This could be of benefit in, e.g., microstructure
informed white matter fiber tractography (e.g., \citep{Girard2015a}) of
fiber populations with distinct g-ratios.  ``Jumping'' from one fiber
population to another is very common in tractography
\citep{Campbell13b,Descoteaux2016a,Maier-Hein2016a}, and constraining
tractography to pathways with consistent microstructural features
could help reduce false positives in regions of closely intermingling
tract systems.

It may be possible, conceptually, to estimate the g-ratio distribution
via a 2D diffusion-relaxation spectroscopic approach.  While extremely
acquisition intensive, 2D spectroscopy of T$_2$ and the diffusion
coefficient \citep{Callaghan2003a} has been demonstrated recently as a
probe of microstructure \citep{Kim2016a}.  The acquisition involves
making all diffusion measurements at different echo times.  If a
distribution of a myelin volume sensitive metric (here, T$_2$) can be
estimated simultaneously in 2D with a distribution of a
diffusion-based metric sensitive to the axon volume, it may be
possible to infer the distribution of g-ratios.

\vspace{1 cm}

This has been an incomplete but useful list of pitfalls.  Now, we will consider the promise of imaging the aggregate g-ratio weighted metric, despite its pitfalls.  g-Ratio imaging is being explored in many different contexts, described below. 

\section{\label{pubs}The promise: g-ratio imaging studies}

The promise of g-ratio imaging is its potential to provide us with valuable \emph{in vivo} estimates of relative myelination.  In the last few years, studies showing the potential of this framework have begun to emerge.

\subsection{Healthy white matter}

Fig. \ref{healthyg} shows an image of g$_{MRI}$ in healthy white
matter using our qMT and NODDI g-ratio protocol (see section
\ref{acq}).  With our protocol, in healthy subjects, the
  g$_{MRI}$ map is relatively flat, with a mean g$_{MRI}$ of 0.76 (SD=0.05).
Other groups have explored these and other MVF and AVF sensitive
contrasts for g-ratio mapping in healthy white matter.  These include
a study of the effects of age and gender in a population of subjects
aged 20 to 76 using qMT and NODDI \citep{Cercignani2016c}, studies of
healthy adults using MT$_{sat}$ and the TFD \citep{Mohammadi2015a} and
MTV and DTI \citep{Berman2017a}, and a study of healthy subjects using
the ViSTa myelin water imaging technique and NODDI \citep{Jung2016a}.

\begin{figure}
\begin{center}
  \includegraphics[width=0.35\textwidth]{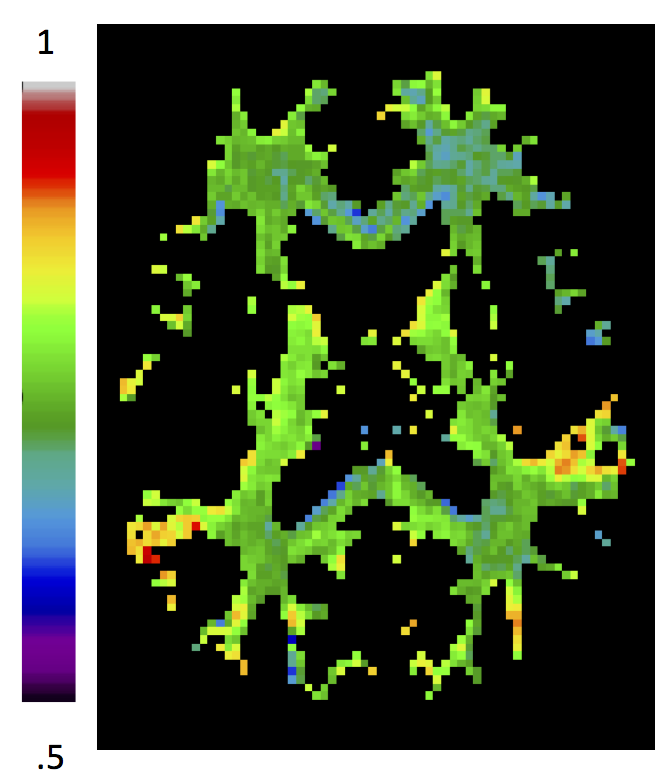}
  \caption{\label{healthyg} g$_{MRI}$ in healthy white matter, imaged using qMT and NODDI.}
\end{center}
\end{figure}

A variation of the g-ratio with age appears to be detectable with this
methodology \citep{Cercignani2016c}.  A variation with gender has not
been seen, and if it exists in adolescence \citep{Paus2009a}, a study
designed for sufficient statistical power at a precise age will be
required to detect it.  In addition to exploring the effect of age and
gender, spatial variability of the g-ratio has been investigated.  An
elevated g-ratio at the splenium of the corpus callosum has been seen
\citep{Stikov2015a,Mohammadi2015a}.  The splenium has been reported to
contain axons of very large diameter \citep{lamantia1990a}, and
these would be expected, due to the nonlinearity of the g-ratio
\citep{Hildebrand78}, to have relatively thinner myelin sheaths.
Electron microscopy in the macaque \citep{Stikov2015a} (see
Fig. \ref{splenium}) confirms this; the ``super-axons'' dominate the
aggregate g-ratio measure, which was seen to be elevated in the
splenium using both EM measurements and MRI of the same tissue
\citep{Stikov2015a}.

\begin{figure}
\begin{center}
  \includegraphics[width=0.5\textwidth]{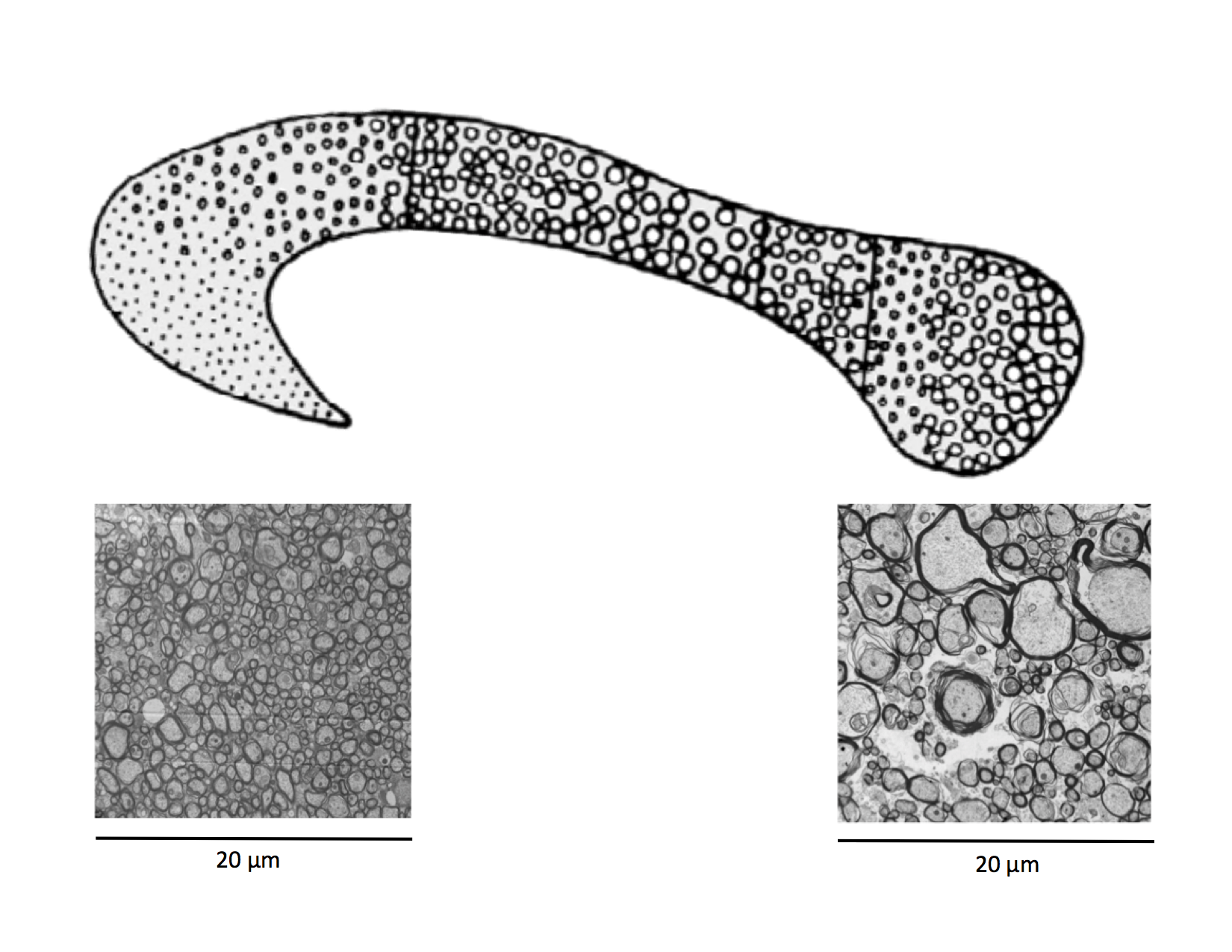}
  \caption{\label{splenium} Existence of very large axons in the splenium of the corpus callosum.  Top: drawing based on histology by Aboitiz \emph{et al.} (reproduced from \citep{Aboitiz2003a}), showing large diameter at the splenium.  Bottom: EM of the g-ratio in the cynamolgus macaque showing one sample from the genu (left) and one sample from the splenium (right).  The splenium contains much larger diameter axons, and these will dominate the aggregate g-ratio measure, which was elevated in the splenium using both EM measurements and MRI of the same tissue shown here \citep{Stikov2015a}. }
\end{center}
\end{figure}

g-Ratio imaging has also been performed in the healthy human spinal
cord \citep{Duval15a}, where there are considerable technical
challenges, such as motion, susceptibility, and the need for
significantly higher resolution than we have described for cerebral
applications.  Duval \emph{et al.} acquired g-ratio data at 0.8 mm x 0.8 mm
inplane voxel size.  This study used the CHARMED model of diffusion,
more accessible on scanners with high gradient strength, on a
CONNECTOM scanner.  It used the MTV myelin marker.  Of interest,
the g-ratio was not found to vary significantly across white matter
tracts in the spinal cord, while the diffusion metrics (restricted
fraction, diffusivity of the hindered compartment, and axon diameter)
and the MTV metric did vary across tracts.  This is expected, as
heterogeneity in packing and axon diameter is expected to be greater
than heterogeneity of the g-ratio, and the g-ratio is also robust to
partial voluming effects.

Multiple groups have studied the g-ratio \emph{in vivo} in the
developing brain \citep{Dean2016a,Melbourne16a}.  Axon growth outpaces
myelination during development, and therefore a decreasing g-ratio is
expected as myelination reaches maturity, as was seen in these
studies.

\subsection{Multiple sclerosis}

Imaging the g-ratio \emph{in vivo} in multiple sclerosis has been
explored by several groups \citep{Stikov2015a,Cercignani15a,Hori2016a}
and is of interest for several reasons.  It can possibly help assess
disease evolution, and can help monitor response to treatment. It has
the potential to aid in the development of new therapies for
remyelination.  It can also help us understand which therapies might
be more fruitful avenues of research.  While currently available
therapies are immunomodulatory or immunosuppressive in nature, several
remyelinating therapies are in clinical trials. Preventing MS-related
demyelination (via immunomodulatory therapies) is always preferred at
earlier disease stages. Once demyelination has occurred, remyelinating
therapies may help protect demyelinated (but not transected) axons
from delayed degeneration due to the loss of trophic factors no longer
received from myelin. Given that remyelination can occur only when
demyelinated but still viable axons are present, interpretation of MRI
markers of remyelination would be improved by including a marker of
axonal integrity. Thus, the g-ratio framework would be useful in the
evaluation of remyelinating therapies. In particular, dynamic changes
of g-ratio over time could be measured within new and chronic lesions
(as detected on conventional MRI) and the (temporally aligned)
timecourses compared between treatment arms.

Despite the promise of imaging the g-ratio \emph{in vivo} with MS, it
is important to remember the pitfall of specificity and miscalibration
of the myelin metric when interpreting g-ratio estimates in MS.  In
our own preliminary experience imaging MS, the relationship between
the g-ratio in lesions and in NAWM appears to be complex.  In section
\ref{cal}, we detailed our observations in one MS patient for whom we
acquired a full qMT protocol.  In a total of four subjects for whom we
have acquired MT$_{sat}$ data, we observed heterogeneity in the
g$_{MRI}$ values within and across subjects.  The average ($\pm$
standard deviation) g-ratio in lesions compared to NAWM was not always
higher than that in NAWM, with
(g$_{MRI}$(lesions),g$_{MRI}$(NAWM))=(0.80$\pm0.07$,0.76$\pm$0.05),
(0.72$\pm$0.05,0.77$\pm$0.03), \\(0.74$\pm$0.06,0.75$\pm$0.05),
(0.75$\pm$0.08,0.78$\pm$0.04) for the four subjects.  On average in
these patients, g$_{MRI}$ was 0.75 in lesions and 0.76 in NAWM. This
variability
of g$_{MRI}$ in lesions compared to NAWM is consistent with
preliminary findings from other groups using MT contrast for MVF
\citep{Cercignani15a}, and could indicate that variable levels of
non-myelin macromolecular content may confound the g-ratio
metric.  The g-ratio itself is not expected to be lower than in NAWM,
but may appear so because other macromolecular content is confounding
the MT measurements.  

Further studies of MS are ongoing, including pediatric populations,
optic neuritis, and studies investigating whether gadolinium enhancing
lesions have a distinct g-ratio.

\subsection{Other potential applications}

g-Ratio imaging has potential to aid in the understanding and
treatment of multiple other diseases.  White matter abnormalities may
underlie many developmental disorders.  These include Pelizaeus
Merzbacher disease and Sturge-Weber syndrome \citep{Hori2016b}. An
increased apparent g-ratio could result from axonal changes that occur
with intact myelin (for example, axonal swelling due to
infarction). g-Ratio differences have been seen in schizophrenia using
electron microscopy \citep{Uranova2001a}, and researchers hope to be
able to study such changes \emph{in vivo} in schizophrenia and other
psychiatric disorders.  Another potential application of g-ratio
imaging is bridging the gap between microstructure and large-scale
functional measures such as conduction delays.  Adding a g-ratio
measure to the human connectome could result in a framework for
evaluating delays.  Finally, g-ratio imaging offers the possibility to
quantify plastic changes in myelin thickness due to learning and
adaptation to injury or disease.  MRI has been used to measure such
changes, with the hypothesis that myelin thickness is changing
\citep{Reid2016a}, but quantitative g-ratio measurement could help
describe the neural changes in plasticity in more detail.

Of note, when examining the g-ratio compared to conduction velocity or
functional metrics, linear correlation would not be expected to be a
good statistic.  The g-ratio has an optimal value, hence, both
decreases and increases from this optimum would be expected to
decrease function.  In the future, assuming sufficient reproducibility,
establishment of an age-dependent atlas of normal g-ratio values could
help determine whether an individual lies within the normal range.

\subsection{\emph{Ex vivo} g-ratio imaging}

One major application of \emph{ex vivo} g-ratio imaging is validation
of the technique for use \emph{in vivo}, which should help elucidate the
true promise of g-ratio imaging.  \emph{Ex vivo} validation has been
performed, in order to investigate the g-ratio explicitly
\citep{Stikov2015a,Stikov2015c,West2016b}, or one or both of the individual metrics
used to compute it
\citep{Duval2016a,Jelescu2016b,Jespersen2010a,West2017d,West2014a,Schmierer2008a,Scherrer2016a,Sepehrband2015a,Wood2016a}.
These studies compare \emph{in vivo} or \emph{ex vivo} MRI metrics to
electron microscopy, optical microscopy, myelin staining,
immunohistochemistry, and coherent anti-Stokes Raman
spectroscopy (CARS). While no microscopy technique is perfect, microscopy
provides a reasonable validation for imaging techniques, taking into
account the possibility for tissue shrinkage and distortion,
limitations in contrast and resolution, and segmentation techniques \citep{Zaimi2016a}.

Interpretation of findings of demyelinating models should take into
account the particularities of the demyelinating challenge.  Jelescu
\emph{et al.} have shown that the extra-axonal diffusivity
perpendicular to axons correlates with the g-ratio in a cuprizone
demyelinating model in mice \citep{Jelescu2016b}.  This is probably
driven by a fiber volume fraction decrease, because little axon loss
would be expected in this model.  In other words, the extra-axonal
diffusivity is not specific to the g-ratio per se, but to the
fiber volume fraction (i.e., size of the extracellular space), but
these two quantities correlate highly in this particular case.
Similarly, West \emph{et al.} have shown a correlation between the
discrepancy between F and MWF and the g-ratio in a knockout model in
mice \citep{West2014a}. This is probably a correlation with absolute
myelin thickness, via exchange effects, as opposed to the g-ratio per
se.

\emph{Ex vivo} g-ratio imaging could also be of use for the study of
neural tissue independent of validation of \emph{in vivo} imaging. MRI
is far more suited to imaging large samples of tissue than are other
\emph{ex vivo} imaging techniques, such as electron microscopy.

\section{Conclusion}

Computing a g-ratio metric is an effective way to interpret any
combination of myelin-weighted and axon/fiber-weighted MR data. In
this article, we have discussed the considerable promise of g-ratio
imaging to help us understand disease, develop therapies, and monitor
disease progression. Additionally, we have shown how the g-ratio
framework can provide a window onto the study of normal brain
variability, development, aging, plasticity, and functional
dynamics. We have also explored the pitfalls of g-ratio imaging, which
include MR artifacts, lack of specificity, low spatial resolution, and
long acquisition times. Keeping these confounds in mind, it is clear
that what we are currently measuring is an \emph{aggregate} g-ratio
\emph{weighted} metric that is strongly dependent on the MRI markers
used to compute it. The framework described in this article provides
information on two quantities: the fiber density and the g-ratio
(equivalently, the myelin and axon volume fractions), and attempts to
decouple these two quantities to the best of the ability of our
current imaging technology.  To improve upon this description of the
microstructure, one needs to fully decouple the g-ratio from the fiber
density, and to provide precise and accurate measures of the myelin
and the axon volume fractions. This task is at the frontier of
microstructural MRI, showing the way for the future of multi-modal
brain imaging.

\section*{Acknowledgements}
The authors would like to thank Tomas Paus, Robert Dougherty, Eva
Alonso-Ortiz, J.F. Cabana, Christine Tardif, Jessica Dubois, Dmitry
Novikov, Ofer Pasternak, Atef Badji, Robert Brown, Masaaki Hori, and
David Rudko for their insights and contributions to this work.  This
work was supported by grants from Campus Alberta Innovates, the
Canadian Institutes for Health Research (GBP, FDN-143290, and JCA,
FDN-143263), the Natural Science and Engineering Research Council of
Canada (NS, 2016-06774, and JCA, 435897-2013), the Montreal Heart
Institute Foundation, the Fonds de Recherche du Qu\'{e}bec - Sant\'{e}
(JCA, 28826), the Quebec BioImaging Network (NS, 8436-0501), the
Canada Research Chair in Quantitative Magnetic Resonance Imaging
(JCA), and the Fonds de Recherche du Qu\'{e}bec - Nature et
Technologies (JCA, 2015-PR-182754).

\bibliographystyle{model5-names}


\section*{References}

\begin{thebibliography}{170}
\expandafter\ifx\csname natexlab\endcsname\relax\def\natexlab#1{#1}\fi
\providecommand{\url}[1]{\texttt{#1}}
\providecommand{\href}[2]{#2}
\providecommand{\path}[1]{#1}
\providecommand{\DOIprefix}{doi:}
\providecommand{\ArXivprefix}{arXiv:}
\providecommand{\URLprefix}{URL: }
\providecommand{\Pubmedprefix}{pmid:}
\providecommand{\doi}[1]{\href{http://dx.doi.org/#1}{\path{#1}}}
\providecommand{\Pubmed}[1]{\href{pmid:#1}{\path{#1}}}
\providecommand{\bibinfo}[2]{#2}
\ifx\xfnm\relax \def\xfnm[#1]{\unskip,\space#1}\fi
\bibitem[{Aboitiz \& Montiel(2003)}]{Aboitiz2003a}
\bibinfo{author}{Aboitiz, F.}, \& \bibinfo{author}{Montiel, J.}
  (\bibinfo{year}{2003}).
\newblock \bibinfo{title}{One hundred million years of interhemispheric
  communication: the history of the corpus callosum}.
\newblock {\it \bibinfo{journal}{Brazilian Journal of Medical and Biological
  Research}\/},  {\it \bibinfo{volume}{36}\/}. \URLprefix
  \url{http://dx.doi.org/10.1590/s0100-879x2003000400002}.
  \DOIprefix\doi{10.1590/s0100-879x2003000400002}.
\bibitem[{Aboitiz et~al.(1992)Aboitiz, Scheibel, Fisher \&
  Zaidel}]{Aboitiz1992a}
\bibinfo{author}{Aboitiz, F.}, \bibinfo{author}{Scheibel, A.~B.},
  \bibinfo{author}{Fisher, R.~S.}, \& \bibinfo{author}{Zaidel, E.}
  (\bibinfo{year}{1992}).
\newblock \bibinfo{title}{Fiber composition of the human corpus callosum}.
\newblock {\it \bibinfo{journal}{Brain Research}\/},  {\it
  \bibinfo{volume}{598}\/}, \bibinfo{pages}{143--153}. \URLprefix
  \url{http://dx.doi.org/10.1016/0006-8993(92)90178-c}.
  \DOIprefix\doi{10.1016/0006-8993(92)90178-c}.
\bibitem[{Albert et~al.(2007)Albert, Antel, Bruck \& Stadelmann}]{Albert07}
\bibinfo{author}{Albert, M.}, \bibinfo{author}{Antel, J.},
  \bibinfo{author}{Bruck, W.}, \& \bibinfo{author}{Stadelmann, C.}
  (\bibinfo{year}{2007}).
\newblock \bibinfo{title}{Extensive cortical remyelination in patients with
  chronic multiple sclerosis.}
\newblock {\it \bibinfo{journal}{Brain Pathol}\/},  {\it
  \bibinfo{volume}{17}\/}, \bibinfo{pages}{129--138}.
  \DOIprefix\doi{10.1111/j.1750-3639.2006.00043.x}.
\bibitem[{Alonso-Ortiz et~al.(2016)Alonso-Ortiz, Levesque, Paquin \&
  Pike}]{Alonso2016a}
\bibinfo{author}{Alonso-Ortiz, E.}, \bibinfo{author}{Levesque, I.~R.},
  \bibinfo{author}{Paquin, R.}, \& \bibinfo{author}{Pike, G.~B.}
  (\bibinfo{year}{2016}).
\newblock \bibinfo{title}{Field inhomogeneity correction for gradient echo
  myelin water fraction imaging}.
\newblock {\it \bibinfo{journal}{Magn. Reson. Med.}\/},  (pp.
  \bibinfo{pages}{49--57}). \URLprefix
  \url{http://dx.doi.org/10.1002/mrm.26334}. \DOIprefix\doi{10.1002/mrm.26334}.
\bibitem[{Andersson et~al.(2003)Andersson, Skare \& Ashburner}]{Andersson2003a}
\bibinfo{author}{Andersson, J.~L.}, \bibinfo{author}{Skare, S.}, \&
  \bibinfo{author}{Ashburner, J.} (\bibinfo{year}{2003}).
\newblock \bibinfo{title}{How to correct susceptibility distortions in
  spin-echo echo-planar images: application to diffusion tensor imaging.}
\newblock {\it \bibinfo{journal}{NeuroImage}\/},  {\it \bibinfo{volume}{20}\/},
  \bibinfo{pages}{870--888}. \URLprefix
  \url{http://dx.doi.org/10.1016/s1053-8119(03)00336-7}.
  \DOIprefix\doi{10.1016/s1053-8119(03)00336-7}.
\bibitem[{Assaf \& Basser(2005)}]{Assaf05a}
\bibinfo{author}{Assaf, Y.}, \& \bibinfo{author}{Basser, P.~J.}
  (\bibinfo{year}{2005}).
\newblock \bibinfo{title}{{Composite hindered and restricted model of diffusion
  (CHARMED) MR imaging of the human brain }}.
\newblock {\it \bibinfo{journal}{NeuroImage}\/},  {\it \bibinfo{volume}{27}\/},
  \bibinfo{pages}{48--58.}
\bibitem[{Assaf et~al.(2008)Assaf, Blumenfeld-Katzir, Yovel \&
  Basser}]{Assaf2008a}
\bibinfo{author}{Assaf, Y.}, \bibinfo{author}{Blumenfeld-Katzir, T.},
  \bibinfo{author}{Yovel, Y.}, \& \bibinfo{author}{Basser, P.~J.}
  (\bibinfo{year}{2008}).
\newblock \bibinfo{title}{{AxCaliber}: a method for measuring axon diameter
  distribution from diffusion {MRI}.}
\newblock {\it \bibinfo{journal}{Magn. Reson. Med.}\/},  {\it
  \bibinfo{volume}{59}\/}, \bibinfo{pages}{1347--1354}. \URLprefix
  \url{http://dx.doi.org/10.1002/mrm.21577}. \DOIprefix\doi{10.1002/mrm.21577}.
\bibitem[{Avram et~al.(2013)Avram, \"{O}zarslan, Sarlls \& Basser}]{avram2013a}
\bibinfo{author}{Avram, A.~V.}, \bibinfo{author}{\"{O}zarslan, E.},
  \bibinfo{author}{Sarlls, J.~E.}, \& \bibinfo{author}{Basser, P.~J.}
  (\bibinfo{year}{2013}).
\newblock \bibinfo{title}{In vivo detection of microscopic anisotropy using
  quadruple pulsed-field gradient ({qPFG}) diffusion {MRI} on a clinical
  scanner.}
\newblock {\it \bibinfo{journal}{NeuroImage}\/},  {\it \bibinfo{volume}{64}\/},
  \bibinfo{pages}{229--239}. \URLprefix
  \url{http://dx.doi.org/10.1016/j.neuroimage.2012.08.048}.
  \DOIprefix\doi{10.1016/j.neuroimage.2012.08.048}.
\bibitem[{Basser et~al.(1994)Basser, Mattiello \& Le~Bihan}]{Basser1994a}
\bibinfo{author}{Basser, P.}, \bibinfo{author}{Mattiello, J.}, \&
  \bibinfo{author}{Le~Bihan, D.} (\bibinfo{year}{1994}).
\newblock \bibinfo{title}{Estimation of the effective self-diffusion tensor
  from the {NMR} spin echo.}
\newblock {\it \bibinfo{journal}{Journal of Magnetic Resonance}\/},  {\it
  \bibinfo{volume}{103}\/}, \bibinfo{pages}{247--254}.
\bibitem[{Behrens et~al.(2007)Behrens, Berg, Jbabdi, Rushworth \&
  Woolrich}]{Behrens07a}
\bibinfo{author}{Behrens, T.~E.}, \bibinfo{author}{Berg, H.~J.},
  \bibinfo{author}{Jbabdi, S.}, \bibinfo{author}{Rushworth, M.~F.}, \&
  \bibinfo{author}{Woolrich, M.~W.} (\bibinfo{year}{2007}).
\newblock \bibinfo{title}{Probabilistic diffusion tractography with multiple
  fibre orientations: What can we gain?}
\newblock {\it \bibinfo{journal}{NeuroImage.}\/},  {\it
  \bibinfo{volume}{34}\/}, \bibinfo{pages}{144--55.}
\bibitem[{Bells et~al.(2011)Bells, Cercignani, Deoni, Assaf, Pasternak, Evans,
  Leemans \& Jones}]{Bells2011a}
\bibinfo{author}{Bells, S.}, \bibinfo{author}{Cercignani, M.},
  \bibinfo{author}{Deoni, S.}, \bibinfo{author}{Assaf, Y.},
  \bibinfo{author}{Pasternak, O.}, \bibinfo{author}{Evans, C.~J.},
  \bibinfo{author}{Leemans, A.}, \& \bibinfo{author}{Jones, D.~K.}
  (\bibinfo{year}{2011}).
\newblock \bibinfo{title}{Tractometry: Comprehensive multi-modal quantitative
  assessment of white matter along specific tracts}.
\newblock In {\it \bibinfo{booktitle}{ISMRM 2011}\/} (p. \bibinfo{pages}{678}).
\bibitem[{Bengio(2009)}]{Bengioreview2009}
\bibinfo{author}{Bengio, Y.} (\bibinfo{year}{2009}).
\newblock \bibinfo{title}{Learning deep architectures for {AI}}.
\newblock {\it \bibinfo{journal}{Foundations and Trends® in Machine
  Learning}\/},  {\it \bibinfo{volume}{2}\/}, \bibinfo{pages}{1--127}.
  \URLprefix \url{http://dx.doi.org/10.1561/2200000006}.
  \DOIprefix\doi{10.1561/2200000006}.
\bibitem[{Benninger et~al.(2006)Benninger, Colognato, Thurnherr, Franklin,
  Leone, Atanasoski, Nave, ffrench Constant, Suter \& Relvas}]{Benninger2006a}
\bibinfo{author}{Benninger, Y.}, \bibinfo{author}{Colognato, H.},
  \bibinfo{author}{Thurnherr, T.}, \bibinfo{author}{Franklin, R. J.~M.},
  \bibinfo{author}{Leone, D.~P.}, \bibinfo{author}{Atanasoski, S.},
  \bibinfo{author}{Nave, K.-A.}, \bibinfo{author}{ffrench Constant, C.},
  \bibinfo{author}{Suter, U.}, \& \bibinfo{author}{Relvas, J.~B.}
  (\bibinfo{year}{2006}).
\newblock \bibinfo{title}{{β1-Integrin} signaling mediates premyelinating
  oligodendrocyte survival but is not required for {CNS} myelination and
  remyelination}.
\newblock {\it \bibinfo{journal}{Journal of Neuroscience}\/},  {\it
  \bibinfo{volume}{26}\/}, \bibinfo{pages}{7665--7673}. \URLprefix
  \url{http://dx.doi.org/10.1523/jneurosci.0444-06.2006}.
  \DOIprefix\doi{10.1523/jneurosci.0444-06.2006}.
\bibitem[{Berman et~al.(2017)Berman, West, Does, Yeatman \&
  Mezer}]{Berman2017a}
\bibinfo{author}{Berman, S.}, \bibinfo{author}{West, K.~L.},
  \bibinfo{author}{Does, M.~D.}, \bibinfo{author}{Yeatman, J.~D.}, \&
  \bibinfo{author}{Mezer, A.~A.} (\bibinfo{year}{2017}).
\newblock \bibinfo{title}{Evaluating g-ratio weighted changes in the corpus
  callosum as a function of age and sex.}
\newblock {\it \bibinfo{journal}{NeuroImage}\/}, . \URLprefix
  \url{http://view.ncbi.nlm.nih.gov/pubmed/28673882}.
\bibitem[{Berthold et~al.(1983)Berthold, Nilsson \& Rydmark}]{Berthold83}
\bibinfo{author}{Berthold, C.~H.}, \bibinfo{author}{Nilsson, I.}, \&
  \bibinfo{author}{Rydmark, M.} (\bibinfo{year}{1983}).
\newblock \bibinfo{title}{Axon diameter and myelin sheath thickness in nerve
  fibres of the ventral spinal root of the seventh lumbar nerve of the adult
  and developing cat}.
\newblock {\it \bibinfo{journal}{J Anat}\/},  {\it \bibinfo{volume}{136}\/},
  \bibinfo{pages}{483--508}.
\bibitem[{Bjarnason et~al.(2005)Bjarnason, Vavasour, Chia \&
  MacKay}]{Bjarnason2005a}
\bibinfo{author}{Bjarnason, T.~A.}, \bibinfo{author}{Vavasour, I.~M.},
  \bibinfo{author}{Chia, C.~L.}, \& \bibinfo{author}{MacKay, A.~L.}
  (\bibinfo{year}{2005}).
\newblock \bibinfo{title}{Characterization of the {NMR} behavior of white
  matter in bovine brain.}
\newblock {\it \bibinfo{journal}{Magn. Reson. Med.}\/},  {\it
  \bibinfo{volume}{54}\/}, \bibinfo{pages}{1072--1081}. \URLprefix
  \url{http://dx.doi.org/10.1002/mrm.20680}. \DOIprefix\doi{10.1002/mrm.20680}.
\bibitem[{Boudreau et~al.(2017)Boudreau, Tardif, Stikov, Sled, Lee \&
  Pike}]{Boudreau2017a}
\bibinfo{author}{Boudreau, M.}, \bibinfo{author}{Tardif, C.~L.},
  \bibinfo{author}{Stikov, N.}, \bibinfo{author}{Sled, J.~G.},
  \bibinfo{author}{Lee, W.}, \& \bibinfo{author}{Pike, G.~B.}
  (\bibinfo{year}{2017}).
\newblock \bibinfo{title}{{B1} mapping for bias-correction in quantitative {T1}
  imaging of the brain at {3T} using standard pulse sequences.}
\newblock {\it \bibinfo{journal}{Journal of magnetic resonance imaging}\/}, .
  \URLprefix \url{http://view.ncbi.nlm.nih.gov/pubmed/28301086}.
\bibitem[{Bouhrara et~al.(2016)Bouhrara, Reiter, Celik, Fishbein, Kijowski \&
  Spencer}]{Bouhrara2016a}
\bibinfo{author}{Bouhrara, M.}, \bibinfo{author}{Reiter, D.~A.},
  \bibinfo{author}{Celik, H.}, \bibinfo{author}{Fishbein, K.~W.},
  \bibinfo{author}{Kijowski, R.}, \& \bibinfo{author}{Spencer, R.~G.}
  (\bibinfo{year}{2016}).
\newblock \bibinfo{title}{Analysis of {mcDESPOT}- and {CPMG}-derived parameter
  estimates for two-component nonexchanging systems}.
\newblock {\it \bibinfo{journal}{Magn. Reson. Med.}\/},  {\it
  \bibinfo{volume}{75}\/}, \bibinfo{pages}{2406--2420}. \URLprefix
  \url{http://dx.doi.org/10.1002/mrm.25801}. \DOIprefix\doi{10.1002/mrm.25801}.
\bibitem[{Bouyagoub et~al.(2016)Bouyagoub, Dowell, Hurley, Wood \&
  Cercignani}]{Bouyagoub2016a}
\bibinfo{author}{Bouyagoub, S.}, \bibinfo{author}{Dowell, N.~G.},
  \bibinfo{author}{Hurley, S.~A.}, \bibinfo{author}{Wood, T.~C.}, \&
  \bibinfo{author}{Cercignani, M.} (\bibinfo{year}{2016}).
\newblock \bibinfo{title}{{Overestimation of CSF fraction in NODDI: possible
  correction techniques and the effect on neurite density and orientation
  dispersion measures}}.
\newblock In {\it \bibinfo{booktitle}{ISMRM 2016}\/} (p.
  \bibinfo{pages}{0007}).
\bibitem[{Cabana et~al.(2015)Cabana, Gu, Boudreau, Levesque, Atchia, Sled,
  Narayanan, Arnold, Pike, Cohen-Adad, Duval, Vuong \& Stikov}]{Cabana2015a}
\bibinfo{author}{Cabana, J.-F.}, \bibinfo{author}{Gu, Y.},
  \bibinfo{author}{Boudreau, M.}, \bibinfo{author}{Levesque, I.~R.},
  \bibinfo{author}{Atchia, Y.}, \bibinfo{author}{Sled, J.~G.},
  \bibinfo{author}{Narayanan, S.}, \bibinfo{author}{Arnold, D.~L.},
  \bibinfo{author}{Pike, G.~B.}, \bibinfo{author}{Cohen-Adad, J.},
  \bibinfo{author}{Duval, T.}, \bibinfo{author}{Vuong, M.-T.}, \&
  \bibinfo{author}{Stikov, N.} (\bibinfo{year}{2015}).
\newblock \bibinfo{title}{Quantitative magnetization transfer imaging made easy
  with {qMTLab}: Software for data simulation, analysis, and visualization}.
\newblock {\it \bibinfo{journal}{Concepts Magn. Reson.}\/},  {\it
  \bibinfo{volume}{44A}\/}, \bibinfo{pages}{263--277}. \URLprefix
  \url{http://dx.doi.org/10.1002/cmr.a.21357}.
  \DOIprefix\doi{10.1002/cmr.a.21357}.
\bibitem[{Callaghan et~al.(2003)Callaghan, Godefroy \& Ryland}]{Callaghan2003a}
\bibinfo{author}{Callaghan, P.~T.}, \bibinfo{author}{Godefroy, S.}, \&
  \bibinfo{author}{Ryland, B.~N.} (\bibinfo{year}{2003}).
\newblock \bibinfo{title}{Diffusion–relaxation correlation in simple pore
  structures}.
\newblock {\it \bibinfo{journal}{Journal of Magnetic Resonance}\/},  {\it
  \bibinfo{volume}{162}\/}, \bibinfo{pages}{320--327}. \URLprefix
  \url{http://dx.doi.org/10.1016/s1090-7807(03)00056-9}.
  \DOIprefix\doi{10.1016/s1090-7807(03)00056-9}.
\bibitem[{Campbell et~al.(2016)Campbell, Leppert, Boudreau, Narayanan, Duval,
  Cohen-Adad, Pike \& Stikov}]{Campbell2016b}
\bibinfo{author}{Campbell, J.~S.}, \bibinfo{author}{Leppert, I.~R.},
  \bibinfo{author}{Boudreau, M.}, \bibinfo{author}{Narayanan, S.},
  \bibinfo{author}{Duval, T.}, \bibinfo{author}{Cohen-Adad, J.},
  \bibinfo{author}{Pike, G.~B.}, \& \bibinfo{author}{Stikov, N.}
  (\bibinfo{year}{2016}).
\newblock \bibinfo{title}{Caveats of miscalibration of myelin metrics for
  g-ratio imaging}.
\newblock In {\it \bibinfo{booktitle}{OHBM 2016}\/} (p. \bibinfo{pages}{1804}).
\bibitem[{Campbell et~al.(2014)Campbell, Stikov, Dougherty \&
  Pike}]{Campbell14a}
\bibinfo{author}{Campbell, J.~S.}, \bibinfo{author}{Stikov, N.},
  \bibinfo{author}{Dougherty, R.~F.}, \& \bibinfo{author}{Pike, G.~B.}
  (\bibinfo{year}{2014}).
\newblock \bibinfo{title}{Combined {NODDI} and q{MT} for full-brain g-ratio
  mapping with complex subvoxel microstructure}.
\newblock In {\it \bibinfo{booktitle}{ISMRM 2014}\/} (p. \bibinfo{pages}{393}).
\bibitem[{Campbell \& Pike(2013)}]{Campbell13b}
\bibinfo{author}{Campbell, J. S.~W.}, \& \bibinfo{author}{Pike, G.~B.}
  (\bibinfo{year}{2013}).
\newblock \bibinfo{title}{Potential and limitations of diffusion {MRI}
  tractography for the study of language}.
\newblock {\it \bibinfo{journal}{Brain and Language}\/},  {\it
  \bibinfo{volume}{131}\/}, \bibinfo{pages}{65--73}. \URLprefix
  \url{http://dx.doi.org/10.1016/j.bandl.2013.06.007}.
  \DOIprefix\doi{10.1016/j.bandl.2013.06.007}.
\bibitem[{Cercignani et~al.(2016{\natexlab{a}})Cercignani, Giulietti, Dowell,
  Spano, Harrison \& Bozzali}]{Cercignani2016b}
\bibinfo{author}{Cercignani, M.}, \bibinfo{author}{Giulietti, G.},
  \bibinfo{author}{Dowell, N.}, \bibinfo{author}{Spano, B.},
  \bibinfo{author}{Harrison, N.}, \& \bibinfo{author}{Bozzali, M.}
  (\bibinfo{year}{2016}{\natexlab{a}}).
\newblock \bibinfo{title}{A simple method to scale the macromolecular pool size
  ratio for computing the g-ratio in vivo}.
\newblock In {\it \bibinfo{booktitle}{ISMRM 2016}\/} (p.
  \bibinfo{pages}{3369}).
\bibitem[{Cercignani et~al.(2016{\natexlab{b}})Cercignani, Giulietti, Dowell,
  Gabel, Broad, Leigh, Harrison \& Bozzali}]{Cercignani2016c}
\bibinfo{author}{Cercignani, M.}, \bibinfo{author}{Giulietti, G.},
  \bibinfo{author}{Dowell, N.~G.}, \bibinfo{author}{Gabel, M.},
  \bibinfo{author}{Broad, R.}, \bibinfo{author}{Leigh, P.~N.},
  \bibinfo{author}{Harrison, N.~A.}, \& \bibinfo{author}{Bozzali, M.}
  (\bibinfo{year}{2016}{\natexlab{b}}).
\newblock \bibinfo{title}{Characterizing axonal myelination within the healthy
  population: a tract-by-tract mapping of effects of age and gender on the
  fiber g-ratio}.
\newblock {\it \bibinfo{journal}{Neurobiology of Aging}\/}, . \URLprefix
  \url{http://dx.doi.org/10.1016/j.neurobiolaging.2016.09.016}.
  \DOIprefix\doi{10.1016/j.neurobiolaging.2016.09.016}.
\bibitem[{Cercignani et~al.(2015)Cercignani, Giulietti, Span\'{o} \&
  Bozzali}]{Cercignani15a}
\bibinfo{author}{Cercignani, M.}, \bibinfo{author}{Giulietti, G.},
  \bibinfo{author}{Span\'{o}, B.}, \& \bibinfo{author}{Bozzali, M.}
  (\bibinfo{year}{2015}).
\newblock \bibinfo{title}{{Mapping the g-ratio within {MS} lesions}}.
\newblock In {\it \bibinfo{booktitle}{ISMRM 2015}\/} (p.
  \bibinfo{pages}{1402}).
\bibitem[{Chomiak \& Hu(2009)}]{Chomiak2009a}
\bibinfo{author}{Chomiak, T.}, \& \bibinfo{author}{Hu, B.}
  (\bibinfo{year}{2009}).
\newblock \bibinfo{title}{What is the optimal value of the g-ratio for
  myelinated fibers in the rat {CNS}? a theoretical approach}.
\newblock {\it \bibinfo{journal}{PLoS ONE}\/},  {\it \bibinfo{volume}{4}\/},
  \bibinfo{pages}{e7754+}. \URLprefix
  \url{http://dx.doi.org/10.1371/journal.pone.0007754}.
  \DOIprefix\doi{10.1371/journal.pone.0007754}.
\bibitem[{Cohen-Adad(2014)}]{Cohen-Adad2014a}
\bibinfo{author}{Cohen-Adad, J.} (\bibinfo{year}{2014}).
\newblock \bibinfo{title}{What can we learn from {T2*} maps of the cortex?}
\newblock {\it \bibinfo{journal}{NeuroImage}\/},  {\it \bibinfo{volume}{93}\/},
  \bibinfo{pages}{189--200}. \URLprefix
  \url{http://dx.doi.org/10.1016/j.neuroimage.2013.01.023}.
  \DOIprefix\doi{10.1016/j.neuroimage.2013.01.023}.
\bibitem[{Daducci et~al.(2015)Daducci, Canales-Rodr\'{\i}guez, Zhang, Dyrby,
  Alexander \& Thiran}]{Daducci2015a}
\bibinfo{author}{Daducci, A.}, \bibinfo{author}{Canales-Rodr\'{\i}guez, E.~J.},
  \bibinfo{author}{Zhang, H.}, \bibinfo{author}{Dyrby, T.~B.},
  \bibinfo{author}{Alexander, D.~C.}, \& \bibinfo{author}{Thiran, J.-P.}
  (\bibinfo{year}{2015}).
\newblock \bibinfo{title}{Accelerated microstructure imaging via convex
  optimization ({AMICO}) from diffusion {MRI} data}.
\newblock {\it \bibinfo{journal}{NeuroImage}\/},  {\it
  \bibinfo{volume}{105}\/}, \bibinfo{pages}{32--44}. \URLprefix
  \url{http://dx.doi.org/10.1016/j.neuroimage.2014.10.026}.
  \DOIprefix\doi{10.1016/j.neuroimage.2014.10.026}.
\bibitem[{De~Santis et~al.(2016{\natexlab{a}})De~Santis, Barazany, Jones \&
  Assaf}]{DeSantis16a}
\bibinfo{author}{De~Santis, S.}, \bibinfo{author}{Barazany, D.},
  \bibinfo{author}{Jones, D.~K.}, \& \bibinfo{author}{Assaf, Y.}
  (\bibinfo{year}{2016}{\natexlab{a}}).
\newblock \bibinfo{title}{Resolving relaxometry and diffusion properties within
  the same voxel in the presence of crossing fibres by combining inversion
  recovery and diffusion-weighted acquisitions}.
\newblock {\it \bibinfo{journal}{Magn. Reson. Med.}\/},  {\it
  \bibinfo{volume}{75}\/}, \bibinfo{pages}{372--380}. \URLprefix
  \url{http://dx.doi.org/10.1002/mrm.25644}. \DOIprefix\doi{10.1002/mrm.25644}.
\bibitem[{De~Santis et~al.(2016{\natexlab{b}})De~Santis, Jones \&
  Roebroeck}]{desantis2016b}
\bibinfo{author}{De~Santis, S.}, \bibinfo{author}{Jones, D.~K.}, \&
  \bibinfo{author}{Roebroeck, A.} (\bibinfo{year}{2016}{\natexlab{b}}).
\newblock \bibinfo{title}{Including diffusion time dependence in the
  extra-axonal space improves in vivo estimates of axonal diameter and density
  in human white matter}.
\newblock {\it \bibinfo{journal}{NeuroImage}\/},  {\it
  \bibinfo{volume}{130}\/}, \bibinfo{pages}{91--103}. \URLprefix
  \url{http://dx.doi.org/10.1016/j.neuroimage.2016.01.047}.
  \DOIprefix\doi{10.1016/j.neuroimage.2016.01.047}.
\bibitem[{Dean et~al.(2016)Dean, O'Muircheartaigh, Dirks, Travers, Adluru,
  Alexander \& Deoni}]{Dean2016a}
\bibinfo{author}{Dean, D.~C.}, \bibinfo{author}{O'Muircheartaigh, J.},
  \bibinfo{author}{Dirks, H.}, \bibinfo{author}{Travers, B.~G.},
  \bibinfo{author}{Adluru, N.}, \bibinfo{author}{Alexander, A.~L.}, \&
  \bibinfo{author}{Deoni, S. C.~L.} (\bibinfo{year}{2016}).
\newblock \bibinfo{title}{Mapping an index of the myelin g-ratio in infants
  using magnetic resonance imaging}.
\newblock {\it \bibinfo{journal}{NeuroImage}\/},  {\it
  \bibinfo{volume}{132}\/}, \bibinfo{pages}{225--237}. \URLprefix
  \url{http://dx.doi.org/10.1016/j.neuroimage.2016.02.040}.
  \DOIprefix\doi{10.1016/j.neuroimage.2016.02.040}.
\bibitem[{Deoni et~al.(2008)Deoni, Rutt, Arun, Pierpaoli \& Jones}]{Deoni2008a}
\bibinfo{author}{Deoni, S. C.~L.}, \bibinfo{author}{Rutt, B.~K.},
  \bibinfo{author}{Arun, T.}, \bibinfo{author}{Pierpaoli, C.}, \&
  \bibinfo{author}{Jones, D.~K.} (\bibinfo{year}{2008}).
\newblock \bibinfo{title}{Gleaning multicomponent {T1 and T2} information from
  steady-state imaging data}.
\newblock {\it \bibinfo{journal}{Magn. Reson. Med.}\/},  {\it
  \bibinfo{volume}{60}\/}, \bibinfo{pages}{1372--1387}. \URLprefix
  \url{http://dx.doi.org/10.1002/mrm.21704}. \DOIprefix\doi{10.1002/mrm.21704}.
\bibitem[{Descoteaux et~al.(2016)Descoteaux, Sidhu, Garyfallidis, Houde, Neher,
  Stieltjes \& Maier-Hein}]{Descoteaux2016a}
\bibinfo{author}{Descoteaux, M.}, \bibinfo{author}{Sidhu, J.},
  \bibinfo{author}{Garyfallidis, E.}, \bibinfo{author}{Houde, J.-C.},
  \bibinfo{author}{Neher, P.}, \bibinfo{author}{Stieltjes, B.}, \&
  \bibinfo{author}{Maier-Hein, K.~H.} (\bibinfo{year}{2016}).
\newblock \bibinfo{title}{False positive bundles in tractography}.
\newblock In {\it \bibinfo{booktitle}{ISMRM 2016}\/} (p. \bibinfo{pages}{790}).
\bibitem[{Does \& Gore(2000)}]{Does2000a}
\bibinfo{author}{Does, M.~D.}, \& \bibinfo{author}{Gore, J.~C.}
  (\bibinfo{year}{2000}).
\newblock \bibinfo{title}{Rapid acquisition transverse relaxometric imaging.}
\newblock {\it \bibinfo{journal}{Journal of magnetic resonance}\/},  {\it
  \bibinfo{volume}{147}\/}, \bibinfo{pages}{116--120}. \URLprefix
  \url{http://view.ncbi.nlm.nih.gov/pubmed/11042054}.
\bibitem[{Du et~al.(2014)Du, Ma, Li, Carl, Szeverenyi, VandenBerg, Corey-Bloom
  \& Bydder}]{Du2014a}
\bibinfo{author}{Du, J.}, \bibinfo{author}{Ma, G.}, \bibinfo{author}{Li, S.},
  \bibinfo{author}{Carl, M.}, \bibinfo{author}{Szeverenyi, N.~M.},
  \bibinfo{author}{VandenBerg, S.}, \bibinfo{author}{Corey-Bloom, J.}, \&
  \bibinfo{author}{Bydder, G.~M.} (\bibinfo{year}{2014}).
\newblock \bibinfo{title}{Ultrashort echo time ({UTE}) magnetic resonance
  imaging of the short t2 components in white matter of the brain using a
  clinical {3T} scanner.}
\newblock {\it \bibinfo{journal}{NeuroImage}\/},  {\it \bibinfo{volume}{87}\/},
  \bibinfo{pages}{32--41}. \URLprefix
  \url{http://dx.doi.org/10.1016/j.neuroimage.2013.10.053}.
  \DOIprefix\doi{10.1016/j.neuroimage.2013.10.053}.
\bibitem[{Du et~al.(2007)Du, Chu, Hwang, Brown, Kleinschmidt-DeMasters, Singel
  \& Simon}]{Du2007a}
\bibinfo{author}{Du, Y.~P.}, \bibinfo{author}{Chu, R.}, \bibinfo{author}{Hwang,
  D.}, \bibinfo{author}{Brown, M.~S.}, \bibinfo{author}{Kleinschmidt-DeMasters,
  B.~K.}, \bibinfo{author}{Singel, D.}, \& \bibinfo{author}{Simon, J.~H.}
  (\bibinfo{year}{2007}).
\newblock \bibinfo{title}{Fast multislice mapping of the myelin water fraction
  using multicompartment analysis of {T2*} decay at {3T}: a preliminary
  postmortem study.}
\newblock {\it \bibinfo{journal}{Magn. Reson. Med.}\/},  {\it
  \bibinfo{volume}{58}\/}, \bibinfo{pages}{865--870}. \URLprefix
  \url{http://dx.doi.org/10.1002/mrm.21409}. \DOIprefix\doi{10.1002/mrm.21409}.
\bibitem[{Dula et~al.(2010)Dula, Gochberg, Valentine, Valentine \&
  Does}]{Dula10}
\bibinfo{author}{Dula, A.~N.}, \bibinfo{author}{Gochberg, D.~F.},
  \bibinfo{author}{Valentine, H.~L.}, \bibinfo{author}{Valentine, W.~M.}, \&
  \bibinfo{author}{Does, M.~D.} (\bibinfo{year}{2010}).
\newblock \bibinfo{title}{Multiexponential {T2}, magnetization transfer, and
  quantitative histology in white matter tracts of rat spinal cord.}
\newblock {\it \bibinfo{journal}{Magn Reson Med}\/},  {\it
  \bibinfo{volume}{63}\/}, \bibinfo{pages}{902--909}.
  \DOIprefix\doi{10.1002/mrm.22267}.
\bibitem[{Duval et~al.(2016{\natexlab{a}})Duval, L\'{e}vy, Stikov, Campbell,
  Mezer, Witzel, Keil, Smith, Wald, Klawiter \& Cohen-Adad}]{Duval2016c}
\bibinfo{author}{Duval, T.}, \bibinfo{author}{L\'{e}vy, S.},
  \bibinfo{author}{Stikov, N.}, \bibinfo{author}{Campbell, J.},
  \bibinfo{author}{Mezer, A.}, \bibinfo{author}{Witzel, T.},
  \bibinfo{author}{Keil, B.}, \bibinfo{author}{Smith, V.},
  \bibinfo{author}{Wald, L.~L.}, \bibinfo{author}{Klawiter, E.}, \&
  \bibinfo{author}{Cohen-Adad, J.} (\bibinfo{year}{2016}{\natexlab{a}}).
\newblock \bibinfo{title}{{g-Ratio} weighted imaging of the human spinal cord
  in vivo}.
\newblock {\it \bibinfo{journal}{NeuroImage}\/}, . \URLprefix
  \url{http://dx.doi.org/10.1016/j.neuroimage.2016.09.018}.
  \DOIprefix\doi{10.1016/j.neuroimage.2016.09.018}.
\bibitem[{Duval et~al.(2015)Duval, Levy, Stikov, Mezer, Witzel, Keil, Smith,
  Wald, Klawiter \& Cohen-Adad}]{Duval15a}
\bibinfo{author}{Duval, T.}, \bibinfo{author}{Levy, S.},
  \bibinfo{author}{Stikov, N.}, \bibinfo{author}{Mezer, A.},
  \bibinfo{author}{Witzel, T.}, \bibinfo{author}{Keil, B.},
  \bibinfo{author}{Smith, V.}, \bibinfo{author}{Wald, L.~L.},
  \bibinfo{author}{Klawiter, E.~C.}, \& \bibinfo{author}{Cohen-Adad, J.}
  (\bibinfo{year}{2015}).
\newblock \bibinfo{title}{In vivo mapping of myelin g-ratio in the human spinal
  cord}.
\newblock In {\it \bibinfo{booktitle}{ISMRM 2015}\/} (p.~\bibinfo{pages}{5}).
\bibitem[{Duval et~al.(2016{\natexlab{b}})Duval, Perraud, Vuong, Rios, Stikov
  \& Cohen-Adad}]{Duval2016a}
\bibinfo{author}{Duval, T.}, \bibinfo{author}{Perraud, B.},
  \bibinfo{author}{Vuong, M.-T.}, \bibinfo{author}{Rios, N.~L.},
  \bibinfo{author}{Stikov, N.}, \& \bibinfo{author}{Cohen-Adad, J.}
  (\bibinfo{year}{2016}{\natexlab{b}}).
\newblock \bibinfo{title}{Validation of quantitative {MRI} metrics using full
  slice histology with automatic axon segmentation}.
\newblock In {\it \bibinfo{booktitle}{ISMRM 2016}\/} (p. \bibinfo{pages}{396}).
\bibitem[{Edwards et~al.(2016)Edwards, Pine, Weiskopf \&
  Mohammadi}]{Edwards2016a}
\bibinfo{author}{Edwards, L.~J.}, \bibinfo{author}{Pine, K.~J.},
  \bibinfo{author}{Weiskopf, N.}, \& \bibinfo{author}{Mohammadi, S.}
  (\bibinfo{year}{2016}).
\newblock \bibinfo{title}{{NODDI}-{DTI}: biophysical parameters from {DTI}
  data}.
\newblock In {\it \bibinfo{booktitle}{ISMRM Workshop on Breaking the Barriers
  of Diffusion MRI}\/} (p.~\bibinfo{pages}{8}).
\bibitem[{Ellerbrock \& Mohammadi(2017)}]{Ellerbrock2017a}
\bibinfo{author}{Ellerbrock, I.}, \& \bibinfo{author}{Mohammadi, S.}
  (\bibinfo{year}{2017}).
\newblock \bibinfo{title}{Comparing in vivo {MR} g-ratio mapping methods:
  accuracy and precision at the group level}.
\newblock In {\it \bibinfo{booktitle}{ISMRM 2017}\/} (p. \bibinfo{pages}{311}).
\bibitem[{Ferizi et~al.(2015)Ferizi, Schneider, Witzel, Wald, Zhang,
  Wheeler-Kingshott \& Alexander}]{Ferizi2015a}
\bibinfo{author}{Ferizi, U.}, \bibinfo{author}{Schneider, T.},
  \bibinfo{author}{Witzel, T.}, \bibinfo{author}{Wald, L.~L.},
  \bibinfo{author}{Zhang, H.}, \bibinfo{author}{Wheeler-Kingshott, C. A.~M.},
  \& \bibinfo{author}{Alexander, D.~C.} (\bibinfo{year}{2015}).
\newblock \bibinfo{title}{White matter compartment models for in vivo diffusion
  {MRI} at {300mT}/m}.
\newblock {\it \bibinfo{journal}{NeuroImage}\/},  {\it
  \bibinfo{volume}{118}\/}, \bibinfo{pages}{468--483}. \URLprefix
  \url{http://dx.doi.org/10.1016/j.neuroimage.2015.06.027}.
  \DOIprefix\doi{10.1016/j.neuroimage.2015.06.027}.
\bibitem[{Fieremans et~al.(2016)Fieremans, Burcaw, Lee, Lemberskiy, Veraart \&
  Novikov}]{Fieremans2016a}
\bibinfo{author}{Fieremans, E.}, \bibinfo{author}{Burcaw, L.~M.},
  \bibinfo{author}{Lee, H.-H.}, \bibinfo{author}{Lemberskiy, G.},
  \bibinfo{author}{Veraart, J.}, \& \bibinfo{author}{Novikov, D.~S.}
  (\bibinfo{year}{2016}).
\newblock \bibinfo{title}{In vivo observation and biophysical interpretation of
  time-dependent diffusion in human white matter}.
\newblock {\it \bibinfo{journal}{NeuroImage}\/},  {\it
  \bibinfo{volume}{129}\/}, \bibinfo{pages}{414--427}. \URLprefix
  \url{http://dx.doi.org/10.1016/j.neuroimage.2016.01.018}.
  \DOIprefix\doi{10.1016/j.neuroimage.2016.01.018}.
\bibitem[{Fieremans et~al.(2008)Fieremans, De~Deene, Delputte, \"{O}zdemir,
  D'Asseler, Vlassenbroeck, Deblaere, Achten \& Lemahieu}]{Fieremans2008a}
\bibinfo{author}{Fieremans, E.}, \bibinfo{author}{De~Deene, Y.},
  \bibinfo{author}{Delputte, S.}, \bibinfo{author}{\"{O}zdemir, M.~S.},
  \bibinfo{author}{D'Asseler, Y.}, \bibinfo{author}{Vlassenbroeck, J.},
  \bibinfo{author}{Deblaere, K.}, \bibinfo{author}{Achten, E.}, \&
  \bibinfo{author}{Lemahieu, I.} (\bibinfo{year}{2008}).
\newblock \bibinfo{title}{Simulation and experimental verification of the
  diffusion in an anisotropic fiber phantom}.
\newblock {\it \bibinfo{journal}{Journal of Magnetic Resonance}\/},  {\it
  \bibinfo{volume}{190}\/}, \bibinfo{pages}{189--199}. \URLprefix
  \url{http://dx.doi.org/10.1016/j.jmr.2007.10.014}.
  \DOIprefix\doi{10.1016/j.jmr.2007.10.014}.
\bibitem[{Fieremans et~al.(2011)Fieremans, Jensen \& Helpern}]{Fieremans2011a}
\bibinfo{author}{Fieremans, E.}, \bibinfo{author}{Jensen, J.~H.}, \&
  \bibinfo{author}{Helpern, J.~A.} (\bibinfo{year}{2011}).
\newblock \bibinfo{title}{White matter characterization with diffusional
  kurtosis imaging}.
\newblock {\it \bibinfo{journal}{NeuroImage}\/},  {\it \bibinfo{volume}{58}\/},
  \bibinfo{pages}{177--188}. \URLprefix
  \url{http://dx.doi.org/10.1016/j.neuroimage.2011.06.006}.
  \DOIprefix\doi{10.1016/j.neuroimage.2011.06.006}.
\bibitem[{Fram et~al.(1987)Fram, Herfkens, Johnson, Glover, Karis, Shimakawa,
  Perkins \& Pelc}]{Fram1987a}
\bibinfo{author}{Fram, E.~K.}, \bibinfo{author}{Herfkens, R.~J.},
  \bibinfo{author}{Johnson, G.~A.}, \bibinfo{author}{Glover, G.~H.},
  \bibinfo{author}{Karis, J.~P.}, \bibinfo{author}{Shimakawa, A.},
  \bibinfo{author}{Perkins, T.~G.}, \& \bibinfo{author}{Pelc, N.~J.}
  (\bibinfo{year}{1987}).
\newblock \bibinfo{title}{Rapid calculation of {T1} using variable flip angle
  gradient refocused imaging.}
\newblock {\it \bibinfo{journal}{Magnetic resonance imaging}\/},  {\it
  \bibinfo{volume}{5}\/}, \bibinfo{pages}{201--208}. \URLprefix
  \url{http://view.ncbi.nlm.nih.gov/pubmed/3626789}.
\bibitem[{Garcia et~al.(2012)Garcia, Gloor, Radue, Stippich, Wetzel, Scheffler
  \& Bieri}]{Garcia2012}
\bibinfo{author}{Garcia, M.}, \bibinfo{author}{Gloor, M.},
  \bibinfo{author}{Radue, E.-W.~W.}, \bibinfo{author}{Stippich, C.~h.},
  \bibinfo{author}{Wetzel, S.~G.}, \bibinfo{author}{Scheffler, K.}, \&
  \bibinfo{author}{Bieri, O.} (\bibinfo{year}{2012}).
\newblock \bibinfo{title}{Fast high-resolution brain imaging with balanced
  {SSFP}: Interpretation of quantitative magnetization transfer towards simple
  {MTR}.}
\newblock {\it \bibinfo{journal}{NeuroImage}\/},  {\it \bibinfo{volume}{59}\/},
  \bibinfo{pages}{202--211}. \URLprefix
  \url{http://view.ncbi.nlm.nih.gov/pubmed/21820061}.
\bibitem[{Gareau et~al.(2000)Gareau, Rutt, Karlik \& Mitchell}]{Gareau2000a}
\bibinfo{author}{Gareau, P.~J.}, \bibinfo{author}{Rutt, B.~K.},
  \bibinfo{author}{Karlik, S.~J.}, \& \bibinfo{author}{Mitchell, J.~R.}
  (\bibinfo{year}{2000}).
\newblock \bibinfo{title}{Magnetization transfer and multicomponent t2
  relaxation measurements with histopathologic correlation in an experimental
  model of {MS}}.
\newblock {\it \bibinfo{journal}{J. Magn. Reson. Imaging}\/},  {\it
  \bibinfo{volume}{11}\/}, \bibinfo{pages}{586--595}. \URLprefix
  \url{http://dx.doi.org/10.1002/1522-2586(200006)11:6\%3C586::
  aid-jmri3\%3E3.0.co;2-v}.
  \DOIprefix\doi{10.1002/1522-2586(200006)11:6\%3C586::aid-jmri3\%3E3.0.co;2-v}.
\bibitem[{Ghosh et~al.(2016)Ghosh, Alexander \& Zhang}]{Ghosh2016a}
\bibinfo{author}{Ghosh, A.}, \bibinfo{author}{Alexander, D.~C.}, \&
  \bibinfo{author}{Zhang, H.} (\bibinfo{year}{2016}).
\newblock \bibinfo{title}{To be dispersed or not to be dispersed: A study using
  {HCP} data}.
\newblock In {\it \bibinfo{booktitle}{ISMRM 2016}\/} (p. \bibinfo{pages}{792}).
\bibitem[{Girard et~al.(2015)Girard, Fick, Descoteaux, Deriche \&
  Wassermann}]{Girard2015a}
\bibinfo{author}{Girard, G.}, \bibinfo{author}{Fick, R.},
  \bibinfo{author}{Descoteaux, M.}, \bibinfo{author}{Deriche, R.}, \&
  \bibinfo{author}{Wassermann, D.} (\bibinfo{year}{2015}).
\newblock \bibinfo{title}{{AxTract}: {Microstructure-Driven} tractography based
  on the ensemble average propagator.}
\newblock {\it \bibinfo{journal}{IPMI 2015}\/},  {\it \bibinfo{volume}{24}\/},
  \bibinfo{pages}{675--686}. \URLprefix
  \url{http://view.ncbi.nlm.nih.gov/pubmed/26221712}.
\bibitem[{Glasser \& Van~Essen(2011)}]{Glasser2011a}
\bibinfo{author}{Glasser, M.~F.}, \& \bibinfo{author}{Van~Essen, D.~C.}
  (\bibinfo{year}{2011}).
\newblock \bibinfo{title}{Mapping human cortical areas in vivo based on myelin
  content as revealed by t1- and t2-weighted {MRI}.}
\newblock {\it \bibinfo{journal}{J. Neurosci.}\/},  {\it
  \bibinfo{volume}{31}\/}, \bibinfo{pages}{11597--11616}. \URLprefix
  \url{http://dx.doi.org/10.1523/jneurosci.2180-11.2011}.
  \DOIprefix\doi{10.1523/jneurosci.2180-11.2011}.
\bibitem[{Grussu et~al.(2014)Grussu, Schneider, Zhang, Alexander \&
  Wheeler-Kingshott}]{GrussuISMRM14}
\bibinfo{author}{Grussu, F.}, \bibinfo{author}{Schneider, T.},
  \bibinfo{author}{Zhang, H.}, \bibinfo{author}{Alexander, D.~C.}, \&
  \bibinfo{author}{Wheeler-Kingshott, C. A.~M.} (\bibinfo{year}{2014}).
\newblock \bibinfo{title}{Single shell diffusion {MRI NODDI} with in vivo
  cervical cord data}.
\newblock In {\it \bibinfo{booktitle}{ISMRM 2014}\/} (p.
  \bibinfo{pages}{1716}).
\bibitem[{Gupta et~al.(2003)Gupta, Rao, Kasiviswanathan, Chawla, Kumar \&
  Venkatesan.}]{Gupta2003a}
\bibinfo{author}{Gupta, R.}, \bibinfo{author}{Rao, A.},
  \bibinfo{author}{Kasiviswanathan, A.}, \bibinfo{author}{Chawla, S.},
  \bibinfo{author}{Kumar, R.}, \& \bibinfo{author}{Venkatesan., R.}
  (\bibinfo{year}{2003}).
\newblock \bibinfo{title}{Diffusion weighted {EPI} with magnetization transfer
  contrast}.
\newblock In {\it \bibinfo{booktitle}{ISMRM 2003}\/} (p.~\bibinfo{pages}{68}).
\bibitem[{Harkins et~al.(2012)Harkins, Dula \& Does}]{Harkins2012a}
\bibinfo{author}{Harkins, K.~D.}, \bibinfo{author}{Dula, A.~N.}, \&
  \bibinfo{author}{Does, M.~D.} (\bibinfo{year}{2012}).
\newblock \bibinfo{title}{Effect of intercompartmental water exchange on the
  apparent myelin water fraction in multiexponential {T2} measurements of rat
  spinal cord.}
\newblock {\it \bibinfo{journal}{Magn. Reson. Med.}\/},  {\it
  \bibinfo{volume}{67}\/}, \bibinfo{pages}{793--800}. \URLprefix
  \url{http://dx.doi.org/10.1002/mrm.23053}. \DOIprefix\doi{10.1002/mrm.23053}.
\bibitem[{Harkins et~al.(2016)Harkins, Xu, Dula, Li, Valentine, Gochberg, Gore
  \& Does}]{Harkins2016a}
\bibinfo{author}{Harkins, K.~D.}, \bibinfo{author}{Xu, J.},
  \bibinfo{author}{Dula, A.~N.}, \bibinfo{author}{Li, K.},
  \bibinfo{author}{Valentine, W.~M.}, \bibinfo{author}{Gochberg, D.~F.},
  \bibinfo{author}{Gore, J.~C.}, \& \bibinfo{author}{Does, M.~D.}
  (\bibinfo{year}{2016}).
\newblock \bibinfo{title}{The microstructural correlates of {T1} in white
  matter}.
\newblock {\it \bibinfo{journal}{Magn. Reson. Med.}\/},  {\it
  \bibinfo{volume}{75}\/}, \bibinfo{pages}{1341--1345}. \URLprefix
  \url{http://dx.doi.org/10.1002/mrm.25709}. \DOIprefix\doi{10.1002/mrm.25709}.
\bibitem[{Helms et~al.(2008)Helms, Dathe, Kallenberg \& Dechent}]{Helms2008}
\bibinfo{author}{Helms, G.}, \bibinfo{author}{Dathe, H.},
  \bibinfo{author}{Kallenberg, K.}, \& \bibinfo{author}{Dechent, P.}
  (\bibinfo{year}{2008}).
\newblock \bibinfo{title}{High-resolution maps of magnetization transfer with
  inherent correction for {RF} inhomogeneity and {T1} relaxation obtained from
  {3D} {FLASH} {MRI}}.
\newblock {\it \bibinfo{journal}{Magn. Reson. Med.}\/},  {\it
  \bibinfo{volume}{60}\/}, \bibinfo{pages}{1396--1407}. \URLprefix
  \url{http://dx.doi.org/10.1002/mrm.21732}. \DOIprefix\doi{10.1002/mrm.21732}.
\bibitem[{Helms et~al.(2010)Helms, Dathe, Kallenberg \&
  Dechent}]{Helms2010erratum}
\bibinfo{author}{Helms, G.}, \bibinfo{author}{Dathe, H.},
  \bibinfo{author}{Kallenberg, K.}, \& \bibinfo{author}{Dechent, P.}
  (\bibinfo{year}{2010}).
\newblock \bibinfo{title}{{Erratum to: Helms, Dathe, Kallenberg and Dechent,
  High-resolution maps of magnetization transfer with inherent correction for
  RF inhomogeneity and {T1} relaxation obtained from {3D} {FLASH} {MRI}.
  \emph{Magn. Reson. Med.} 2008; 60(6):1396-1407}}.
\newblock {\it \bibinfo{journal}{Magn. Reson. Med.}\/},  {\it
  \bibinfo{volume}{64}\/}, \bibinfo{pages}{1856}. \URLprefix
  \url{http://dx.doi.org/10.1002/mrm.22607}. \DOIprefix\doi{10.1002/mrm.22607}.
\bibitem[{Hildebrand \& Hahn(1978)}]{Hildebrand78}
\bibinfo{author}{Hildebrand, C.}, \& \bibinfo{author}{Hahn, R.}
  (\bibinfo{year}{1978}).
\newblock \bibinfo{title}{Relation between myelin sheath thickness and axon
  size in spinal cord white matter of some vertebrate species.}
\newblock {\it \bibinfo{journal}{J Neurol Sci}\/},  {\it
  \bibinfo{volume}{38}\/}, \bibinfo{pages}{421--434}.
\bibitem[{Hori(2016)}]{Hori2016b}
\bibinfo{author}{Hori, M.} (\bibinfo{year}{2016}).
\newblock \bibinfo{title}{Toward clinically feasible acquisition protocol for
  g-ratio}.
\newblock In {\it \bibinfo{booktitle}{{QBIN} Workshop: Toward a super-big
  brain: promises and pitfalls of microstructural imaging}\/}.
\bibitem[{Hori et~al.(2016)Hori, Stikov, Nojiri, Murata, Ishigame, Kamiya,
  Suzuki, Kamagata \& Aoki}]{Hori2016a}
\bibinfo{author}{Hori, M.}, \bibinfo{author}{Stikov, N.},
  \bibinfo{author}{Nojiri, Y., Ryuji ad~Tsurushima}, \bibinfo{author}{Murata,
  K.}, \bibinfo{author}{Ishigame, K.}, \bibinfo{author}{Kamiya, K.},
  \bibinfo{author}{Suzuki, Y.}, \bibinfo{author}{Kamagata, K.}, \&
  \bibinfo{author}{Aoki, S.} (\bibinfo{year}{2016}).
\newblock \bibinfo{title}{Magnetic resonance myelin g-ratio mapping for the
  brain and cervical spinal cord: 10 minutes protocol for clinical
  application}.
\newblock In {\it \bibinfo{booktitle}{ISMRM 2016}\/} (p.
  \bibinfo{pages}{3377}).
\bibitem[{Hutchinson et~al.(2016)Hutchinson, Avram, Komlosh, Irfanoglu,
  Barnett, Ozarslan, Schwerin, Radomski, Juliano \&
  Pierpaoli}]{Hutchinson2016a}
\bibinfo{author}{Hutchinson, E.~B.}, \bibinfo{author}{Avram, A.},
  \bibinfo{author}{Komlosh, M.}, \bibinfo{author}{Irfanoglu, M.~O.},
  \bibinfo{author}{Barnett, A.}, \bibinfo{author}{Ozarslan, E.},
  \bibinfo{author}{Schwerin, S.}, \bibinfo{author}{Radomski, K.},
  \bibinfo{author}{Juliano, S.}, \& \bibinfo{author}{Pierpaoli, C.}
  (\bibinfo{year}{2016}).
\newblock \bibinfo{title}{A systematic comparative study of dti and higher
  order diffusion models in brain fixed tissue}.
\newblock In {\it \bibinfo{booktitle}{ISMRM 2016}\/} (p.
  \bibinfo{pages}{1048}).
\bibitem[{Hwang et~al.(2010)Hwang, Kim \& Du}]{Hwang2010a}
\bibinfo{author}{Hwang, D.}, \bibinfo{author}{Kim, D.-H.}, \&
  \bibinfo{author}{Du, Y.~P.} (\bibinfo{year}{2010}).
\newblock \bibinfo{title}{{In vivo multi-slice mapping of myelin water content
  using T2* decay}}.
\newblock {\it \bibinfo{journal}{NeuroImage}\/},  {\it \bibinfo{volume}{52}\/},
  \bibinfo{pages}{198--204}. \URLprefix
  \url{http://dx.doi.org/10.1016/j.neuroimage.2010.04.023}.
  \DOIprefix\doi{10.1016/j.neuroimage.2010.04.023}.
\bibitem[{Jelescu et~al.(2015)Jelescu, Veraart, Adisetiyo, Milla, Novikov \&
  Fieremans}]{Jelescu2015a}
\bibinfo{author}{Jelescu, I.~O.}, \bibinfo{author}{Veraart, J.},
  \bibinfo{author}{Adisetiyo, V.}, \bibinfo{author}{Milla, S.~S.},
  \bibinfo{author}{Novikov, D.~S.}, \& \bibinfo{author}{Fieremans, E.}
  (\bibinfo{year}{2015}).
\newblock \bibinfo{title}{One diffusion acquisition and different white matter
  models: How does microstructure change in human early development based on
  {WMTI} and {NODDI}?}
\newblock {\it \bibinfo{journal}{NeuroImage}\/},  {\it
  \bibinfo{volume}{107}\/}, \bibinfo{pages}{242--256}. \URLprefix
  \url{http://dx.doi.org/10.1016/j.neuroimage.2014.12.009}.
  \DOIprefix\doi{10.1016/j.neuroimage.2014.12.009}.
\bibitem[{Jelescu et~al.(2016{\natexlab{a}})Jelescu, Veraart, Fieremans \&
  Novikov}]{Jelescu2016a}
\bibinfo{author}{Jelescu, I.~O.}, \bibinfo{author}{Veraart, J.},
  \bibinfo{author}{Fieremans, E.}, \& \bibinfo{author}{Novikov, D.~S.}
  (\bibinfo{year}{2016}{\natexlab{a}}).
\newblock \bibinfo{title}{Degeneracy in model parameter estimation for
  multi-compartmental diffusion in neuronal tissue}.
\newblock {\it \bibinfo{journal}{NMR Biomed.}\/},  {\it
  \bibinfo{volume}{29}\/}, \bibinfo{pages}{33--47}. \URLprefix
  \url{http://dx.doi.org/10.1002/nbm.3450}. \DOIprefix\doi{10.1002/nbm.3450}.
\bibitem[{Jelescu et~al.(2016{\natexlab{b}})Jelescu, Zurek, Winters, Veraart,
  Rajaratnam, Kim, Babb, Shepherd, Novikov, Kim \& Fieremans}]{Jelescu2016b}
\bibinfo{author}{Jelescu, I.~O.}, \bibinfo{author}{Zurek, M.},
  \bibinfo{author}{Winters, K.~V.}, \bibinfo{author}{Veraart, J.},
  \bibinfo{author}{Rajaratnam, A.}, \bibinfo{author}{Kim, N.~S.},
  \bibinfo{author}{Babb, J.~S.}, \bibinfo{author}{Shepherd, T.~M.},
  \bibinfo{author}{Novikov, D.~S.}, \bibinfo{author}{Kim, S.~G.}, \&
  \bibinfo{author}{Fieremans, E.} (\bibinfo{year}{2016}{\natexlab{b}}).
\newblock \bibinfo{title}{In vivo quantification of demyelination and recovery
  using compartment-specific diffusion {MRI} metrics validated by electron
  microscopy}.
\newblock {\it \bibinfo{journal}{NeuroImage}\/},  {\it
  \bibinfo{volume}{132}\/}, \bibinfo{pages}{104--114}. \URLprefix
  \url{http://dx.doi.org/10.1016/j.neuroimage.2016.02.004}.
  \DOIprefix\doi{10.1016/j.neuroimage.2016.02.004}.
\bibitem[{Jespersen et~al.(2010)Jespersen, Bjarkam, Nyengaard, Chakravarty,
  Hansen, Vosegaard, {\O}stergaard, Yablonskiy, Nielsen \&
  Vestergaard-Poulsen}]{Jespersen2010a}
\bibinfo{author}{Jespersen, S.~N.}, \bibinfo{author}{Bjarkam, C.~R.},
  \bibinfo{author}{Nyengaard, J.~R.}, \bibinfo{author}{Chakravarty, M.~M.},
  \bibinfo{author}{Hansen, B.}, \bibinfo{author}{Vosegaard, T.},
  \bibinfo{author}{{\O}stergaard, L.}, \bibinfo{author}{Yablonskiy, D.},
  \bibinfo{author}{Nielsen, N. C.~C.}, \& \bibinfo{author}{Vestergaard-Poulsen,
  P.} (\bibinfo{year}{2010}).
\newblock \bibinfo{title}{Neurite density from magnetic resonance diffusion
  measurements at ultrahigh field: comparison with light microscopy and
  electron microscopy.}
\newblock {\it \bibinfo{journal}{NeuroImage}\/},  {\it \bibinfo{volume}{49}\/},
  \bibinfo{pages}{205--216}. \URLprefix
  \url{http://dx.doi.org/10.1016/j.neuroimage.2009.08.053}.
  \DOIprefix\doi{10.1016/j.neuroimage.2009.08.053}.
\bibitem[{Jeurissen et~al.(2010)Jeurissen, Leemans, Tournier, Jones \&
  Sijbers}]{Jeurissen2010a}
\bibinfo{author}{Jeurissen, B.}, \bibinfo{author}{Leemans, A.},
  \bibinfo{author}{Tournier, J.-D.}, \bibinfo{author}{Jones, D.~K.}, \&
  \bibinfo{author}{Sijbers, J.} (\bibinfo{year}{2010}).
\newblock \bibinfo{title}{Estimating the number of fiber orientations in
  diffusion {MRI} voxels: a constrained spherical deconvolution study.}
\newblock In {\it \bibinfo{booktitle}{ISMRM 2010}\/} (p. \bibinfo{pages}{573}).
\bibitem[{Jones et~al.(2004)Jones, Xiang, Whittall \& MacKay}]{JonesCK2004a}
\bibinfo{author}{Jones, C.~K.}, \bibinfo{author}{Xiang, Q.-S.~S.},
  \bibinfo{author}{Whittall, K.~P.}, \& \bibinfo{author}{MacKay, A.~L.}
  (\bibinfo{year}{2004}).
\newblock \bibinfo{title}{Linear combination of multiecho data: short t2
  component selection.}
\newblock {\it \bibinfo{journal}{Magn. Reson. Med.}\/},  {\it
  \bibinfo{volume}{51}\/}, \bibinfo{pages}{495--502}. \URLprefix
  \url{http://dx.doi.org/10.1002/mrm.10713}. \DOIprefix\doi{10.1002/mrm.10713}.
\bibitem[{Jung et~al.(2016)Jung, Nam, Zhang \& Lee}]{Jung2016a}
\bibinfo{author}{Jung, W.}, \bibinfo{author}{Nam, Y.}, \bibinfo{author}{Zhang,
  H.}, \& \bibinfo{author}{Lee, J.} (\bibinfo{year}{2016}).
\newblock \bibinfo{title}{{Whole brain in-vivo g-ratio mapping using neurite
  orientation dispersion and density imaging (NODDI) and GRE myelin water
  imaging (GRE-MWI)}}.
\newblock In {\it \bibinfo{booktitle}{ISMRM 2016}\/} (p.
  \bibinfo{pages}{1112}).
\bibitem[{Kaden et~al.(2016)Kaden, Kelm, Carson, Does \&
  Alexander}]{Kaden2016a}
\bibinfo{author}{Kaden, E.}, \bibinfo{author}{Kelm, N.~D.},
  \bibinfo{author}{Carson, R.~P.}, \bibinfo{author}{Does, M.~D.}, \&
  \bibinfo{author}{Alexander, D.~C.} (\bibinfo{year}{2016}).
\newblock \bibinfo{title}{Multi-compartment microscopic diffusion imaging.}
\newblock {\it \bibinfo{journal}{NeuroImage}\/},  {\it
  \bibinfo{volume}{139}\/}, \bibinfo{pages}{346--359}. \URLprefix
  \url{http://view.ncbi.nlm.nih.gov/pubmed/27282476}.
\bibitem[{Graf~von Keyserlingk \& Schramm(1984)}]{Graf1984}
\bibinfo{author}{Graf~von Keyserlingk, D.}, \& \bibinfo{author}{Schramm, U.}
  (\bibinfo{year}{1984}).
\newblock \bibinfo{title}{Diameter of axons and thickness of myelin sheaths of
  the pyramidal tract fibres in the adult human medullary pyramid.}
\newblock {\it \bibinfo{journal}{Anatomischer Anzeiger}\/},  {\it
  \bibinfo{volume}{157}\/}, \bibinfo{pages}{97--111}. \URLprefix
  \url{http://view.ncbi.nlm.nih.gov/pubmed/6507887}.
\bibitem[{Kim et~al.(2016)Kim, Kim,  \& Haldar}]{Kim2016a}
\bibinfo{author}{Kim, D.}, \bibinfo{author}{Kim, J.~H.}, , \&
  \bibinfo{author}{Haldar, J.~P.} (\bibinfo{year}{2016}).
\newblock \bibinfo{title}{Diffusion-relaxation correlation spectroscopic
  imaging ({DR-CSI}): An enhanced approach to imaging microstructure}.
\newblock In {\it \bibinfo{booktitle}{ISMRM 2016}\/} (p. \bibinfo{pages}{660}).
\bibitem[{Kiselev \& Il'yasov(2007)}]{Kiselev2007a}
\bibinfo{author}{Kiselev, V.~G.}, \& \bibinfo{author}{Il'yasov, K.~A.}
  (\bibinfo{year}{2007}).
\newblock \bibinfo{title}{Is the ” biexponential diffusion” biexponential?}
\newblock {\it \bibinfo{journal}{Magn. Reson. Med.}\/},  {\it
  \bibinfo{volume}{57}\/}, \bibinfo{pages}{464--469}. \URLprefix
  \url{http://dx.doi.org/10.1002/mrm.21164}. \DOIprefix\doi{10.1002/mrm.21164}.
\bibitem[{Kucharczyk et~al.(1994)Kucharczyk, Macdonald, Stanisz \&
  Henkelman}]{Kucharczyk1994}
\bibinfo{author}{Kucharczyk, W.}, \bibinfo{author}{Macdonald, P.~M.},
  \bibinfo{author}{Stanisz, G.~J.}, \& \bibinfo{author}{Henkelman, R.~M.}
  (\bibinfo{year}{1994}).
\newblock \bibinfo{title}{Relaxivity and magnetization transfer of white matter
  lipids at {MR} imaging: importance of cerebrosides and {pH}.}
\newblock {\it \bibinfo{journal}{Radiology}\/},  {\it \bibinfo{volume}{192}\/},
  \bibinfo{pages}{521--529}. \URLprefix
  \url{http://view.ncbi.nlm.nih.gov/pubmed/8029426}.
\bibitem[{LaMantia \& Rakic(1990)}]{lamantia1990a}
\bibinfo{author}{LaMantia, A.~S.}, \& \bibinfo{author}{Rakic, P.}
  (\bibinfo{year}{1990}).
\newblock \bibinfo{title}{Cytological and quantitative characteristics of four
  cerebral commissures in the rhesus monkey.}
\newblock {\it \bibinfo{journal}{The Journal of comparative neurology}\/},
  {\it \bibinfo{volume}{291}\/}, \bibinfo{pages}{520--537}. \URLprefix
  \url{http://dx.doi.org/10.1002/cne.902910404}.
  \DOIprefix\doi{10.1002/cne.902910404}.
\bibitem[{Lampinen et~al.(2017)Lampinen, Szczepankiewicz, M\r{a}rtensson, van
  Westen, Sundgren \& Nilsson}]{Lampinen2017a}
\bibinfo{author}{Lampinen, B.}, \bibinfo{author}{Szczepankiewicz, F.},
  \bibinfo{author}{M\r{a}rtensson, J.}, \bibinfo{author}{van Westen, D.},
  \bibinfo{author}{Sundgren, P.~C.}, \& \bibinfo{author}{Nilsson, M.}
  (\bibinfo{year}{2017}).
\newblock \bibinfo{title}{Neurite density imaging versus imaging of microscopic
  anisotropy in diffusion {MRI}: A model comparison using spherical tensor
  encoding}.
\newblock {\it \bibinfo{journal}{NeuroImage}\/},  {\it
  \bibinfo{volume}{147}\/}, \bibinfo{pages}{517--531}. \URLprefix
  \url{http://dx.doi.org/10.1016/j.neuroimage.2016.11.053}.
  \DOIprefix\doi{10.1016/j.neuroimage.2016.11.053}.
\bibitem[{Lampron et~al.(2015)Lampron, Larochelle, Laflamme, Pr\'{e}fontaine,
  Plante, S\'{a}nchez, Yong, Stys, Tremblay \& Rivest}]{Lampron2015a}
\bibinfo{author}{Lampron, A.}, \bibinfo{author}{Larochelle, A.},
  \bibinfo{author}{Laflamme, N.}, \bibinfo{author}{Pr\'{e}fontaine, P.},
  \bibinfo{author}{Plante, M.-M.}, \bibinfo{author}{S\'{a}nchez, M.~G.},
  \bibinfo{author}{Yong, V.~W.}, \bibinfo{author}{Stys, P.~K.},
  \bibinfo{author}{Tremblay, M.-E.}, \& \bibinfo{author}{Rivest, S.}
  (\bibinfo{year}{2015}).
\newblock \bibinfo{title}{Inefficient clearance of myelin debris by microglia
  impairs remyelinating processes}.
\newblock {\it \bibinfo{journal}{Journal of Experimental Medicine}\/},  {\it
  \bibinfo{volume}{212}\/}, \bibinfo{pages}{481--495}. \URLprefix
  \url{http://dx.doi.org/10.1084/jem.20141656}.
  \DOIprefix\doi{10.1084/jem.20141656}.
\bibitem[{Lankford \& Does(2013)}]{Lankford2013a}
\bibinfo{author}{Lankford, C.~L.}, \& \bibinfo{author}{Does, M.~D.}
  (\bibinfo{year}{2013}).
\newblock \bibinfo{title}{On the inherent precision of {mcDESPOT}}.
\newblock {\it \bibinfo{journal}{Magn Reson Med}\/},  {\it
  \bibinfo{volume}{69}\/}, \bibinfo{pages}{127--136}. \URLprefix
  \url{http://dx.doi.org/10.1002/mrm.24241}. \DOIprefix\doi{10.1002/mrm.24241}.
\bibitem[{Laule et~al.(2006)Laule, Leung, Lis, Traboulsee, Paty, MacKay \&
  Moore}]{Laule2006a}
\bibinfo{author}{Laule, C.}, \bibinfo{author}{Leung, E.}, \bibinfo{author}{Lis,
  D.~K.}, \bibinfo{author}{Traboulsee, A.~L.}, \bibinfo{author}{Paty, D.~W.},
  \bibinfo{author}{MacKay, A.~L.}, \& \bibinfo{author}{Moore, G.~R.}
  (\bibinfo{year}{2006}).
\newblock \bibinfo{title}{Myelin water imaging in multiple sclerosis:
  quantitative correlations with histopathology.}
\newblock {\it \bibinfo{journal}{Multiple sclerosis}\/},  {\it
  \bibinfo{volume}{12}\/}, \bibinfo{pages}{747--753}. \URLprefix
  \url{http://view.ncbi.nlm.nih.gov/pubmed/17263002}.
\bibitem[{Laule et~al.(2007)Laule, Vavasour, Kolind, Li, Traboulsee, Moore \&
  MacKay}]{Laule2007review}
\bibinfo{author}{Laule, C.}, \bibinfo{author}{Vavasour, I.~M.},
  \bibinfo{author}{Kolind, S.~H.}, \bibinfo{author}{Li, D.~K.},
  \bibinfo{author}{Traboulsee, T.~L.}, \bibinfo{author}{Moore, G.~W.}, \&
  \bibinfo{author}{MacKay, A.~L.} (\bibinfo{year}{2007}).
\newblock \bibinfo{title}{Magnetic resonance imaging of myelin}.
\newblock {\it \bibinfo{journal}{Neurotherapeutics : the journal of the
  American Society for Experimental NeuroTherapeutics}\/},  {\it
  \bibinfo{volume}{4}\/}, \bibinfo{pages}{460--484}. \URLprefix
  \url{http://dx.doi.org/10.1016/j.nurt.2007.05.004}.
  \DOIprefix\doi{10.1016/j.nurt.2007.05.004}.
\bibitem[{Lema et~al.(2017)Lema, Bishop, Malik, Mattoscio, Ali, Nicholas,
  Muraro, Matthews, Waldman \& Newbould}]{Lema2017a}
\bibinfo{author}{Lema, A.}, \bibinfo{author}{Bishop, C.},
  \bibinfo{author}{Malik, O.}, \bibinfo{author}{Mattoscio, M.},
  \bibinfo{author}{Ali, R.}, \bibinfo{author}{Nicholas, R.},
  \bibinfo{author}{Muraro, P.~A.}, \bibinfo{author}{Matthews, P.~M.},
  \bibinfo{author}{Waldman, A.~D.}, \& \bibinfo{author}{Newbould, R.~D.}
  (\bibinfo{year}{2017}).
\newblock \bibinfo{title}{A comparison of magnetization transfer methods to
  assess brain and cervical cord microstructure in multiple sclerosis}.
\newblock {\it \bibinfo{journal}{J Neuroimaging}\/},  {\it
  \bibinfo{volume}{27}\/}, \bibinfo{pages}{221--226}. \URLprefix
  \url{http://dx.doi.org/10.1111/jon.12377}. \DOIprefix\doi{10.1111/jon.12377}.
\bibitem[{Levesque et~al.(2005)Levesque, Sled, Narayanan, Santos, Brass,
  Francis, Arnold \& Pike}]{Levesque2005}
\bibinfo{author}{Levesque, I.}, \bibinfo{author}{Sled, J.~G.},
  \bibinfo{author}{Narayanan, S.}, \bibinfo{author}{Santos, A.~C.},
  \bibinfo{author}{Brass, S.~D.}, \bibinfo{author}{Francis, S.~J.},
  \bibinfo{author}{Arnold, D.~L.}, \& \bibinfo{author}{Pike, G.~B.}
  (\bibinfo{year}{2005}).
\newblock \bibinfo{title}{The role of edema and demyelination in chronic {T1}
  black holes: A quantitative magnetization transfer study}.
\newblock {\it \bibinfo{journal}{J. Magn. Reson. Imaging}\/},  {\it
  \bibinfo{volume}{21}\/}, \bibinfo{pages}{103--110}. \URLprefix
  \url{http://dx.doi.org/10.1002/jmri.20231}.
  \DOIprefix\doi{10.1002/jmri.20231}.
\bibitem[{Levesque \& Pike(2009)}]{Levesque2009a}
\bibinfo{author}{Levesque, I.~R.}, \& \bibinfo{author}{Pike, G.~B.}
  (\bibinfo{year}{2009}).
\newblock \bibinfo{title}{Characterizing healthy and diseased white matter
  using quantitative magnetization transfer and multicomponent {T2}
  relaxometry: A unified view via a four-pool model}.
\newblock {\it \bibinfo{journal}{Magn. Reson. Med.}\/},  {\it
  \bibinfo{volume}{62}\/}, \bibinfo{pages}{1487--1496}. \URLprefix
  \url{http://dx.doi.org/10.1002/mrm.22131}. \DOIprefix\doi{10.1002/mrm.22131}.
\bibitem[{Liu et~al.(2015)Liu, Li, Tong, Yeom \& Kuzminski}]{Liu2015a}
\bibinfo{author}{Liu, C.}, \bibinfo{author}{Li, W.}, \bibinfo{author}{Tong,
  K.~A.}, \bibinfo{author}{Yeom, K.~W.}, \& \bibinfo{author}{Kuzminski, S.}
  (\bibinfo{year}{2015}).
\newblock \bibinfo{title}{Susceptibility-weighted imaging and quantitative
  susceptibility mapping in the brain.}
\newblock {\it \bibinfo{journal}{Journal of magnetic resonance imaging}\/},
  {\it \bibinfo{volume}{42}\/}, \bibinfo{pages}{23--41}. \URLprefix
  \url{http://view.ncbi.nlm.nih.gov/pubmed/25270052}.
\bibitem[{Lustig et~al.(2007)Lustig, Donoho \& Pauly}]{Lustig2007a}
\bibinfo{author}{Lustig, M.}, \bibinfo{author}{Donoho, D.}, \&
  \bibinfo{author}{Pauly, J.~M.} (\bibinfo{year}{2007}).
\newblock \bibinfo{title}{Sparse {MRI}: The application of compressed sensing
  for rapid {MR} imaging.}
\newblock {\it \bibinfo{journal}{Magn. Reson. Med.}\/},  {\it
  \bibinfo{volume}{58}\/}, \bibinfo{pages}{1182--1195}. \URLprefix
  \url{http://dx.doi.org/10.1002/mrm.21391}. \DOIprefix\doi{10.1002/mrm.21391}.
\bibitem[{MacKay et~al.(1994)MacKay, Whittall, Adler, Li, Paty \&
  Graeb}]{MacKay1994a}
\bibinfo{author}{MacKay, A.}, \bibinfo{author}{Whittall, K.},
  \bibinfo{author}{Adler, J.}, \bibinfo{author}{Li, D.}, \bibinfo{author}{Paty,
  D.}, \& \bibinfo{author}{Graeb, D.} (\bibinfo{year}{1994}).
\newblock \bibinfo{title}{In vivo visualization of myelin water in brain by
  magnetic resonance.}
\newblock {\it \bibinfo{journal}{Magn Reson Med}\/},  {\it
  \bibinfo{volume}{31}\/}, \bibinfo{pages}{673--677}. \URLprefix
  \url{http://view.ncbi.nlm.nih.gov/pubmed/8057820}.
\bibitem[{Magnollay et~al.(2014)Magnollay, Grussu, Wheeler-Kingshott, Sethi,
  Zhang, Chard, Miller \& Ciccarelli}]{MagnollayISMRM14}
\bibinfo{author}{Magnollay, L.}, \bibinfo{author}{Grussu, F.},
  \bibinfo{author}{Wheeler-Kingshott, C.}, \bibinfo{author}{Sethi, V.},
  \bibinfo{author}{Zhang, H.}, \bibinfo{author}{Chard, D.},
  \bibinfo{author}{Miller, D.}, \& \bibinfo{author}{Ciccarelli, O.}
  (\bibinfo{year}{2014}).
\newblock \bibinfo{title}{An investigation of brain neurite density and
  dispersion in multiple sclerosis using single shell diffusion imaging}.
\newblock In {\it \bibinfo{booktitle}{ISMRM 2014}\/} (p.
  \bibinfo{pages}{2048}).
\bibitem[{Maier-Hein et~al.(2016)Maier-Hein, Neher \& et~al.}]{Maier-Hein2016a}
\bibinfo{author}{Maier-Hein, K.}, \bibinfo{author}{Neher, P.}, \&
  \bibinfo{author}{et~al.} (\bibinfo{year}{2016}).
\newblock \bibinfo{title}{Tractography-based connectomes are dominated by
  false-positive connections}.
\newblock {\it \bibinfo{journal}{bioRxiv}\/},  (pp. \bibinfo{pages}{084137+}).
  \URLprefix \url{http://dx.doi.org/10.1101/084137}.
  \DOIprefix\doi{10.1101/084137}.
\bibitem[{Mangeat et~al.(2015)Mangeat, Govindarajan, Mainero \&
  Cohen-Adad}]{Mangeat2015a}
\bibinfo{author}{Mangeat, G.}, \bibinfo{author}{Govindarajan, S.~T.},
  \bibinfo{author}{Mainero, C.}, \& \bibinfo{author}{Cohen-Adad, J.}
  (\bibinfo{year}{2015}).
\newblock \bibinfo{title}{{Multivariate combination of magnetization transfer,
  T2* and B0 orientation to study the myelo-architecture of the in vivo human
  cortex}}.
\newblock {\it \bibinfo{journal}{NeuroImage}\/},  {\it
  \bibinfo{volume}{119}\/}, \bibinfo{pages}{89--102}. \URLprefix
  \url{http://dx.doi.org/10.1016/j.neuroimage.2015.06.033}.
  \DOIprefix\doi{10.1016/j.neuroimage.2015.06.033}.
\bibitem[{Manning et~al.(2016)Manning, Chang, MacKay \& Michal}]{Manning2016b}
\bibinfo{author}{Manning, A.~P.}, \bibinfo{author}{Chang, K.~L.},
  \bibinfo{author}{MacKay, A.~L.}, \& \bibinfo{author}{Michal, C.~A.}
  (\bibinfo{year}{2016}).
\newblock \bibinfo{title}{The physical mechanism of ” inhomogeneous”
  magnetization transfer {MRI}}.
\newblock {\it \bibinfo{journal}{Journal of Magnetic Resonance}\/}, .
  \URLprefix \url{http://dx.doi.org/10.1016/j.jmr.2016.11.013}.
  \DOIprefix\doi{10.1016/j.jmr.2016.11.013}.
\bibitem[{Mc{L}ean et~al.(2017)Mc{L}ean, MacDonald, Lebel, Boudreau \&
  Pike}]{McLean2017a}
\bibinfo{author}{Mc{L}ean, M.}, \bibinfo{author}{MacDonald, M.},
  \bibinfo{author}{Lebel, R.~M.}, \bibinfo{author}{Boudreau, M.}, \&
  \bibinfo{author}{Pike, G.~B.} (\bibinfo{year}{2017}).
\newblock \bibinfo{title}{Accelerated z-spectrum imaging}.
\newblock In {\it \bibinfo{booktitle}{ISMRM 2017}\/} (p.
  \bibinfo{pages}{1205}).
\bibitem[{Melbourne et~al.(2016)Melbourne, Eaton-Rosen, Orasanu, Price,
  Bainbridge, Cardoso, Kendall, Robertson, Marlow \& Ourselin}]{Melbourne16a}
\bibinfo{author}{Melbourne, A.}, \bibinfo{author}{Eaton-Rosen, Z.},
  \bibinfo{author}{Orasanu, E.}, \bibinfo{author}{Price, D.},
  \bibinfo{author}{Bainbridge, A.}, \bibinfo{author}{Cardoso, M.~J.},
  \bibinfo{author}{Kendall, G.~S.}, \bibinfo{author}{Robertson, N.~J.},
  \bibinfo{author}{Marlow, N.}, \& \bibinfo{author}{Ourselin, S.}
  (\bibinfo{year}{2016}).
\newblock \bibinfo{title}{Longitudinal development in the preterm thalamus and
  posterior white matter: {MRI} correlations between diffusion weighted imaging
  and {T2} relaxometry}.
\newblock {\it \bibinfo{journal}{Hum. Brain Mapp.}\/},  (p.
  \bibinfo{pages}{n/a}). \URLprefix \url{http://dx.doi.org/10.1002/hbm.23188}.
  \DOIprefix\doi{10.1002/hbm.23188}.
\bibitem[{Mezer et~al.(2013)Mezer, Yeatman, Stikov, Kay, Cho, Dougherty, Perry,
  Parvizi, Hua, Butts-Pauly \& Wandell}]{Mezer2013a}
\bibinfo{author}{Mezer, A.}, \bibinfo{author}{Yeatman, J.~D.},
  \bibinfo{author}{Stikov, N.}, \bibinfo{author}{Kay, K.~N.},
  \bibinfo{author}{Cho, N.-J.}, \bibinfo{author}{Dougherty, R.~F.},
  \bibinfo{author}{Perry, M.~L.}, \bibinfo{author}{Parvizi, J.},
  \bibinfo{author}{Hua, L.~H.}, \bibinfo{author}{Butts-Pauly, K.}, \&
  \bibinfo{author}{Wandell, B.~A.} (\bibinfo{year}{2013}).
\newblock \bibinfo{title}{Quantifying the local tissue volume and composition
  in individual brains with magnetic resonance imaging}.
\newblock {\it \bibinfo{journal}{Nat Med}\/},  {\it \bibinfo{volume}{19}\/},
  \bibinfo{pages}{1667--1672}. \URLprefix
  \url{http://dx.doi.org/10.1038/nm.3390}. \DOIprefix\doi{10.1038/nm.3390}.
\bibitem[{{mincdiffusion}(2013)}]{mincdiffusionurl}
\bibinfo{author}{{mincdiffusion}} (\bibinfo{year}{2013}), .
\newblock \URLprefix
  \url{http://www.bic.mni.mcgill.ca/${\sim}$ilana/ diffusion/diffusion\_tools.htm}.
\newblock \bibinfo{note}{August 2013}.
\bibitem[{Mohammadi et~al.(2015)Mohammadi, Carey, Dick, Diedrichsen, Sereno,
  Reisert, Callaghan \& Weiskopf}]{Mohammadi2015a}
\bibinfo{author}{Mohammadi, S.}, \bibinfo{author}{Carey, D.},
  \bibinfo{author}{Dick, F.}, \bibinfo{author}{Diedrichsen, J.},
  \bibinfo{author}{Sereno, M.~I.}, \bibinfo{author}{Reisert, M.},
  \bibinfo{author}{Callaghan, M.~F.}, \& \bibinfo{author}{Weiskopf, N.}
  (\bibinfo{year}{2015}).
\newblock \bibinfo{title}{Whole-brain in-vivo measurements of the axonal
  g-ratio in a group of 37 healthy volunteers.}
\newblock {\it \bibinfo{journal}{Frontiers in neuroscience}\/},  {\it
  \bibinfo{volume}{9}\/}. \URLprefix
  \url{http://dx.doi.org/10.3389/fnins.2015.00441}.
  \DOIprefix\doi{10.3389/fnins.2015.00441}.
\bibitem[{Molina-Romero et~al.(2016)Molina-Romero, Gomez, Sperl, Jones, Menzel
  \& Menze1}]{Molina-Romero2016a}
\bibinfo{author}{Molina-Romero, M.}, \bibinfo{author}{Gomez, P.~A.},
  \bibinfo{author}{Sperl, J.~I.}, \bibinfo{author}{Jones, D.~K.},
  \bibinfo{author}{Menzel, M.~I.}, \& \bibinfo{author}{Menze1, B.~H.}
  (\bibinfo{year}{2016}).
\newblock \bibinfo{title}{{Tissue microstructure characterisation through
  relaxometry and diffusion MRI using sparse component analysis}}.
\newblock In {\it \bibinfo{booktitle}{{ISMRM} workshop: Breaking the Barriers
  of Diffusion {MRI}}\/} (p.~\bibinfo{pages}{17}).
\bibitem[{Mollink et~al.(2016)Mollink, Kleinnijenhuis, Sotiropoulos, Cottaar,
  van Cappellen~van Walsum, Gamarallage, Ansorge, Jbabdi \&
  Miller}]{Mollink2016a}
\bibinfo{author}{Mollink, J.}, \bibinfo{author}{Kleinnijenhuis, M.},
  \bibinfo{author}{Sotiropoulos, S.~N.}, \bibinfo{author}{Cottaar, M.},
  \bibinfo{author}{van Cappellen~van Walsum, A.-M.},
  \bibinfo{author}{Gamarallage, M.~P.}, \bibinfo{author}{Ansorge, O.},
  \bibinfo{author}{Jbabdi, S.}, \& \bibinfo{author}{Miller, K.~L.}
  (\bibinfo{year}{2016}).
\newblock \bibinfo{title}{Exploring fibre orientation dispersion in the corpus
  callosum: Comparison of diffusion mri, polarized light imaging and
  histology}.
\newblock In {\it \bibinfo{booktitle}{ISMRM 2016}\/} (p. \bibinfo{pages}{795}).
\bibitem[{Mottershead et~al.(2003)Mottershead, Schmierer, Clemence, Thornton,
  Scaravilli, Barker, Tofts, Newcombe, Cuzner, Ordidge, McDonald \&
  Miller}]{Mottershead2003a}
\bibinfo{author}{Mottershead, J.~P.}, \bibinfo{author}{Schmierer, K.},
  \bibinfo{author}{Clemence, M.}, \bibinfo{author}{Thornton, J.~S.},
  \bibinfo{author}{Scaravilli, F.}, \bibinfo{author}{Barker, G.~J.},
  \bibinfo{author}{Tofts, P.~S.}, \bibinfo{author}{Newcombe, J.},
  \bibinfo{author}{Cuzner, M.~L.}, \bibinfo{author}{Ordidge, R.~J.},
  \bibinfo{author}{McDonald, W.~I.}, \& \bibinfo{author}{Miller, D.~H.}
  (\bibinfo{year}{2003}).
\newblock \bibinfo{title}{High field {MRI} correlates of myelin content and
  axonal density in multiple sclerosis--a post-mortem study of the spinal
  cord.}
\newblock {\it \bibinfo{journal}{Journal of neurology}\/},  {\it
  \bibinfo{volume}{250}\/}, \bibinfo{pages}{1293--1301}. \URLprefix
  \url{http://view.ncbi.nlm.nih.gov/pubmed/14648144}.
\bibitem[{Nguyen et~al.(2016)Nguyen, Deh, Monohan, Pandya, Spincemaille, Raj,
  Wang \& Gauthier}]{Nguyen2016a}
\bibinfo{author}{Nguyen, T.~D.}, \bibinfo{author}{Deh, K.},
  \bibinfo{author}{Monohan, E.}, \bibinfo{author}{Pandya, S.},
  \bibinfo{author}{Spincemaille, P.}, \bibinfo{author}{Raj, A.},
  \bibinfo{author}{Wang, Y.}, \& \bibinfo{author}{Gauthier, S.~A.}
  (\bibinfo{year}{2016}).
\newblock \bibinfo{title}{Feasibility and reproducibility of whole brain myelin
  water mapping in 4 minutes using fast acquisition with spiral trajectory and
  adiabatic t2prep ({FAST}-t2) at {3T}}.
\newblock {\it \bibinfo{journal}{Magn. Reson. Med.}\/},  {\it
  \bibinfo{volume}{76}\/}, \bibinfo{pages}{456--465}. \URLprefix
  \url{http://dx.doi.org/10.1002/mrm.25877}. \DOIprefix\doi{10.1002/mrm.25877}.
\bibitem[{Ning et~al.(2016)Ning, Westin \& Rathi}]{Ning2016a}
\bibinfo{author}{Ning, L.}, \bibinfo{author}{Westin, C.-F.}, \&
  \bibinfo{author}{Rathi, Y.} (\bibinfo{year}{2016}).
\newblock \bibinfo{title}{Estimation of bounded and unbounded trajectories in
  diffusion {MRI}}.
\newblock {\it \bibinfo{journal}{Frontiers in Neuroscience}\/},  {\it
  \bibinfo{volume}{10}\/}. \URLprefix
  \url{http://dx.doi.org/10.3389/fnins.2016.00129}.
  \DOIprefix\doi{10.3389/fnins.2016.00129}.
\bibitem[{{NODDI Matlab Toolbox}(2013)}]{NODDIurl}
\bibinfo{author}{{NODDI Matlab Toolbox}} (\bibinfo{year}{2013}).
\newblock
  \bibinfo{howpublished}{{http://cmic.cs.ucl.ac.uk/mig/index.php?n= Tutorial.NODDImatlab}}.
\newblock \bibinfo{note}{May 2013}.
\bibitem[{Nossin-Manor et~al.(2015)Nossin-Manor, Card, Raybaud, Taylor \&
  Sled}]{Nossin-Manor2015a}
\bibinfo{author}{Nossin-Manor, R.}, \bibinfo{author}{Card, D.},
  \bibinfo{author}{Raybaud, C.}, \bibinfo{author}{Taylor, M.~J.}, \&
  \bibinfo{author}{Sled, J.~G.} (\bibinfo{year}{2015}).
\newblock \bibinfo{title}{Cerebral maturation in the early preterm {period-A}
  magnetization transfer and diffusion tensor imaging study using voxel-based
  analysis.}
\newblock {\it \bibinfo{journal}{NeuroImage}\/},  {\it
  \bibinfo{volume}{112}\/}, \bibinfo{pages}{30--42}. \URLprefix
  \url{http://view.ncbi.nlm.nih.gov/pubmed/25731990}.
\bibitem[{Novikov \& Fieremans(2012)}]{Novikov2012a}
\bibinfo{author}{Novikov, D.}, \& \bibinfo{author}{Fieremans, E.}
  (\bibinfo{year}{2012}).
\newblock \bibinfo{title}{Relating extracellular diffusivity to cell size
  distribution and packing density as applied to white matter}.
\newblock In {\it \bibinfo{booktitle}{ISMRM 2012}\/} (p.
  \bibinfo{pages}{1829}).
\bibitem[{Novikov et~al.(2011)Novikov, Fieremans, Jensen \&
  Helpern}]{Novikov2011a}
\bibinfo{author}{Novikov, D.~S.}, \bibinfo{author}{Fieremans, E.},
  \bibinfo{author}{Jensen, J.~H.}, \& \bibinfo{author}{Helpern, J.~A.}
  (\bibinfo{year}{2011}).
\newblock \bibinfo{title}{Random walks with barriers}.
\newblock {\it \bibinfo{journal}{Nature Physics}\/},  {\it
  \bibinfo{volume}{7}\/}, \bibinfo{pages}{508--514}. \URLprefix
  \url{http://dx.doi.org/10.1038/nphys1936}. \DOIprefix\doi{10.1038/nphys1936}.
\bibitem[{Novikov et~al.(2016)Novikov, Veraart, Jelescu \&
  Fieremans}]{Novikov2016a}
\bibinfo{author}{Novikov, D.~S.}, \bibinfo{author}{Veraart, J.},
  \bibinfo{author}{Jelescu, I.~O.}, \& \bibinfo{author}{Fieremans, E.}
  (\bibinfo{year}{2016}).
\newblock \bibinfo{title}{Mapping orientational and microstructural metrics of
  neuronal integrity with in vivo diffusion {MRI}}.
\newblock {\it \bibinfo{journal}{arXiv}\/},  (p. \bibinfo{pages}{1609.09144}).
  \URLprefix \url{http://arxiv.org/abs/1609.09144}.
\bibitem[{Oh et~al.(2006)Oh, Han, Pelletier \& Nelson}]{Oh2006a}
\bibinfo{author}{Oh, J.}, \bibinfo{author}{Han, E.~T.},
  \bibinfo{author}{Pelletier, D.}, \& \bibinfo{author}{Nelson, S.~J.}
  (\bibinfo{year}{2006}).
\newblock \bibinfo{title}{{Measurement of in vivo multi-component T2 relaxation
  times for brain tissue using multi-slice T2 prep at 1.5 and 3 T.}}
\newblock {\it \bibinfo{journal}{Magnetic resonance imaging}\/},  {\it
  \bibinfo{volume}{24}\/}, \bibinfo{pages}{33--43}. \URLprefix
  \url{http://dx.doi.org/10.1016/j.mri.2005.10.016}.
  \DOIprefix\doi{10.1016/j.mri.2005.10.016}.
\bibitem[{Oh et~al.(2013)Oh, Bilello, Schindler, Markowitz, Detre \&
  Lee}]{Oh2015a}
\bibinfo{author}{Oh, S.-H.}, \bibinfo{author}{Bilello, M.},
  \bibinfo{author}{Schindler, M.}, \bibinfo{author}{Markowitz, C.~E.},
  \bibinfo{author}{Detre, J.~A.}, \& \bibinfo{author}{Lee, J.}
  (\bibinfo{year}{2013}).
\newblock \bibinfo{title}{Direct visualization of short transverse relaxation
  time component ({ViSTa})}.
\newblock {\it \bibinfo{journal}{NeuroImage}\/},  {\it \bibinfo{volume}{83}\/},
  \bibinfo{pages}{485--492}. \URLprefix
  \url{http://dx.doi.org/10.1016/j.neuroimage.2013.06.047}.
  \DOIprefix\doi{10.1016/j.neuroimage.2013.06.047}.
\bibitem[{Pampel et~al.(2015)Pampel, M\"{u}ller, Anwander, Marschner \&
  M\"{o}ller}]{Pampel2015a}
\bibinfo{author}{Pampel, A.}, \bibinfo{author}{M\"{u}ller, D.~K.},
  \bibinfo{author}{Anwander, A.}, \bibinfo{author}{Marschner, H.}, \&
  \bibinfo{author}{M\"{o}ller, H.~E.} (\bibinfo{year}{2015}).
\newblock \bibinfo{title}{Orientation dependence of magnetization transfer
  parameters in human white matter}.
\newblock {\it \bibinfo{journal}{NeuroImage}\/},  {\it
  \bibinfo{volume}{114}\/}, \bibinfo{pages}{136--146}. \URLprefix
  \url{http://dx.doi.org/10.1016/j.neuroimage.2015.03.068}.
  \DOIprefix\doi{10.1016/j.neuroimage.2015.03.068}.
\bibitem[{Pasternak et~al.(2009)Pasternak, Sochen, Gur, Intrator \&
  Assaf}]{Pasternak2009a}
\bibinfo{author}{Pasternak, O.}, \bibinfo{author}{Sochen, N.},
  \bibinfo{author}{Gur, Y.}, \bibinfo{author}{Intrator, N.}, \&
  \bibinfo{author}{Assaf, Y.} (\bibinfo{year}{2009}).
\newblock \bibinfo{title}{Free water elimination and mapping from diffusion
  {MRI}.}
\newblock {\it \bibinfo{journal}{Magn. Reson. Med.}\/},  {\it
  \bibinfo{volume}{62}\/}, \bibinfo{pages}{717--730}. \URLprefix
  \url{http://dx.doi.org/10.1002/mrm.22055}. \DOIprefix\doi{10.1002/mrm.22055}.
\bibitem[{Paus \& Toro(2009)}]{Paus2009a}
\bibinfo{author}{Paus, T.}, \& \bibinfo{author}{Toro, R.}
  (\bibinfo{year}{2009}).
\newblock \bibinfo{title}{Could sex differences in white matter be explained by
  g ratio?}
\newblock {\it \bibinfo{journal}{Frontiers in neuroanatomy}\/},  {\it
  \bibinfo{volume}{3}\/}. \URLprefix
  \url{http://dx.doi.org/10.3389/neuro.05.014.2009}.
  \DOIprefix\doi{10.3389/neuro.05.014.2009}.
\bibitem[{Perrin et~al.(2009)Perrin, Leonard, Perron, Pike, Pitiot, Richer,
  Veillette, Pausova \& Paus}]{Perrin09}
\bibinfo{author}{Perrin, J.~S.}, \bibinfo{author}{Leonard, G.},
  \bibinfo{author}{Perron, M.}, \bibinfo{author}{Pike, G.~B.},
  \bibinfo{author}{Pitiot, A.}, \bibinfo{author}{Richer, L.},
  \bibinfo{author}{Veillette, S.}, \bibinfo{author}{Pausova, Z.}, \&
  \bibinfo{author}{Paus, T.} (\bibinfo{year}{2009}).
\newblock \bibinfo{title}{Sex differences in the growth of white matter during
  adolescence.}
\newblock {\it \bibinfo{journal}{NeuroImage}\/},  {\it \bibinfo{volume}{45}\/},
  \bibinfo{pages}{1055--1066}.
  \DOIprefix\doi{10.1016/j.neuroimage.2009.01.023}.
\bibitem[{Pesaresi et~al.(2015)Pesaresi, Soon-Shiong, French, Kaplan, Miller \&
  Paus}]{Pesaresi2015a}
\bibinfo{author}{Pesaresi, M.}, \bibinfo{author}{Soon-Shiong, R.},
  \bibinfo{author}{French, L.}, \bibinfo{author}{Kaplan, D.~R.},
  \bibinfo{author}{Miller, F.~D.}, \& \bibinfo{author}{Paus, T.}
  (\bibinfo{year}{2015}).
\newblock \bibinfo{title}{Axon diameter and axonal transport: In vivo and in
  vitro effects of androgens}.
\newblock {\it \bibinfo{journal}{NeuroImage}\/},  {\it
  \bibinfo{volume}{115}\/}, \bibinfo{pages}{191--201}. \URLprefix
  \url{http://dx.doi.org/10.1016/j.neuroimage.2015.04.048}.
  \DOIprefix\doi{10.1016/j.neuroimage.2015.04.048}.
\bibitem[{Prasloski et~al.(2012)Prasloski, Rauscher, MacKay, Hodgson, Vavasour,
  Laule \& M\"{a}dler}]{Prasloski2012a}
\bibinfo{author}{Prasloski, T.}, \bibinfo{author}{Rauscher, A.},
  \bibinfo{author}{MacKay, A.~L.}, \bibinfo{author}{Hodgson, M.},
  \bibinfo{author}{Vavasour, I.~M.}, \bibinfo{author}{Laule, C.}, \&
  \bibinfo{author}{M\"{a}dler, B.} (\bibinfo{year}{2012}).
\newblock \bibinfo{title}{Rapid whole cerebrum myelin water imaging using a
  {3D} {GRASE} sequence.}
\newblock {\it \bibinfo{journal}{NeuroImage}\/},  {\it \bibinfo{volume}{63}\/},
  \bibinfo{pages}{533--539}. \URLprefix
  \url{http://dx.doi.org/10.1016/j.neuroimage.2012.06.064}.
  \DOIprefix\doi{10.1016/j.neuroimage.2012.06.064}.
\bibitem[{Raffelt et~al.(2012)Raffelt, Tournier, Rose, Ridgway, Henderson,
  Crozier, Salvado \& Connelly}]{Raffelt2012a}
\bibinfo{author}{Raffelt, D.}, \bibinfo{author}{Tournier},
  \bibinfo{author}{Rose, S.}, \bibinfo{author}{Ridgway, G.~R.},
  \bibinfo{author}{Henderson, R.}, \bibinfo{author}{Crozier, S.},
  \bibinfo{author}{Salvado, O.}, \& \bibinfo{author}{Connelly, A.}
  (\bibinfo{year}{2012}).
\newblock \bibinfo{title}{Apparent fibre density: A novel measure for the
  analysis of diffusion-weighted magnetic resonance images}.
\newblock {\it \bibinfo{journal}{NeuroImage}\/},  {\it \bibinfo{volume}{59}\/},
  \bibinfo{pages}{3976--3994}. \URLprefix
  \url{http://dx.doi.org/10.1016/j.neuroimage.2011.10.045}.
  \DOIprefix\doi{10.1016/j.neuroimage.2011.10.045}.
\bibitem[{Ramani et~al.(2002)Ramani, Dalton, Miller, Tofts \&
  Barker}]{Ramani2002a}
\bibinfo{author}{Ramani, A.}, \bibinfo{author}{Dalton, C.},
  \bibinfo{author}{Miller, D.~H.}, \bibinfo{author}{Tofts, P.~S.}, \&
  \bibinfo{author}{Barker, G.~J.} (\bibinfo{year}{2002}).
\newblock \bibinfo{title}{Precise estimate of fundamental in-vivo {MT}
  parameters in human brain in clinically feasible times.}
\newblock {\it \bibinfo{journal}{Magnetic resonance imaging}\/},  {\it
  \bibinfo{volume}{20}\/}, \bibinfo{pages}{721--731}. \URLprefix
  \url{http://view.ncbi.nlm.nih.gov/pubmed/12591568}.
\bibitem[{Reid et~al.(2016)Reid, Sale, Cunnington, Mattingley \&
  Rose}]{Reid2016a}
\bibinfo{author}{Reid, L.~B.}, \bibinfo{author}{Sale, M.~V.},
  \bibinfo{author}{Cunnington, R.}, \bibinfo{author}{Mattingley, J.~B.}, \&
  \bibinfo{author}{Rose, S.~E.} (\bibinfo{year}{2016}).
\newblock \bibinfo{title}{Structural and functional brain changes following
  four weeks of unimanual motor training: evidence from {fMRI}-guided diffusion
  {MRI} tractography}.
\newblock {\it \bibinfo{journal}{bioRxiv}\/},  (pp. \bibinfo{pages}{088328+}).
  \URLprefix \url{http://dx.doi.org/10.1101/088328}.
  \DOIprefix\doi{10.1101/088328}.
\bibitem[{Reisert et~al.(2013)Reisert, Mader, Umarova, Maier, Tebartz~van Elst
  \& Kiselev}]{Reisert2013a}
\bibinfo{author}{Reisert, M.}, \bibinfo{author}{Mader, I.},
  \bibinfo{author}{Umarova, R.}, \bibinfo{author}{Maier, S.},
  \bibinfo{author}{Tebartz~van Elst, L.}, \& \bibinfo{author}{Kiselev, V.~G.}
  (\bibinfo{year}{2013}).
\newblock \bibinfo{title}{Fiber density estimation from single q-shell
  diffusion imaging by tensor divergence}.
\newblock {\it \bibinfo{journal}{NeuroImage}\/},  {\it \bibinfo{volume}{77}\/},
  \bibinfo{pages}{166--176}. \URLprefix
  \url{http://dx.doi.org/10.1016/j.neuroimage.2013.03.032}.
  \DOIprefix\doi{10.1016/j.neuroimage.2013.03.032}.
\bibitem[{Rokem et~al.(2015)Rokem, Yeatman, Pestilli, Kay, Mezer, van~der Walt
  \& Wandell}]{Rokem2015a}
\bibinfo{author}{Rokem, A.}, \bibinfo{author}{Yeatman, J.~D.},
  \bibinfo{author}{Pestilli, F.}, \bibinfo{author}{Kay, K.~N.},
  \bibinfo{author}{Mezer, A.}, \bibinfo{author}{van~der Walt, S.}, \&
  \bibinfo{author}{Wandell, B.~A.} (\bibinfo{year}{2015}).
\newblock \bibinfo{title}{Evaluating the accuracy of diffusion {MRI} models in
  white matter}.
\newblock {\it \bibinfo{journal}{PLoS ONE}\/},  {\it \bibinfo{volume}{10}\/},
  \bibinfo{pages}{e0123272+}. \URLprefix
  \url{http://dx.doi.org/10.1371/journal.pone.0123272}.
  \DOIprefix\doi{10.1371/journal.pone.0123272}.
\bibitem[{Ronen et~al.(2014)Ronen, Budde, Ercan, Annese, Techawiboonwong \&
  Webb}]{Ronen2014a}
\bibinfo{author}{Ronen, I.}, \bibinfo{author}{Budde, M.},
  \bibinfo{author}{Ercan, E.}, \bibinfo{author}{Annese, J.},
  \bibinfo{author}{Techawiboonwong, A.}, \& \bibinfo{author}{Webb, A.}
  (\bibinfo{year}{2014}).
\newblock \bibinfo{title}{Microstructural organization of axons in the human
  corpus callosum quantified by diffusion-weighted magnetic resonance
  spectroscopy of n-acetylaspartate and post-mortem histology.}
\newblock {\it \bibinfo{journal}{Brain structure \& function}\/},  {\it
  \bibinfo{volume}{219}\/}, \bibinfo{pages}{1773--1785}. \URLprefix
  \url{http://view.ncbi.nlm.nih.gov/pubmed/23794120}.
\bibitem[{Rooney et~al.(2007)Rooney, Johnson, Li, Cohen, Kim, Ugurbil \&
  Springer}]{rooney2007a}
\bibinfo{author}{Rooney, W.~D.}, \bibinfo{author}{Johnson, G.},
  \bibinfo{author}{Li, X.}, \bibinfo{author}{Cohen, E.~R.},
  \bibinfo{author}{Kim, S.-G.}, \bibinfo{author}{Ugurbil, K.}, \&
  \bibinfo{author}{Springer, C.~S.} (\bibinfo{year}{2007}).
\newblock \bibinfo{title}{Magnetic field and tissue dependencies of human brain
  longitudinal {1H2O} relaxation in vivo}.
\newblock {\it \bibinfo{journal}{Magn. Reson. Med.}\/},  {\it
  \bibinfo{volume}{57}\/}, \bibinfo{pages}{308--318}. \URLprefix
  \url{http://dx.doi.org/10.1002/mrm.21122}. \DOIprefix\doi{10.1002/mrm.21122}.
\bibitem[{Rudko et~al.(2014)Rudko, Klassen, de~Chickera, Gati, Dekaban \&
  Menon}]{Rudko2014a}
\bibinfo{author}{Rudko, D.~A.}, \bibinfo{author}{Klassen, L.~M.},
  \bibinfo{author}{de~Chickera, S.~N.}, \bibinfo{author}{Gati, J.~S.},
  \bibinfo{author}{Dekaban, G.~A.}, \& \bibinfo{author}{Menon, R.~S.}
  (\bibinfo{year}{2014}).
\newblock \bibinfo{title}{Origins of {R2} orientation dependence in gray and
  white matter}.
\newblock {\it \bibinfo{journal}{PNAS}\/},  {\it \bibinfo{volume}{111}\/},
  \bibinfo{pages}{E159--E167}. \URLprefix
  \url{http://dx.doi.org/10.1073/pnas.1306516111}.
  \DOIprefix\doi{10.1073/pnas.1306516111}.
\bibitem[{Rushton(1951)}]{Rushton51}
\bibinfo{author}{Rushton, W. A.~H.} (\bibinfo{year}{1951}).
\newblock \bibinfo{title}{A theory of the effects of fibre size in medullated
  nerve}.
\newblock {\it \bibinfo{journal}{J Physiol}\/},  {\it \bibinfo{volume}{115}\/},
  \bibinfo{pages}{101--22}.
\bibitem[{Sati et~al.(2013)Sati, van Gelderen, Silva, Reich, Merkle, de~Zwart
  \& Duyn}]{Sati2013a}
\bibinfo{author}{Sati, P.}, \bibinfo{author}{van Gelderen, P.},
  \bibinfo{author}{Silva, A.~C.}, \bibinfo{author}{Reich, D.~S.},
  \bibinfo{author}{Merkle, H.}, \bibinfo{author}{de~Zwart, J.~A.}, \&
  \bibinfo{author}{Duyn, J.~H.} (\bibinfo{year}{2013}).
\newblock \bibinfo{title}{Micro-compartment specific t2* relaxation in the
  brain.}
\newblock {\it \bibinfo{journal}{NeuroImage}\/},  {\it \bibinfo{volume}{77}\/},
  \bibinfo{pages}{268--278}. \URLprefix
  \url{http://dx.doi.org/10.1016/j.neuroimage.2013.03.005}.
  \DOIprefix\doi{10.1016/j.neuroimage.2013.03.005}.
\bibitem[{Scherrer et~al.(2016)Scherrer, Jacobs, Taquet, des Rieux, Macq,
  Prabhu \& Warfield}]{Scherrer2016a}
\bibinfo{author}{Scherrer, B.}, \bibinfo{author}{Jacobs, D.},
  \bibinfo{author}{Taquet, M.}, \bibinfo{author}{des Rieux, A.},
  \bibinfo{author}{Macq, B.}, \bibinfo{author}{Prabhu, S.~P.}, \&
  \bibinfo{author}{Warfield, S.~K.} (\bibinfo{year}{2016}).
\newblock \bibinfo{title}{Measurement of restricted and hindered anisotropic
  diffusion tissue compartments in a rat model of wallerian degeneration}.
\newblock In {\it \bibinfo{booktitle}{ISMRM 2016}\/} (p.
  \bibinfo{pages}{1087}).
\bibitem[{Schmierer et~al.(2004)Schmierer, Scaravilli, Altmann, Barker \&
  Miller}]{Schmierer2004a}
\bibinfo{author}{Schmierer, K.}, \bibinfo{author}{Scaravilli, F.},
  \bibinfo{author}{Altmann, D.~R.}, \bibinfo{author}{Barker, G.~J.}, \&
  \bibinfo{author}{Miller, D.~H.} (\bibinfo{year}{2004}).
\newblock \bibinfo{title}{Magnetization transfer ratio and myelin in postmortem
  multiple sclerosis brain}.
\newblock {\it \bibinfo{journal}{Ann Neurol.}\/},  {\it
  \bibinfo{volume}{56}\/}, \bibinfo{pages}{407--415}. \URLprefix
  \url{http://dx.doi.org/10.1002/ana.20202}. \DOIprefix\doi{10.1002/ana.20202}.
\bibitem[{Schmierer et~al.(2007)Schmierer, Tozer, Scaravilli, Altmann, Barker,
  Tofts \& Miller}]{Schmierer2007a}
\bibinfo{author}{Schmierer, K.}, \bibinfo{author}{Tozer, D.~J.},
  \bibinfo{author}{Scaravilli, F.}, \bibinfo{author}{Altmann, D.~R.},
  \bibinfo{author}{Barker, G.~J.}, \bibinfo{author}{Tofts, P.~S.}, \&
  \bibinfo{author}{Miller, D.~H.} (\bibinfo{year}{2007}).
\newblock \bibinfo{title}{Quantitative magnetization transfer imaging in
  postmortem multiple sclerosis brain.}
\newblock {\it \bibinfo{journal}{Journal of magnetic resonance imaging :
  JMRI}\/},  {\it \bibinfo{volume}{26}\/}, \bibinfo{pages}{41--51}. \URLprefix
  \url{http://dx.doi.org/10.1002/jmri.20984}.
  \DOIprefix\doi{10.1002/jmri.20984}.
\bibitem[{Schmierer et~al.(2008)Schmierer, Wheeler-Kingshott, Tozer, Boulby,
  Parkes, Yousry, Scaravilli, Barker, Tofts \& Miller}]{Schmierer2008a}
\bibinfo{author}{Schmierer, K.}, \bibinfo{author}{Wheeler-Kingshott, C.~A.},
  \bibinfo{author}{Tozer, D.~J.}, \bibinfo{author}{Boulby, P.~A.},
  \bibinfo{author}{Parkes, H.~G.}, \bibinfo{author}{Yousry, T.~A.},
  \bibinfo{author}{Scaravilli, F.}, \bibinfo{author}{Barker, G.~J.},
  \bibinfo{author}{Tofts, P.~S.}, \& \bibinfo{author}{Miller, D.~H.}
  (\bibinfo{year}{2008}).
\newblock \bibinfo{title}{Quantitative magnetic resonance of postmortem
  multiple sclerosis brain before and after fixation.}
\newblock {\it \bibinfo{journal}{Magn Reson Med}\/},  {\it
  \bibinfo{volume}{59}\/}, \bibinfo{pages}{268--277}. \URLprefix
  \url{http://dx.doi.org/10.1002/mrm.21487}. \DOIprefix\doi{10.1002/mrm.21487}.
\bibitem[{Schr{\"o}der et~al.(1988)Schr{\"o}der, Bohl \& von
  Bardeleben}]{Schroder88}
\bibinfo{author}{Schr{\"o}der, J.~M.}, \bibinfo{author}{Bohl, J.}, \&
  \bibinfo{author}{von Bardeleben, U.} (\bibinfo{year}{1988}).
\newblock \bibinfo{title}{Changes of the ratio between myelin thickness and
  axon diameter in human developing sural, femoral, ulnar, facial, and
  trochlear nerves}.
\newblock {\it \bibinfo{journal}{Acta Neuropathol}\/},  {\it
  \bibinfo{volume}{76}\/}, \bibinfo{pages}{471--83}.
\bibitem[{Sepehrband et~al.(2015)Sepehrband, Clark, Ullmann, Kurniawan,
  Leanage, Reutens \& Yang}]{Sepehrband2015a}
\bibinfo{author}{Sepehrband, F.}, \bibinfo{author}{Clark, K.~A.},
  \bibinfo{author}{Ullmann, J. F.~P.}, \bibinfo{author}{Kurniawan, N.~D.},
  \bibinfo{author}{Leanage, G.}, \bibinfo{author}{Reutens, D.~C.}, \&
  \bibinfo{author}{Yang, Z.} (\bibinfo{year}{2015}).
\newblock \bibinfo{title}{Brain tissue compartment density estimated using
  diffusion-weighted {MRI} yields tissue parameters consistent with histology}.
\newblock {\it \bibinfo{journal}{Hum. Brain Mapp.}\/},  {\it
  \bibinfo{volume}{36}\/}, \bibinfo{pages}{3687--3702}. \URLprefix
  \url{http://dx.doi.org/10.1002/hbm.22872}. \DOIprefix\doi{10.1002/hbm.22872}.
\bibitem[{Setsompop et~al.(2012)Setsompop, Gagoski, Polimeni, Witzel, Wedeen \&
  Wald}]{setsompop12a}
\bibinfo{author}{Setsompop, K.}, \bibinfo{author}{Gagoski, B.~A.},
  \bibinfo{author}{Polimeni, J.~R.}, \bibinfo{author}{Witzel, T.},
  \bibinfo{author}{Wedeen, V.~J.}, \& \bibinfo{author}{Wald, L.~L.}
  (\bibinfo{year}{2012}).
\newblock \bibinfo{title}{Blipped-controlled aliasing in parallel imaging for
  simultaneous multislice echo planar imaging with reduced g-factor penalty}.
\newblock {\it \bibinfo{journal}{Magn. Reson. Med.}\/},  {\it
  \bibinfo{volume}{67}\/}, \bibinfo{pages}{1210--1224}. \URLprefix
  \url{http://dx.doi.org/10.1002/mrm.23097}. \DOIprefix\doi{10.1002/mrm.23097}.
\bibitem[{Shemesh et~al.(2012)Shemesh, Barazany, Sadan, Bar, Zur, Barhum,
  Sochen, Offen, Assaf \& Cohen}]{Shemesh2012a}
\bibinfo{author}{Shemesh, N.}, \bibinfo{author}{Barazany, D.},
  \bibinfo{author}{Sadan, O.}, \bibinfo{author}{Bar, L.}, \bibinfo{author}{Zur,
  Y.}, \bibinfo{author}{Barhum, Y.}, \bibinfo{author}{Sochen, N.},
  \bibinfo{author}{Offen, D.}, \bibinfo{author}{Assaf, Y.}, \&
  \bibinfo{author}{Cohen, Y.} (\bibinfo{year}{2012}).
\newblock \bibinfo{title}{Mapping apparent eccentricity and residual ensemble
  anisotropy in the gray matter using angular double-pulsed-field-gradient
  {MRI}}.
\newblock {\it \bibinfo{journal}{Magn Reson Med}\/},  {\it
  \bibinfo{volume}{68}\/}, \bibinfo{pages}{794--806}. \URLprefix
  \url{http://dx.doi.org/10.1002/mrm.23300}. \DOIprefix\doi{10.1002/mrm.23300}.
\bibitem[{Sled \& Pike(2001)}]{Sled2001a}
\bibinfo{author}{Sled, J.~G.}, \& \bibinfo{author}{Pike, G.~B.}
  (\bibinfo{year}{2001}).
\newblock \bibinfo{title}{Quantitative imaging of magnetization transfer
  exchange and relaxation properties in vivo using {MRI}}.
\newblock {\it \bibinfo{journal}{Magn. Reson. Med.}\/},  {\it
  \bibinfo{volume}{46}\/}, \bibinfo{pages}{923--931}. \URLprefix
  \url{http://dx.doi.org/10.1002/mrm.1278}. \DOIprefix\doi{10.1002/mrm.1278}.
\bibitem[{Smith et~al.(2004)Smith, Jenkinson, Woolrich, Beckmann, Behrens,
  Johansen-Berg, Bannister, De~Luca, Drobnjak, Flitney, Niazy, Saunders,
  Vickers, Zhang, De~Stefano, Brady \& Matthews}]{Smith2012a}
\bibinfo{author}{Smith, S.~M.}, \bibinfo{author}{Jenkinson, M.},
  \bibinfo{author}{Woolrich, M.~W.}, \bibinfo{author}{Beckmann, C.~F.},
  \bibinfo{author}{Behrens, T.~E.}, \bibinfo{author}{Johansen-Berg, H.},
  \bibinfo{author}{Bannister, P.~R.}, \bibinfo{author}{De~Luca, M.},
  \bibinfo{author}{Drobnjak, I.}, \bibinfo{author}{Flitney, D.~E.},
  \bibinfo{author}{Niazy, R.~K.}, \bibinfo{author}{Saunders, J.},
  \bibinfo{author}{Vickers, J.}, \bibinfo{author}{Zhang, Y.},
  \bibinfo{author}{De~Stefano, N.}, \bibinfo{author}{Brady, J.~M.}, \&
  \bibinfo{author}{Matthews, P.~M.} (\bibinfo{year}{2004}).
\newblock \bibinfo{title}{Advances in functional and structural {MR} image
  analysis and implementation as {FSL}.}
\newblock {\it \bibinfo{journal}{NeuroImage}\/},  {\it \bibinfo{volume}{23
  Suppl 1}\/}, \bibinfo{pages}{S208--S219}. \URLprefix
  \url{http://dx.doi.org/10.1016/j.neuroimage.2004.07.051}.
  \DOIprefix\doi{10.1016/j.neuroimage.2004.07.051}.
\bibitem[{Stanisz et~al.(1997)Stanisz, Szafer, Wright \&
  Henkelman}]{Stanisz1997a}
\bibinfo{author}{Stanisz, G.~J.}, \bibinfo{author}{Szafer, A.},
  \bibinfo{author}{Wright, G.~A.}, \& \bibinfo{author}{Henkelman, R.~M.}
  (\bibinfo{year}{1997}).
\newblock \bibinfo{title}{An analytical model of restricted diffusion in bovine
  optic nerve.}
\newblock {\it \bibinfo{journal}{Magn. Reson. Med.}\/},  {\it
  \bibinfo{volume}{37}\/}, \bibinfo{pages}{103--111}. \URLprefix
  \url{http://view.ncbi.nlm.nih.gov/pubmed/8978638}.
\bibitem[{Stikov et~al.(2015{\natexlab{a}})Stikov, Campbell, Stroh,
  Lavel\'{e}e, Frey, Novek, Nuara, Ho, Bedell, Dougherty, Leppert, Boudreau,
  Narayanan, Duval, Cohen-Adad, Picard, Gasecka, C\^{o}t\'{e} \&
  Pike}]{Stikov2015c}
\bibinfo{author}{Stikov, N.}, \bibinfo{author}{Campbell, J.~S.},
  \bibinfo{author}{Stroh, T.}, \bibinfo{author}{Lavel\'{e}e, M.},
  \bibinfo{author}{Frey, S.}, \bibinfo{author}{Novek, J.},
  \bibinfo{author}{Nuara, S.}, \bibinfo{author}{Ho, M.-K.~K.},
  \bibinfo{author}{Bedell, B.~J.}, \bibinfo{author}{Dougherty, R.~F.},
  \bibinfo{author}{Leppert, I.~R.}, \bibinfo{author}{Boudreau, M.},
  \bibinfo{author}{Narayanan, S.}, \bibinfo{author}{Duval, T.},
  \bibinfo{author}{Cohen-Adad, J.}, \bibinfo{author}{Picard, P.-A.~A.},
  \bibinfo{author}{Gasecka, A.}, \bibinfo{author}{C\^{o}t\'{e}, D.}, \&
  \bibinfo{author}{Pike, G.~B.} (\bibinfo{year}{2015}{\natexlab{a}}).
\newblock \bibinfo{title}{Quantitative analysis of the myelin g-ratio from
  electron microscopy images of the macaque corpus callosum.}
\newblock {\it \bibinfo{journal}{Data in brief}\/},  {\it
  \bibinfo{volume}{4}\/}, \bibinfo{pages}{368--373}. \URLprefix
  \url{http://dx.doi.org/10.1016/j.dib.2015.05.019}.
  \DOIprefix\doi{10.1016/j.dib.2015.05.019}.
\bibitem[{Stikov et~al.(2015{\natexlab{b}})Stikov, Campbell, Stroh,
  Lavel\'{e}e, Frey, Novek, Nuara, Ho, Bedell, Dougherty, Leppert, Boudreau,
  Narayanan, Duval, Cohen-Adad, Picard, Gasecka, C\^{o}t\'{e} \&
  Bruce~Pike}]{Stikov2015a}
\bibinfo{author}{Stikov, N.}, \bibinfo{author}{Campbell, J. S.~W.},
  \bibinfo{author}{Stroh, T.}, \bibinfo{author}{Lavel\'{e}e, M.},
  \bibinfo{author}{Frey, S.}, \bibinfo{author}{Novek, J.},
  \bibinfo{author}{Nuara, S.}, \bibinfo{author}{Ho, M.-K.},
  \bibinfo{author}{Bedell, B.~J.}, \bibinfo{author}{Dougherty, R.~F.},
  \bibinfo{author}{Leppert, I.~R.}, \bibinfo{author}{Boudreau, M.},
  \bibinfo{author}{Narayanan, S.}, \bibinfo{author}{Duval, T.},
  \bibinfo{author}{Cohen-Adad, J.}, \bibinfo{author}{Picard, P.},
  \bibinfo{author}{Gasecka, A.}, \bibinfo{author}{C\^{o}t\'{e}, D.}, \&
  \bibinfo{author}{Bruce~Pike, G.} (\bibinfo{year}{2015}{\natexlab{b}}).
\newblock \bibinfo{title}{In vivo histology of the myelin g-ratio with magnetic
  resonance imaging}.
\newblock {\it \bibinfo{journal}{NeuroImage}\/}, . \URLprefix
  \url{http://dx.doi.org/10.1016/j.neuroimage.2015.05.023}.
  \DOIprefix\doi{10.1016/j.neuroimage.2015.05.023}.
\bibitem[{Stikov et~al.(2011)Stikov, Perry, Mezer, Rykhlevskaia, Wandell, Pauly
  \& Dougherty}]{Stikov2011}
\bibinfo{author}{Stikov, N.}, \bibinfo{author}{Perry, L.~M.},
  \bibinfo{author}{Mezer, A.}, \bibinfo{author}{Rykhlevskaia, E.},
  \bibinfo{author}{Wandell, B.~A.}, \bibinfo{author}{Pauly, J.~M.}, \&
  \bibinfo{author}{Dougherty, R.~F.} (\bibinfo{year}{2011}).
\newblock \bibinfo{title}{Bound pool fractions complement diffusion measures to
  describe white matter micro and macrostructure}.
\newblock {\it \bibinfo{journal}{NeuroImage}\/},  {\it \bibinfo{volume}{54}\/},
  \bibinfo{pages}{1112--1121}. \URLprefix
  \url{http://dx.doi.org/10.1016/j.neuroimage.2010.08.068}.
  \DOIprefix\doi{10.1016/j.neuroimage.2010.08.068}.
\bibitem[{Stollberger \& Wach(1996)}]{Stollberger1996a}
\bibinfo{author}{Stollberger, R.}, \& \bibinfo{author}{Wach, P.}
  (\bibinfo{year}{1996}).
\newblock \bibinfo{title}{Imaging of the active {B1} field in vivo}.
\newblock {\it \bibinfo{journal}{Magn. Reson. Med.}\/},  {\it
  \bibinfo{volume}{35}\/}, \bibinfo{pages}{246--251}. \URLprefix
  \url{http://dx.doi.org/10.1002/mrm.1910350217}.
  \DOIprefix\doi{10.1002/mrm.1910350217}.
\bibitem[{St\"{u}ber et~al.(2014)St\"{u}ber, Morawski, Sch\"{a}fer, Labadie,
  W\"{a}hnert, Leuze, Streicher, Barapatre, Reimann, Geyer, Spemann \&
  Turner}]{Stuber2014a}
\bibinfo{author}{St\"{u}ber, C.}, \bibinfo{author}{Morawski, M.},
  \bibinfo{author}{Sch\"{a}fer, A.}, \bibinfo{author}{Labadie, C.},
  \bibinfo{author}{W\"{a}hnert, M.}, \bibinfo{author}{Leuze, C.},
  \bibinfo{author}{Streicher, M.}, \bibinfo{author}{Barapatre, N.},
  \bibinfo{author}{Reimann, K.}, \bibinfo{author}{Geyer, S.},
  \bibinfo{author}{Spemann, D.}, \& \bibinfo{author}{Turner, R.}
  (\bibinfo{year}{2014}).
\newblock \bibinfo{title}{Myelin and iron concentration in the human brain: a
  quantitative study of {MRI} contrast.}
\newblock {\it \bibinfo{journal}{NeuroImage}\/},  {\it \bibinfo{volume}{93 Pt
  1}\/}, \bibinfo{pages}{95--106}. \URLprefix
  \url{http://view.ncbi.nlm.nih.gov/pubmed/24607447}.
\bibitem[{Sveinsson \& Dougherty(2011)}]{Sveinsson2011a}
\bibinfo{author}{Sveinsson, B.}, \& \bibinfo{author}{Dougherty, R.~F.}
  (\bibinfo{year}{2011}).
\newblock {\it \bibinfo{title}{{dSim}: Simulating Diffusion in Biologically
  Realistic Tissue Models}\/}.
\newblock \bibinfo{type}{Technical Report} \bibinfo{number}{CNITR-001} Stanford
  Center for Cognitive and Neurobiological Imaging.
\bibitem[{Szafer et~al.(1995)Szafer, Zhong \& Gore}]{Szafer1995a}
\bibinfo{author}{Szafer, A.}, \bibinfo{author}{Zhong, J.}, \&
  \bibinfo{author}{Gore, J.~C.} (\bibinfo{year}{1995}).
\newblock \bibinfo{title}{Theoretical model for water diffusion in tissues}.
\newblock {\it \bibinfo{journal}{Magn. Reson. Med.}\/},  {\it
  \bibinfo{volume}{33}\/}, \bibinfo{pages}{697--712}. \URLprefix
  \url{http://dx.doi.org/10.1002/mrm.1910330516}.
  \DOIprefix\doi{10.1002/mrm.1910330516}.
\bibitem[{Thiessen et~al.(2013)Thiessen, Zhang, Zhang, Wang, Buist, Del~Bigio,
  Kong, Li \& Martin}]{Thiessen2013a}
\bibinfo{author}{Thiessen, J.~D.}, \bibinfo{author}{Zhang, Y.},
  \bibinfo{author}{Zhang, H.}, \bibinfo{author}{Wang, L.},
  \bibinfo{author}{Buist, R.}, \bibinfo{author}{Del~Bigio, M.~R.},
  \bibinfo{author}{Kong, J.}, \bibinfo{author}{Li, X.-M.}, \&
  \bibinfo{author}{Martin, M.} (\bibinfo{year}{2013}).
\newblock \bibinfo{title}{Quantitative {MRI} and ultrastructural examination of
  the cuprizone mouse model of demyelination}.
\newblock {\it \bibinfo{journal}{NMR Biomed.}\/},  {\it
  \bibinfo{volume}{26}\/}, \bibinfo{pages}{1562--1581}. \URLprefix
  \url{http://dx.doi.org/10.1002/nbm.2992}. \DOIprefix\doi{10.1002/nbm.2992}.
\bibitem[{Uranova et~al.(2001)Uranova, Orlovskaya, Vikhreva, Zimina, Kolomeets,
  Vostrikov \& Rachmanova}]{Uranova2001a}
\bibinfo{author}{Uranova, N.}, \bibinfo{author}{Orlovskaya, D.},
  \bibinfo{author}{Vikhreva, O.}, \bibinfo{author}{Zimina, I.},
  \bibinfo{author}{Kolomeets, N.}, \bibinfo{author}{Vostrikov, V.}, \&
  \bibinfo{author}{Rachmanova, V.} (\bibinfo{year}{2001}).
\newblock \bibinfo{title}{Electron microscopy of oligodendroglia in severe
  mental illness.}
\newblock {\it \bibinfo{journal}{Brain research bulletin}\/},  {\it
  \bibinfo{volume}{55}\/}, \bibinfo{pages}{597--610}. \URLprefix
  \url{http://view.ncbi.nlm.nih.gov/pubmed/11576756}.
\bibitem[{Varma et~al.(2015)Varma, Duhamel, de~Bazelaire \& Alsop}]{Varma2015a}
\bibinfo{author}{Varma, G.}, \bibinfo{author}{Duhamel, G.},
  \bibinfo{author}{de~Bazelaire, C.}, \& \bibinfo{author}{Alsop, D.~C.}
  (\bibinfo{year}{2015}).
\newblock \bibinfo{title}{Magnetization transfer from inhomogeneously broadened
  lines: A potential marker for myelin.}
\newblock {\it \bibinfo{journal}{Magn Reson Med}\/},  {\it
  \bibinfo{volume}{73}\/}, \bibinfo{pages}{614--622}. \URLprefix
  \url{http://view.ncbi.nlm.nih.gov/pubmed/24604578}.
\bibitem[{Vavasour et~al.(2011)Vavasour, Laule, Li, Traboulsee \&
  MacKay}]{Vavasour2011a}
\bibinfo{author}{Vavasour, I.~M.}, \bibinfo{author}{Laule, C.},
  \bibinfo{author}{Li, D. K.~B.}, \bibinfo{author}{Traboulsee, A.~L.}, \&
  \bibinfo{author}{MacKay, A.~L.} (\bibinfo{year}{2011}).
\newblock \bibinfo{title}{Is the magnetization transfer ratio a marker for
  myelin in multiple sclerosis?}
\newblock {\it \bibinfo{journal}{J. Magn. Reson. Imaging}\/},  {\it
  \bibinfo{volume}{33}\/}, \bibinfo{pages}{710--718}. \URLprefix
  \url{http://dx.doi.org/10.1002/jmri.22441}.
  \DOIprefix\doi{10.1002/jmri.22441}.
\bibitem[{Vidarsson et~al.(2005)Vidarsson, Conolly, Lim, Gold \&
  Pauly}]{Vidarsson2005a}
\bibinfo{author}{Vidarsson, L.}, \bibinfo{author}{Conolly, S.~M.},
  \bibinfo{author}{Lim, K.~O.}, \bibinfo{author}{Gold, G.~E.}, \&
  \bibinfo{author}{Pauly, J.~M.} (\bibinfo{year}{2005}).
\newblock \bibinfo{title}{Echo time optimization for linear combination myelin
  imaging.}
\newblock {\it \bibinfo{journal}{Magn. Reson. Med.}\/},  {\it
  \bibinfo{volume}{53}\/}, \bibinfo{pages}{398--407}. \URLprefix
  \url{http://dx.doi.org/10.1002/mrm.20360}. \DOIprefix\doi{10.1002/mrm.20360}.
\bibitem[{Volz et~al.(2010)Volz, N\"{o}th, Rotarska-Jagiela \&
  Deichmann}]{Volz2010a}
\bibinfo{author}{Volz, S.}, \bibinfo{author}{N\"{o}th, U.},
  \bibinfo{author}{Rotarska-Jagiela, A.}, \& \bibinfo{author}{Deichmann, R.}
  (\bibinfo{year}{2010}).
\newblock \bibinfo{title}{{A fast B1-mapping method for the correction and
  normalization of magnetization transfer ratio maps at 3 T.}}
\newblock {\it \bibinfo{journal}{NeuroImage}\/},  {\it \bibinfo{volume}{49}\/},
  \bibinfo{pages}{3015--3026}. \URLprefix
  \url{http://view.ncbi.nlm.nih.gov/pubmed/19948229}.
\bibitem[{Wang et~al.(2011)Wang, Wang, Haldar, Yeh, Xie, Sun, Tu, Trinkaus,
  Klein, Cross \& Song}]{Wang2011a}
\bibinfo{author}{Wang, Y.}, \bibinfo{author}{Wang, Q.},
  \bibinfo{author}{Haldar, J.~P.}, \bibinfo{author}{Yeh, F.-C.},
  \bibinfo{author}{Xie, M.}, \bibinfo{author}{Sun, P.}, \bibinfo{author}{Tu,
  T.-W.}, \bibinfo{author}{Trinkaus, K.}, \bibinfo{author}{Klein, R.~S.},
  \bibinfo{author}{Cross, A.~H.}, \& \bibinfo{author}{Song, S.-K.}
  (\bibinfo{year}{2011}).
\newblock \bibinfo{title}{{Quantification of increased cellularity during
  inflammatory demyelination}}.
\newblock {\it \bibinfo{journal}{Brain}\/},  {\it \bibinfo{volume}{134}\/},
  \bibinfo{pages}{3590--3601}. \URLprefix
  \url{http://dx.doi.org/10.1093/brain/awr307}.
  \DOIprefix\doi{10.1093/brain/awr307}.
\bibitem[{Waxman(1975)}]{Waxman75}
\bibinfo{author}{Waxman, S.~G.} (\bibinfo{year}{1975}).
\newblock \bibinfo{title}{Integrative properties and design principles of
  axons}.
\newblock {\it \bibinfo{journal}{Int Rev Neurobiol}\/},  {\it
  \bibinfo{volume}{18}\/}, \bibinfo{pages}{1--40}.
\bibitem[{Weiskopf et~al.(2013)Weiskopf, Suckling, Williams, Correia, Inkster,
  Tait, Ooi, Bullmore \& Lutti}]{Weiskopf13a}
\bibinfo{author}{Weiskopf, N.}, \bibinfo{author}{Suckling, J.},
  \bibinfo{author}{Williams, G.}, \bibinfo{author}{Correia, M.~M.},
  \bibinfo{author}{Inkster, B.}, \bibinfo{author}{Tait, R.},
  \bibinfo{author}{Ooi, C.}, \bibinfo{author}{Bullmore, E.~T.}, \&
  \bibinfo{author}{Lutti, A.} (\bibinfo{year}{2013}).
\newblock \bibinfo{title}{{Quantitative multi-parameter mapping of R1, PD*, MT,
  and R2* at 3T: a multi-center validation.}}
\newblock {\it \bibinfo{journal}{Frontiers in neuroscience}\/},  {\it
  \bibinfo{volume}{7}\/}. \URLprefix
  \url{http://view.ncbi.nlm.nih.gov/pubmed/23772204}.
\bibitem[{West et~al.(2016{\natexlab{a}})West, Kelm, Carson \&
  Does}]{West2016b}
\bibinfo{author}{West, K.~L.}, \bibinfo{author}{Kelm, N.~D.},
  \bibinfo{author}{Carson, R.~P.}, \& \bibinfo{author}{Does, M.~D.}
  (\bibinfo{year}{2016}{\natexlab{a}}).
\newblock \bibinfo{title}{Quantitative assessment of g-ratio from {MRI}}.
\newblock In {\it \bibinfo{booktitle}{{QBIN} Workshop: Toward a super-big
  brain: promises and pitfalls of microstructural imaging}\/}
  (p.~\bibinfo{pages}{28}).
\bibitem[{West et~al.(2016{\natexlab{b}})West, Kelm, Carson \& Does}]{West15a}
\bibinfo{author}{West, K.~L.}, \bibinfo{author}{Kelm, N.~D.},
  \bibinfo{author}{Carson, R.~P.}, \& \bibinfo{author}{Does, M.~D.}
  (\bibinfo{year}{2016}{\natexlab{b}}).
\newblock \bibinfo{title}{A revised model for estimating g-ratio from {MRI}}.
\newblock {\it \bibinfo{journal}{NeuroImage}\/}, . \URLprefix
  \url{http://dx.doi.org/10.1016/j.neuroimage.2015.08.017}.
  \DOIprefix\doi{10.1016/j.neuroimage.2015.08.017}.
\bibitem[{West et~al.(2017)West, Kelm, Carson, Gochberg, Ess \&
  Does}]{West2017d}
\bibinfo{author}{West, K.~L.}, \bibinfo{author}{Kelm, N.~D.},
  \bibinfo{author}{Carson, R.~P.}, \bibinfo{author}{Gochberg, D.~F.},
  \bibinfo{author}{Ess, K.~C.}, \& \bibinfo{author}{Does, M.~D.}
  (\bibinfo{year}{2017}).
\newblock \bibinfo{title}{Myelin volume fraction imaging with {MRI}}.
\newblock {\it \bibinfo{journal}{NeuroImage}\/}, . \URLprefix
  \url{http://dx.doi.org/10.1016/j.neuroimage.2016.12.067}.
  \DOIprefix\doi{10.1016/j.neuroimage.2016.12.067}.
\bibitem[{West et~al.(2014)West, Kelm, Gochberg, Carson, Ess \&
  Does}]{West2014a}
\bibinfo{author}{West, K.~L.}, \bibinfo{author}{Kelm, N.~D.},
  \bibinfo{author}{Gochberg, D.~F.}, \bibinfo{author}{Carson, R.~P.},
  \bibinfo{author}{Ess, K.~C.}, \& \bibinfo{author}{Does, M.~D.}
  (\bibinfo{year}{2014}).
\newblock \bibinfo{title}{Multiexponential {T2} and quantitative magnetization
  transfer in rodent brain models of hypomyelination}.
\newblock In {\it \bibinfo{booktitle}{ISMRM 2014}\/} (p.
  \bibinfo{pages}{2088}).
\bibitem[{White et~al.(2013)White, Leergaard, D'Arceuil, Bjaalie \&
  Dale}]{White2013a}
\bibinfo{author}{White, N.~S.}, \bibinfo{author}{Leergaard, T.~B.},
  \bibinfo{author}{D'Arceuil, H.}, \bibinfo{author}{Bjaalie, J.~G.}, \&
  \bibinfo{author}{Dale, A.~M.} (\bibinfo{year}{2013}).
\newblock \bibinfo{title}{{Probing tissue microstructure with restriction
  spectrum imaging: Histological and theoretical validation}}.
\newblock {\it \bibinfo{journal}{Hum. Brain Mapp.}\/},  {\it
  \bibinfo{volume}{34}\/}, \bibinfo{pages}{327--346}. \URLprefix
  \url{http://dx.doi.org/10.1002/hbm.21454}. \DOIprefix\doi{10.1002/hbm.21454}.
\bibitem[{Wilhelm et~al.(2012)Wilhelm, Ong, Wehrli, Li, Tsai, Hackney \&
  Wehrli}]{Wilhelm2012a}
\bibinfo{author}{Wilhelm, M.~J.}, \bibinfo{author}{Ong, H.~H.},
  \bibinfo{author}{Wehrli, S.~L.}, \bibinfo{author}{Li, C.},
  \bibinfo{author}{Tsai, P.-H.}, \bibinfo{author}{Hackney, D.~B.}, \&
  \bibinfo{author}{Wehrli, F.~W.} (\bibinfo{year}{2012}).
\newblock \bibinfo{title}{Direct magnetic resonance detection of myelin and
  prospects for quantitative imaging of myelin density}.
\newblock {\it \bibinfo{journal}{Proceedings of the National Academy of
  Sciences}\/},  {\it \bibinfo{volume}{109}\/}, \bibinfo{pages}{9605--9610}.
  \URLprefix \url{http://dx.doi.org/10.1073/pnas.1115107109}.
  \DOIprefix\doi{10.1073/pnas.1115107109}.
\bibitem[{Wolff et~al.(1991)Wolff, Eng \& Balaban}]{Wolff1991a}
\bibinfo{author}{Wolff, S.}, \bibinfo{author}{Eng, J.}, \&
  \bibinfo{author}{Balaban, R.} (\bibinfo{year}{1991}).
\newblock \bibinfo{title}{Magnetization transfer contrast: Method for improving
  contrast in gradient-recalled-echo images}.
\newblock {\it \bibinfo{journal}{Radiology}\/},  {\it \bibinfo{volume}{179}\/},
  \bibinfo{pages}{133--137}.
\bibitem[{Wood et~al.(2016)Wood, Simmons, Hurley, Vernon, Torres, Dell'Acqua,
  Williams \& Cash}]{Wood2016a}
\bibinfo{author}{Wood, T.~C.}, \bibinfo{author}{Simmons, C.},
  \bibinfo{author}{Hurley, S.~A.}, \bibinfo{author}{Vernon, A.~C.},
  \bibinfo{author}{Torres, J.}, \bibinfo{author}{Dell'Acqua, F.},
  \bibinfo{author}{Williams, S. C.~R.}, \& \bibinfo{author}{Cash, D.}
  (\bibinfo{year}{2016}).
\newblock \bibinfo{title}{Whole-brain ex-vivo quantitative {MRI} of the
  cuprizone mouse model}.
\newblock {\it \bibinfo{journal}{PeerJ}\/},  {\it \bibinfo{volume}{4}\/},
  \bibinfo{pages}{e2632+}. \URLprefix
  \url{http://dx.doi.org/10.7717/peerj.2632}.
  \DOIprefix\doi{10.7717/peerj.2632}.
\bibitem[{Wu et~al.(2016)Wu, He, Chen, Chen, Liu, Du \& Zhong}]{Wu2016a}
\bibinfo{author}{Wu, Z.}, \bibinfo{author}{He, H.}, \bibinfo{author}{Chen, Y.},
  \bibinfo{author}{Chen, S.}, \bibinfo{author}{Liu, H.}, \bibinfo{author}{Du,
  Y.~P.}, \& \bibinfo{author}{Zhong, J.} (\bibinfo{year}{2016}).
\newblock \bibinfo{title}{Feasibility study of high resolution mapping for
  myelin water fraction and frequency shift using tissue susceptibility}.
\newblock In {\it \bibinfo{booktitle}{ISMRM 2016}\/} (p.~\bibinfo{pages}{31}).
\bibitem[{Xu et~al.(2014)Xu, Li, Harkins, Jiang, Xie, Kang, Does \&
  Gore}]{Xu2014a}
\bibinfo{author}{Xu, J.}, \bibinfo{author}{Li, H.}, \bibinfo{author}{Harkins,
  K.~D.}, \bibinfo{author}{Jiang, X.}, \bibinfo{author}{Xie, J.},
  \bibinfo{author}{Kang, H.}, \bibinfo{author}{Does, M.~D.}, \&
  \bibinfo{author}{Gore, J.~C.} (\bibinfo{year}{2014}).
\newblock \bibinfo{title}{Mapping mean axon diameter and axonal volume fraction
  by {MRI} using temporal diffusion spectroscopy}.
\newblock {\it \bibinfo{journal}{NeuroImage}\/},  {\it
  \bibinfo{volume}{103}\/}, \bibinfo{pages}{10--19}. \URLprefix
  \url{http://dx.doi.org/10.1016/j.neuroimage.2014.09.006}.
  \DOIprefix\doi{10.1016/j.neuroimage.2014.09.006}.
\bibitem[{Yarnykh(2002)}]{Yarnykh2002a}
\bibinfo{author}{Yarnykh, V.~L.} (\bibinfo{year}{2002}).
\newblock \bibinfo{title}{Pulsed z-spectroscopic imaging of cross-relaxation
  parameters in tissues for human {MRI}: Theory and clinical applications}.
\newblock {\it \bibinfo{journal}{Magn. Reson. Med.}\/},  {\it
  \bibinfo{volume}{47}\/}, \bibinfo{pages}{929--939}. \URLprefix
  \url{http://dx.doi.org/10.1002/mrm.10120}. \DOIprefix\doi{10.1002/mrm.10120}.
\bibitem[{Yarnykh(2012)}]{Yarnykh2012a}
\bibinfo{author}{Yarnykh, V.~L.} (\bibinfo{year}{2012}).
\newblock \bibinfo{title}{Fast macromolecular proton fraction mapping from a
  single off-resonance magnetization transfer measurement.}
\newblock {\it \bibinfo{journal}{Magn Reson Med}\/},  {\it
  \bibinfo{volume}{68}\/}, \bibinfo{pages}{166--178}. \URLprefix
  \url{http://view.ncbi.nlm.nih.gov/pubmed/22190042}.
\bibitem[{Yarnykh \& Khodanovich(2015)}]{Yarnykh2015a}
\bibinfo{author}{Yarnykh, V.~L.}, \& \bibinfo{author}{Khodanovich}
  (\bibinfo{year}{2015}).
\newblock \bibinfo{title}{Analytical method of correction of {B1} errors in
  mapping of magnetization transfer ratio in highfield magnetic resonance
  tomography}.
\newblock {\it \bibinfo{journal}{Russian Physics Journal}\/},  {\it
  \bibinfo{volume}{57}\/}, \bibinfo{pages}{1784--1788}. \URLprefix
  \url{http://dx.doi.org/10.1007/s11182-015-0451-7}.
  \DOIprefix\doi{10.1007/s11182-015-0451-7}.
\bibitem[{Zaimi et~al.(2016)Zaimi, Duval, Gasecka, C\^{o}t\'{e}, Stikov \&
  Cohen-Adad}]{Zaimi2016a}
\bibinfo{author}{Zaimi, A.}, \bibinfo{author}{Duval, T.},
  \bibinfo{author}{Gasecka, A.}, \bibinfo{author}{C\^{o}t\'{e}, D.},
  \bibinfo{author}{Stikov, N.}, \& \bibinfo{author}{Cohen-Adad, J.}
  (\bibinfo{year}{2016}).
\newblock \bibinfo{title}{{AxonSeg}: Open source software for axon and myelin
  segmentation and morphometric analysis}.
\newblock {\it \bibinfo{journal}{Frontiers in Neuroinformatics}\/},  {\it
  \bibinfo{volume}{10}\/}. \URLprefix
  \url{http://dx.doi.org/10.3389/fninf.2016.00037}.
  \DOIprefix\doi{10.3389/fninf.2016.00037}.
\bibitem[{Zhang et~al.(2011)Zhang, Hubbard, Parker \& Alexander}]{Zhang2011a}
\bibinfo{author}{Zhang, H.}, \bibinfo{author}{Hubbard, P.~L.},
  \bibinfo{author}{Parker, G. J.~M.}, \& \bibinfo{author}{Alexander, D.~C.}
  (\bibinfo{year}{2011}).
\newblock \bibinfo{title}{Axon diameter mapping in the presence of orientation
  dispersion with diffusion {MRI}}.
\newblock {\it \bibinfo{journal}{NeuroImage}\/},  {\it \bibinfo{volume}{56}\/},
  \bibinfo{pages}{1301--1315}. \URLprefix
  \url{http://dx.doi.org/10.1016/j.neuroimage.2011.01.084}.
  \DOIprefix\doi{10.1016/j.neuroimage.2011.01.084}.
\bibitem[{Zhang et~al.(2012)Zhang, Schneider, Wheeler-Kingshott \&
  Alexander}]{Zhang2012a}
\bibinfo{author}{Zhang, H.}, \bibinfo{author}{Schneider, T.},
  \bibinfo{author}{Wheeler-Kingshott, C.~A.}, \& \bibinfo{author}{Alexander,
  D.~C.} (\bibinfo{year}{2012}).
\newblock \bibinfo{title}{{NODDI}: Practical in vivo neurite orientation
  dispersion and density imaging of the human brain}.
\newblock {\it \bibinfo{journal}{NeuroImage}\/},  {\it \bibinfo{volume}{61}\/},
  \bibinfo{pages}{1000--1016}. \URLprefix
  \url{http://dx.doi.org/10.1016/j.neuroimage.2012.03.072}.
  \DOIprefix\doi{10.1016/j.neuroimage.2012.03.072}.
\bibitem[{Zhou et~al.(2013)Zhou, Walsh \& Laidlaw}]{Zhou13a}
\bibinfo{author}{Zhou, W.}, \bibinfo{author}{Walsh, E.}, \&
  \bibinfo{author}{Laidlaw, D.} (\bibinfo{year}{2013}).
\newblock \bibinfo{title}{{DoubleAx}: In-vivo axon measurement in the human
  corpus callosum using angular double-{PFG MRI}}.
\newblock In {\it \bibinfo{booktitle}{OHBM 2013}\/} (p. \bibinfo{pages}{2222}).

\end{thebibliography}

\end{document}